\documentclass[a4paper,11pt]{article}
\pdfoutput=1

\usepackage{jheppub}
\usepackage[T1]{fontenc}

\usepackage{booktabs,tabularx,xspace}

\usepackage{defs}

\newcolumntype{R}[1]{>{\raggedleft\let\newline\\\arraybackslash\hspace{0pt}}m{#1}}

\usepackage{hyphenat}
\title{An EWPD SMEFT likelihood for the LHC -- and how to improve it with measurements of W and Z boson properties}
\author{Hannes Mildner}
\affiliation{Johannes Gutenberg-Universität Mainz,\\
  Staudingerweg 7, 55128 Mainz, Germany}

\emailAdd{hannes.mildner@cern.ch}

\abstract{This paper presents a computer code for analyzing electroweak precision data
(EWPD) in the framework of the Standard Model Effective Field Theory (SMEFT),
highlights the importance of recent ATLAS and CMS precision measurements, and
introduces a novel analysis of the forward--backward asymmetry at the LHC. The
computer code provides the likelihood of SMEFT Wilson coefficients based on
precision measurements of $W$ and $Z$ pole observables, interpolation formulas
for Standard Model predictions, and modular SMEFT parametrizations. SMEFT
predictions including next-to-leading-order (NLO) effects in perturbative and
SMEFT expansion are available and five alternative electroweak input parameter
schemes are supported. The likelihood addresses shortcomings of previous
formulations in the treatment of parametric uncertainties and can be
straightforwardly included in SMEFT fits of LHC data. The input parameter scheme
dependence and role of NLO corrections is studied for the EWPD fit in the SMEFT.
Furthermore, the impact of recent ATLAS and CMS measurements -- of
the $W$ boson mass and width, of the lepton flavour universality (LFU) of $W$
branching fractions, and the effective leptonic weak mixing angle, \st\ -- is
analyzed. A test of LFU that surpasses the precision of existing
measurements is proposed based on the \st measurement. Finally, an ATLAS Drell--Yan
triple-differential cross-section measurement is reinterpreted in the SMEFT and
combined with the EWPD likelihood. This analysis demonstrates the feasibility
of the LFU precision test, improves constraints on muon couplings with respect
to the world average, and determines a combination of the quark-coupling
asymmetry parameters $A_u$ and $A_d$ with a precision comparable to that of the heavy
flavour parameters $A_c$ and $A_b$.
}
\begin{document}
\maketitle
\section*{Introduction}

The Standard Model (SM) of particle physics is an elegant, predictive, and very
successful theory. However, it cannot explain certain phenomena, e.g., dark
matter or neutrino masses, suggesting the existence of  physics beyond the SM (BSM).
One of the main goals of the LHC is to discover new particles and with them
direct proof of BSM physics. So far this search has been
unsuccessful, a likely explanation being that the energy required to produce
them in sufficient numbers is beyond the reach of the LHC. However, it may be
possible to observe new particles indirectly, as their existence will subtly
influence the production of known particles at lower energies, resulting in
systematic deviations from SM predictions.

The Standard Model Effective Field Theory (SMEFT~\cite{Buchmuller:1985jz,Grzadkowski:2010es}, see e.g.\
Ref.~\cite{Brivio:2017vri} for a comprehensive review) systematically
parametrizes these potential deviations. It provides predictions for
experimental observables in terms of an expansion in $E/\Lambda$ and $v/\Lambda$,
where $E$ is the typical energy exchanged in the process, $v$ the Higgs field's vacuum expectation value,
and $\Lambda$ the scale of BSM physics.
This is achieved by extending the SM Lagrangian by a series of
operators ${\mathcal{O}_i^{(d)}}$ that consist of gauge invariant combinations
of SM fields with an energy dimension $d$ greater than four:
\begin{equation}
	\label{eq:smeft}
	\mathcal{L}_\textrm{SMEFT} = \mathcal{L}_\textrm{SM} + \sum\limits_i \frac{c^{(5)}_i}{\Lambda} \mathcal{O}_i^{(5)} +  \sum\limits_i \frac{c^{(6)}_i}{\Lambda^2} \mathcal{O}_i^{(6)} + \dots\,.
\end{equation}
These operator are multiplied by dimensionless Wilson coefficients $c_i^{(d)}$,
which are, along with $\Lambda$, the unknown parameters of the theory and reflect
the strength of BSM interactions.
At dimension-five there is only one type of operator, which violates lepton number
conservation. The leading effects on collider physics observables are described by
dimension-six operators, while the effect of operators of higher order are
suppressed by increasing powers of $1/\Lambda$.

The measurement of electroweak precision observables (EWPOs), i.e., the $Z$
pole observables precisely measured at LEP and SLD~\cite{ALEPH:2005ab} as well
as measurements of the $W$ boson mass and partial widths, provide important
constraints in the SMEFT. The high precision of measurements translates into
stringent limits on Wilson coefficients. Since EWPO measurements involve
relatively low energy scales $E$ (compared to high-mass searches at the
LHC), higher-order corrections in $E/\Lambda$ to the observables are small.
This makes the leading terms of the SMEFT expansion a good approximation of
many BSM models. The inclusion of these tight and fairly model-independent
constraints in any global SMEFT interpretation is thus crucial.

At the LHC, it is possible to extend and refine electroweak precision data (EWPD),
for example by precisely measuring the effective leptonic weak mixing
angle \st or through the measurement of $W$ boson properties, such as its mass,
width, and branching fractions. In addition to advancing our knowledge of
precision observables, the LHC allows for measurements of electroweak processes
at high energies and studies of processes involving top quarks or Higgs bosons,
potentially revealing anomalies that cannot be detected with EWPD alone.
A thorough understanding of the constraints already imposed by precision data is
essential for evaluating the impact of LHC measurements in constraining or
discovering BSM physics within the SMEFT framework. It sharpens the focus of LHC
data analysis on effects not already excluded by previous experiments.

To facilitate the combined analysis of LHC and electroweak precision data, a
computer code that calculates the likelihood of Wilson coefficients based on
precision measurements is presented in Section~\ref{sec:likelihood} of this
paper.
Section~\ref{sec:impact} discussed the impact of recent LHC measurements
-- of the $W$ boson mass and width~\cite{ATLAS:2024erm}, the lepton flavour
universality (LFU) of $W$ branching fractions~\cite{ATLAS:2024tlf}, and the
effective leptonic weak mixing angle~\cite{CMS:2024ony} -- on this likelihood.
It is demonstrated that the measurement of the the effective leptonic weak
mixing angle, which is based on the forward-backward asymmetry in Drell--Yan
events, can be reinterpreted as one of the most precise LFU tests to date. To
investigate the importance of forward--backward asymmetry measurements more
thoroughly, an ATLAS Drell--Yan triple-differential cross-section
measurement~\cite{ATLAS:2017rue} is interpreted in a more general SMEFT
framework in Section~\ref{sec:Z3D}. This interpretation provides more accurate
limits on lepton couplings than the estimates obtained in the previous section
and also constrains quark couplings.

\section{An EWPD SMEFT likelihood for the LHC}
\label{sec:likelihood}

The likelihood of electroweak precision data as a function of SMEFT Wilson
coefficients, has been formulated in various ways in the
literature~\citep{Han:2004az,Pomarol:2013zra,Falkowski:2014tna,Efrati:2015eaa,Berthier:2015gja,deBlas:2017wmn,daSilvaAlmeida:2018iqo,Biekotter:2018ohn,Aebischer:2018iyb,Ellis:2018gqa,Falkowski:2019hvp,Ellis:2020unq,Corbett:2021eux,Bellafronte:2023amz,Celada:2024mcf}.
This paper introduces a new \texttt{python} tool named
\ewpdlhc\footnote{Available at
	\href{https://github.com/ewpd4lhc/ewpd4lhc}{https://github.com/ewpd4lhc/ewpd4lhc}}
that is specifically designed for the combination with SMEFT interpretations of
LHC data, like the ATLAS and CMS global EFT
fits~\cite{ATLAS:2022xyx,CMS:2024kvw}. The tool is easily configurable and
returns numerical outputs in both text file format and as a
\texttt{Roofit}~\cite{Verkerke:2003ir} workspace, the latter being the primary
format for combining likelihoods from the LHC experiments.

It provides SM and SMEFT predictions not only in the \{$\alpha$,$M_Z$,$G_\mu$\}
scheme used in most existing EWPD analyses but also in four alternative
electroweak input parameter schemes, as listed in Table~\ref{tab:schemes}.
Schemes using \MW as an input parameter, in particular the
\{$M_W$,$M_Z$,$G_\mu$\} scheme, are preferred for interpretations of LHC
data~\cite{Brivio:2021yjb}.

SM predictions of observables in all schemes are dynamically calculated via
interpolation formulas based on state-of-the art theory predictions. The tool
accurately models the correlated impact of input parameter uncertainties on SM
predictions, which are substantial in schemes treating the less precisely known
observables \MW or \st as input parameters.

Baseline parametrizations are based on
\texttt{SMEFTsim}~\cite{Brivio:2017btx,Brivio:2020onw}, with the option to
include contributions quadratic in dimension-six Wilson coefficients,
consistent with analyses from the ATLAS and CMS collaborations.
Next-to-leading-order (NLO)
perturbative~\cite{Dawson:2019clf,Dawson:2022bxd,Bellafronte:2023amz,Biekotter:2023xle,Biekotter:2023vbh}
corrections in the SMEFT as well as dimension-eight SMEFT
contributions~\cite{Helset:2020yio,Corbett:2021eux} are available, too.
Perturbative NLO corrections are in particular relevant for LHC analyses as
large contributions from top quark operators arise at loop level, making EWPD
constraints as important as measurements of top quark production at the
LHC~\cite{Liu:2022vgo}. The three types of parametrizations are listed in
Table~\ref{tab:schemes}, too. For all parametrizations, notations and operator
definitions are in line with \texttt{SMEFTsim} conventions.

SMEFT parametrizations included in the tool are compatible with the symmetry
assumptions favoured for LHC analyses: fully flavour symmetric scenarios as
well as the $U(2)_q\times U(2)_u \times U(2)_d$ scenario for top
physics~\cite{Aguilar-Saavedra:2018ksv}, both with and without lepton flavour
universality assumption. Available symmetry assumptions are also listed in
Table~\ref{tab:schemes}.

\begin{table}
	\centering
	\begin{tabular}{@{}lll@{}}
		\toprule
		\ewpdlhc option  & Internal notation       & Explanation                                                               \\
		\midrule
		Input scheme     & \texttt{alpha}          & \{$\alpha$,$M_Z$,$G_\mu$\}                                                \\
		                 & \texttt{MW}             & \{$M_W$,$M_Z$,$G_\mu$\}                                                   \\
		                 & \texttt{alphaMW}        & \{$\alpha$,$M_W$,$M_Z$\}                                                  \\
		                 & \texttt{sin2theta}      & \{$\st$,$M_Z$,$G_\mu$\}                                                   \\
		                 & \texttt{alphasin2theta} & \{$\alpha$,$\st$,$M_Z$\}                                                  \\
		\midrule
		Parametrizations & \texttt{SMEFTsim}       & Linear and quadr. dim.-six dependence~\cite{Brivio:2020onw}               \\
		                 & \texttt{EWPDatNLO}      & Linear and quadr. dim.-six dependence, NLO~\cite{Biekotter:2023vbh}       \\
		                 & \texttt{EWPD2dim8}      & Full $O(\Lambda^{-4})$ dependence, incl dim. eight~\cite{Corbett:2021eux} \\
		\midrule
		Symmetries       & \texttt{general}        & General but massless light quarks, diag. CKM                              \\
		                 & \texttt{top}            & $U(2)_q\times U(2)_u \times U(2)_d$                                       \\
		                 & \texttt{topU3l}         & $U(2)_q\times U(2)_u \times U(2)_d\times U(3)_l \times U(3)_e$            \\
		                 & \texttt{U35}            & $U(3)_q\times U(3)_u \times U(3)_d\times U(3)_l \times U(3)_e$            \\
		\bottomrule
	\end{tabular}
	\caption{Electroweak input parameter schemes, SMEFT parametrizations, and symmetry assumptions included in \ewpdlhc. The electroweak parameters are the fine-structure constant $\alpha$, the $W$ and $Z$ boson masses, \MW and \MZ, the Fermi constant from muon decays, \Gmu, and the effective weak mixing angle, \st. The symmetry assumptions are described in detail in Ref.~\cite{Brivio:2020onw}.}
	\label{tab:schemes}
\end{table}

Section~\ref{sec:observables} introduces precision $W$ and $Z$ pole observables
in \ewpdlhc and their prediction in the SM is described in
Section~\ref{sec:predictions}. The modelling of the effect of uncertainties in
input parameters as well as theory uncertainties is discussed in
Section~\ref{sec:uncertainties}. The implementation of the uncertainty model is
validated by performing an electroweak fit described in
Section~\ref{sec:SMfit}. Section~\ref{sec:SMEFTpara} outlines the derivation of
SMEFT corrections to the SM predictions and a fit of EWPD in the SMEFT is
presented in Section~\ref{sec:SMEFTfit}, which is compared to existing fits in
Section~\ref{sec:smefit}.

\subsection{Input observables}
\label{sec:observables}
The \ewpdlhc code incorporates observables sensitive to couplings of the $Z$ and $W$ boson to all lepton flavours, charm quarks, and bottom quarks.
It also considers the total width and hadronic branching fractions of the $W$ and $Z$ bosons, which constrain combinations of fermion couplings.
However, observables that distinguish the three lightest quark flavours are not included due to insufficient measurement precision.
Measurements of \st derived from the charge asymmetry $Q^\text{had}_{FB}$ at LEP and the forward--backward asymmetry at hadron colliders, \AFB, are also not considered due to their model-dependent extraction, requiring for example SM-like quark couplings. A more accurate interpretation of this type of measurement is discussed in Section~\ref{sec:Z3D}.

The (pseudo) observables \GammaZ, \sigmahad, \Rl (replaced by \Rel, \Rmu, and
\Rtau in SMEFT scenarios without LFU), \Rc, and \Rb, constrain combinations of
$Z$ boson partial widths:
\begin{equation}
	\sigmahad=\frac{12\pi}{\MZ^2}\frac{\Gamma_{Z\rightarrow e^+e^-}\Gamma_{Z\rightarrow \textrm{had}}}{\GammaZ^2}, \ \Rl=\frac{\Gamma_{Z\rightarrow \textrm{had}}}{\Gamma_{Z\rightarrow \ell^+\ell^-}}, \ \Rb=\frac{\Gamma_{Z\rightarrow b\bar b}}{\Gamma_{Z\rightarrow \textrm{had}}}, \ \Rc=\frac{\Gamma_{Z\rightarrow c\bar c}}{\Gamma_{Z\rightarrow \textrm{had}}}\,.
\end{equation}
They are sensitive to the combined effect of the $Z$ boson couplings to left-handed and right-handed fermions (or equivalently, vector and axial vector couplings)
and have been precisely measured at LEP and SLD~\cite{ALEPH:2005ab}. The values
used in \ewpdlhc include an updated estimate of the LEP
luminosity~\cite{Janot:2019oyi}, as recommended by the particle data group
(PDG)~\cite{ParticleDataGroup:2024cfk}.

The results of asymmetry measurements utilizing the polarized beams of SLC --
\AlSLD (\AeSLD, \AmuSLD, and \AtauSLD if not assuming LFU), \Ac, and \Ab\ --
as well as $\tau$ polarization measurements at LEP, \AlLEP (\AeLEP and
\AtauLEP), are also included for both lepton flavour universal and
non-universal cases. These measurements allow distinguishing couplings of
left-handed and right-handed fermions, as the following relationship holds:
\begin{equation}
	A_f =  \frac{\Gamma_{Z\rightarrow f_L\bar f_L} - \Gamma_{Z\rightarrow f_R\bar f_R}}{\Gamma_{Z\rightarrow f\bar f}}\,.
\end{equation}
The forward--backward asymmetries on the $Z$ peak measured by LEP, \AFBl (\AFBe,  \AFBmu, and \AFBtau without LFU), \AFBc, and \AFBb also serve this purpose.
They are related to the asymmetry parameters $A_f$ by:
\begin{equation}
	A^{0,f}_\textrm{FB} = \frac{3}{4}A_fA_e\,.
	\label{eq:AFBdef}
\end{equation}

The PDG values of $W$ boson branching fractions, \BrWhad, (\BrWe, \BrWmu, and
\BrWtau if not assuming LFU), based on LEP, and the total width, derived from a
LEP+Tevatron combination, \GammaW, are used to constrain the $W$ couplings to
(left-handed) fermions. An LHC average for ratios of $W$ branching fractions,
\begin{equation}
	\RWmue=\frac{\BrWmu}{\BrWe},\ \RWtaue=\frac{\BrWtau}{\BrWe},\ \RWtaumu=\frac{\BrWtau}{\BrWmu},
	\label{eq:RWdef}
\end{equation}
which excludes contributions from LEP (and Tevatron) data but is  based on
ATLAS~\cite{ATLAS:2016nqi,ATLAS:2020xea}, CMS~\cite{CMS:2022mhs}, and LHCb~\cite{LHCb:2016zpq}
measurements, is also included. The role of the latest ATLAS measurement of LFU
in $W$ boson decays~\cite{ATLAS:2024tlf} is more subtle and discussed in Section~\ref{sec:impact}.

To predict these observables in the SM, knowledge of three electroweak input
parameters is required, usually taken to be three of the set \{$\alpha$,
$G_\mu$, $M_W$, $M_Z$, \st\}. They enable the calculation of all other
electroweak parameters. In the SMEFT, the remaining parameters are not solely
determined by the SM input parameter values but also depend on Wilson
coefficients. Like any other observable, their measurement can thus provide
constraints on BSM physics. PDG values are used for \MW (which currently
excludes the CDF result~\cite{CDF:2022hxs} and does not yet incorporate the
preliminary CMS result~\cite{CMS:2024nau}), \MZ, the Fermi coupling determined
from muon decays, $G_\mu$, and the fine-structure constant $\alpha(Q^2\!=\!0)$.

While the electromagnetic coupling at low scales, $\alpha(Q^2\!=\!0)$, is known
precisely, it is the electromagnetic coupling at the $Z$ boson energy scale,
$\alpha(Q^2=M_Z^2)$, that is required for the prediction of $W$ and $Z$ pole
observables, including \MW. The running of the coupling up to $Q^2=M_Z^2$
cannot be reliably calculated using perturbative QCD alone, due to the
existence of hadronic resonances in the intermediate $Q^2$ region. Instead,
$\Deltaalpha=1-\frac{\alpha(Q^2\!=\!0)}{\alpha(Q^2\!=\!M_Z^2)}$ is typically
determined using additional experimental inputs. By default \ewpdlhc uses the
value of Ref.~\cite{Davier:2019can} in combination with the theoretical
calculation of the leptonic contributions~\cite{Steinhauser:1998rq}. In schemes
using \MW or \st instead of $\alpha$ as an input, $\alpha(Q^2=M_Z^2)$ is
predicted by the SM. In that case \ewpdlhc treats \Deltaalpha\ -- implicitly
combined with $\alpha(Q^2\!=\!0)$, which carries negligible uncertainty -- as
an observable. This approach has recently been explored in
Ref.~\cite{Celada:2024cxw}, too, albeit missing the parametric uncertainty
discussed below. An alternative approach, used in~\cite{Brivio:2017bnu},
involves using the value of $\alpha(Q^2\!=\!M_Z^2)$ implied by Bhabha
scattering measurements at LEP, although these measurements are less precise
and introduce additional dependences on four-fermion operators in the SMEFT.

If \st is taken as a SM input, its value is automatically determined in a fit
of all observables that depend on \st but do not carry additional Wilson
coefficient dependence, i.e., \AlSLD, \AlLEP and \AFBl. Observables that depend
on additional couplings, like \AFBb, are not considered at that stage. If LFU
is not assumed, the input parameter is the effective weak mixing angle for
electrons, \ste, determined by \AeSLD, \AeLEP, and \AFBe.

The strong coupling \alphas, also needed for the prediction of precision
observables, is set to the average of the Flavour Lattice Averaging Group
(FLAG)~\cite{FlavourLatticeAveragingGroupFLAG:2021npn}, rather than the PDG
value, as the lattice extraction is more robust against SMEFT
effects~\cite{Trott:2023jrw}.

The Higgs boson and top quark masses, \MH and \mt, contribute to SM predictions
of EWPOs at loop level. The values used in \ewpdlhc correspond to the PDG
average, with an additional 0.5\,\GeV uncertainty on \mt, to account for
ambiguities in the top mass definition~\cite{Hoang:2020iah}.

Correlations in the measurements of \{\MZ, \GammaZ, \sigmahad, \Rl, \AFBl\!\}
(or their counterparts without assuming LFU), \{\AeSLD, \AmuSLD, \AtauSLD\},
\{\AeLEP, \AtauLEP\!\}, \{\Rb, \Rc, \AFBb, \AFBc, \Ab, \Ac\!\}, and \{\BrWe,
\BrWmu, \BrWtau\}, are taken into account within each group indicated by
braces.

The \ewpdlhc tool enables users to transparently set the measured values of
observables, including SM input parameters, and their correlation in a
\texttt{yaml} text file. A subset of observables can be chosen in the main
configuration file. The default central values can be found in
Table~\ref{tab:smfit} (Table~\ref{tab:smeftfit}) assuming (not assuming) LFU,
alongside the fit results discussed in the following sections.

\subsection{SM predictions}
\label{sec:predictions}

Precise SM predictions for the above observables are calculated, typically with
at least two-loop accuracy, using interpolation formulas. These predictions are
later combined with lower-accuracy predictions of the Wilson-coefficient
dependent corrections within the SMEFT framework.

Central predictions for \Rl, \Rc, \Rb, \st, and \stb are calculated from the SM
input observables \MZ, \MH, \mt, \alphas, and \Deltaalpha, using the formulas
of Ref.~\cite{Dubovyk:2019szj}. A correction that accounts for the small impact
of \Gmu variations is implemented at one-loop level. For \stc the formulas of
Ref~\cite{Awramik:2006uz} are used. The effective weak mixing angles \st as
well as \stb and \stc determine the values of \Al, \AFBl, \Ab, \Ac, \AFBc, and
\AFBb, using
\begin{equation}
	A_f = \frac{1-4|Q_f|\stf}{1-4|Q_f|\stf+8(Q_F\stf)^2}
	\label{eq:Adef}
\end{equation}
and Equation~\ref{eq:AFBdef}.
The $W$ boson mass is predicted as per Ref.~\cite{Awramik:2003rn} while a one-loop prediction for the width \GammaW is taken from~\cite{Cho:2011rk}.

To obtain predictions using the four alternative input parameter sets listed in
Table~\ref{tab:schemes}, either $\alpha$ or \Gmu is substituted for \MW or \st
as inputs to the interpolation formulas, following the methodology of
Ref.~\cite{Brivio:2017bnu}. Although these predictions still rely on the
\{$\alpha$,$M_Z$,$G_\mu$\} scheme for SM calculations, this approach enables
the use of different input parameter sets for SMEFT analyses, where alternative
configurations provide advantages~\cite{Brivio:2021yjb}. For instance, to
calculate predictions using the set of \{$M_W$,$M_Z$,$G_\mu$\}, the
relationship $\MW(\Delta\alpha,\dots)$ given in Ref.~\cite{Awramik:2003rn} is
inverted to obtain $\Delta\alpha(\MW,\dots)$, which is subsequently substituted
into the interpolation formulas. Similarly, the inversion of the \MW--\Gmu,
\st--$\alpha$, or \st--\Gmu relationships enables the code to provide
predictions in the remaining three schemes of Table~\ref{tab:schemes}.

\subsection{Treatment of uncertainties}
\label{sec:uncertainties}
The \ewpdlhc model of the likelihood $L$ is based on a multivariate Gaussian distribution,
\begin{equation}
	-2\log L = \chi^2=(\mathbf{\Delta x}^\top V^{-1}\mathbf{\Delta x}),
	\label{eq:likelihood}
\end{equation}
where
\begin{equation}
	\Delta \mathbf x=\mathbf x_\textrm{meas}-\mathbf x_\textrm{pred}
	\label{eq:diff}
\end{equation}
represents the difference between the measured values $\mathbf x_\textrm{meas}$ and predicted values $\mathbf x_\textrm{pred}$ of observables.
The prediction is a combination of the SM prediction $\mathbf x^\text{SM}_\textrm{pred}$ and Wilson-coefficient-dependent SMEFT corrections.
The covariance matrix $V$ encodes uncertainties.

Uncertainties arise not only in the measurement of observable but also in their
prediction. These include ``parametric uncertainties'' in the SM prediction,
which stem from uncertainties in input parameters, as well as ``theory
uncertainties'', which arise for example due to missing higher-order
corrections in theoretical calculations.

The \ewpdlhc tool provides two options to account for parametric uncertainties.
The default method is to include their effect in the covariance $V$, along with
the uncertainty of measurements. In this approach SM predictions $\mathbf
	x^\textrm{SM}_\textrm{pred}$ are fixed to the values implied by the nominal
input parameter values, $\mathbf x^\textrm{in}_\textrm{meas}$. The alternative
is to model the dependence of $\mathbf x^\textrm{SM}_\textrm{pred}$ on SM input
parameters with nuisance parameters $\mathbf x^\textrm{in}$ that are part of
the likelihood, so that $\mathbf x^\textrm{SM}_\textrm{pred}\equiv \mathbf
	x^\textrm{SM}_\textrm{pred}(\mathbf x^\textrm{in})$.

In the input-parameter-free likelihood approach, the the method most commonly
used in analyses of EWPD to date, the difference between measurement and SM
prediction, $\Delta \mathbf x^\textrm{SM}=\mathbf x_\textrm{meas}-\mathbf
	x^\textrm{SM}_\textrm{pred}$, is treated as the random variable within the
multivariate Gaussian model. In this framework, parametric uncertainties of
$\mathbf x^\textrm{SM}_\textrm{pred}$ must be incorporated into the covariance
matrix $V$. This becomes particularly important when \MW or \st are treated as
input parameters, as the parametric uncertainty in the prediction of some
observables can then become similar to or even greater than the uncertainty
associated with the direct measurements.

\begin{figure}
	\includegraphics[width=0.5\textwidth]{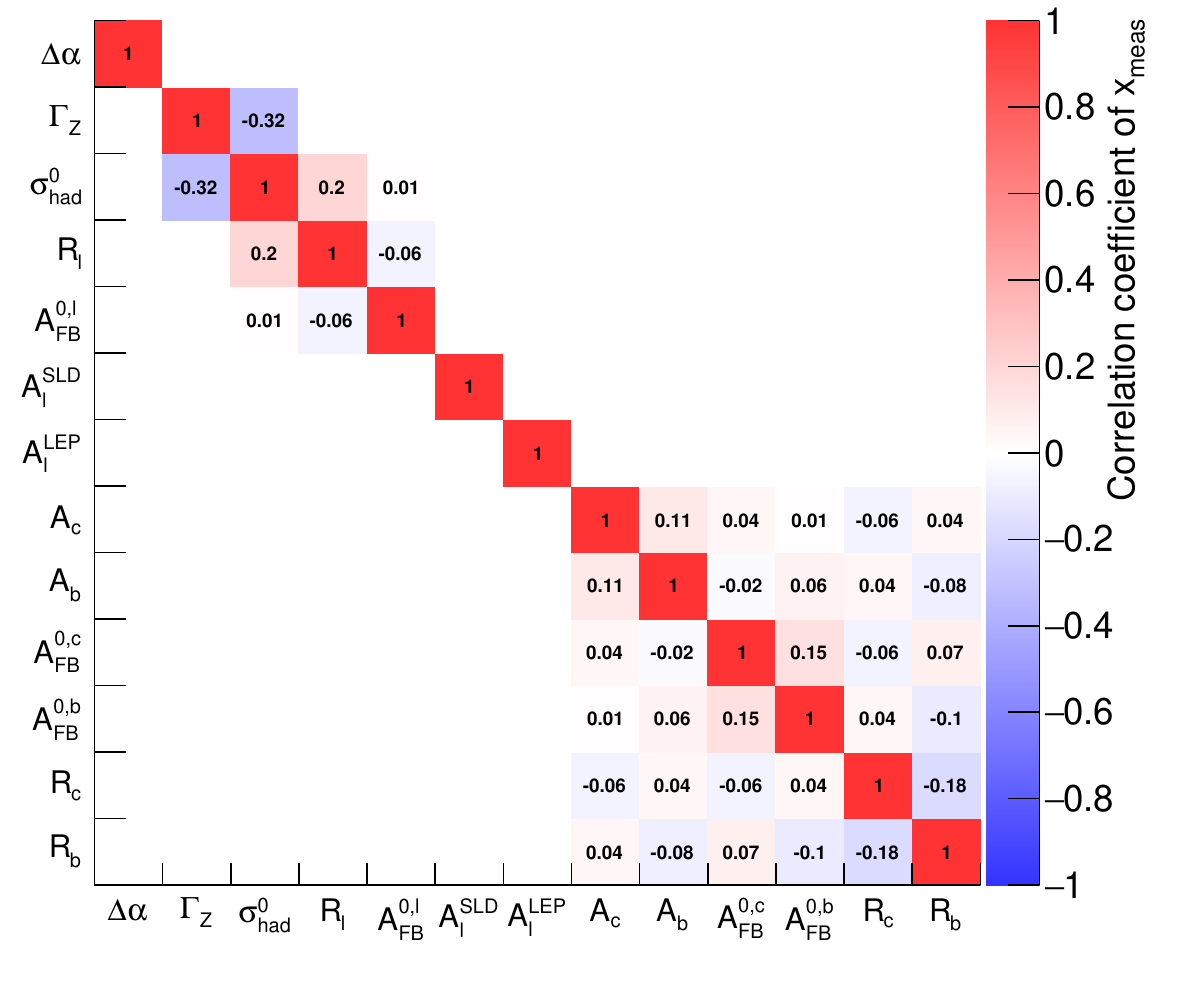}%
	\includegraphics[width=0.5\textwidth]{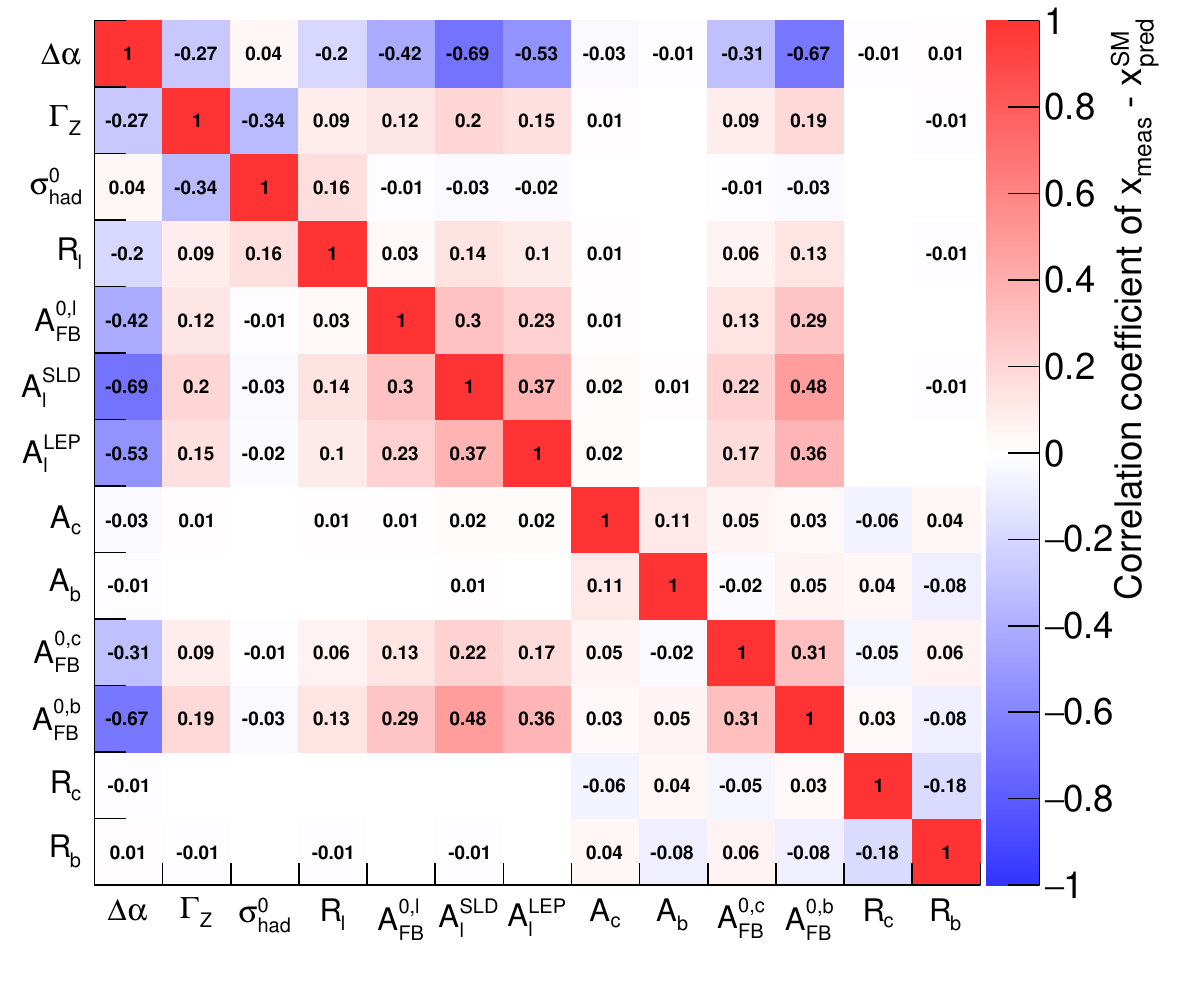}
	\caption{Comparison of the correlation coefficient of various $Z$-pole measurements and \Deltaalpha (left) with the correlation coefficient of the difference of measurement and SM prediction, $\Delta \mathbf x^\textrm{SM}=\mathbf x_\textrm{meas}-\mathbf x^\textrm{SM}_\textrm{pred}$, in the  \{$M_W$,$M_Z$,$G_\mu$\} input parameter scheme (right). The latter is particularly important in the SMEFT analysis, where differences between measurement and prediction may indicate potential signs of BSM physics.}
	\label{fig:correlation}
\end{figure}

The impact of this parametric uncertainties on the prediction of EWPOs is
highly correlated. To illustrate the extent of this correlation, one million
pseudo data sets were generated. These sets include $\Deltaalpha$, $Z$ pole
observables, as well as the SM inputs, taken to be \{$M_W$,$M_Z$,$G_\mu$\},
were sampled randomly according to their experimental covariance. SM
predictions for each pseudo data set were calculated using the formulas
introduced in Section~\ref{sec:predictions}. Figure~\ref{fig:correlation}
presents the sample correlation coefficient for measurements $\mathbf
	x_\textrm{meas}$ and the difference between measurement and prediction, $\Delta
	\mathbf x^\textrm{SM}=\mathbf x_\textrm{meas}-\mathbf
	x^\textrm{SM}_\textrm{pred}$. The correlation of uncertainties in $\Delta
	\mathbf x^\textrm{SM}$ is significantly higher than that of the measured values
$\mathbf x_\textrm{meas}$ alone, highlighting the importance of an accurate
correlation model.

The \ewpdlhc code employs linear error propagation to construct the covariance
matrix rather than relying on the pseudo experiments used to generate
Figure~\ref{fig:correlation}. The contribution to the covariance from
parametric uncertainties, $V^\textrm{param}$, is calculated as
\begin{equation}
	V^\textrm{param}_{ij}=\sum_k
	\frac{\partial x^\textrm{SM}_{\textrm{pred},i}}{\partial x_{k}^\textrm{in}}|_{{x_{k}^\textrm{in}}=x^{\textrm{in}}_{\textrm{meas},k}}
	\times\frac{\partial x^\textrm{SM}_{\textrm{pred},j}}{\partial x_{k}^\textrm{in}}|_{{x_{k}^\textrm{in}}=x^\textrm{in}_{\textrm{meas},k}}
	\times\sigma_{\textrm{meas},k}^2\,.
\end{equation}
Here, $\sigma_{\textrm{meas},k}$ represents the uncertainty in the measured value $x^\textrm{in}_{\textrm{meas}}$.
The Jacobian matrix $\frac{\partial \mathbf x^\textrm{SM}_\textrm{pred}}{\partial \mathbf x^\textrm{in}}$, which is implemented in \ewpdlhc for all input parameters schemes offered, is evaluated dynamically based on the provided measurement values.
The analytically derived covariance shows excellent agreement with the one in Figure~\ref{fig:correlation}, confirming the correct technical implementation and indicating that the impact of uncertainty propagation beyond the linear contribution is negligible.

\begin{figure}
	\includegraphics[width=0.5\textwidth]{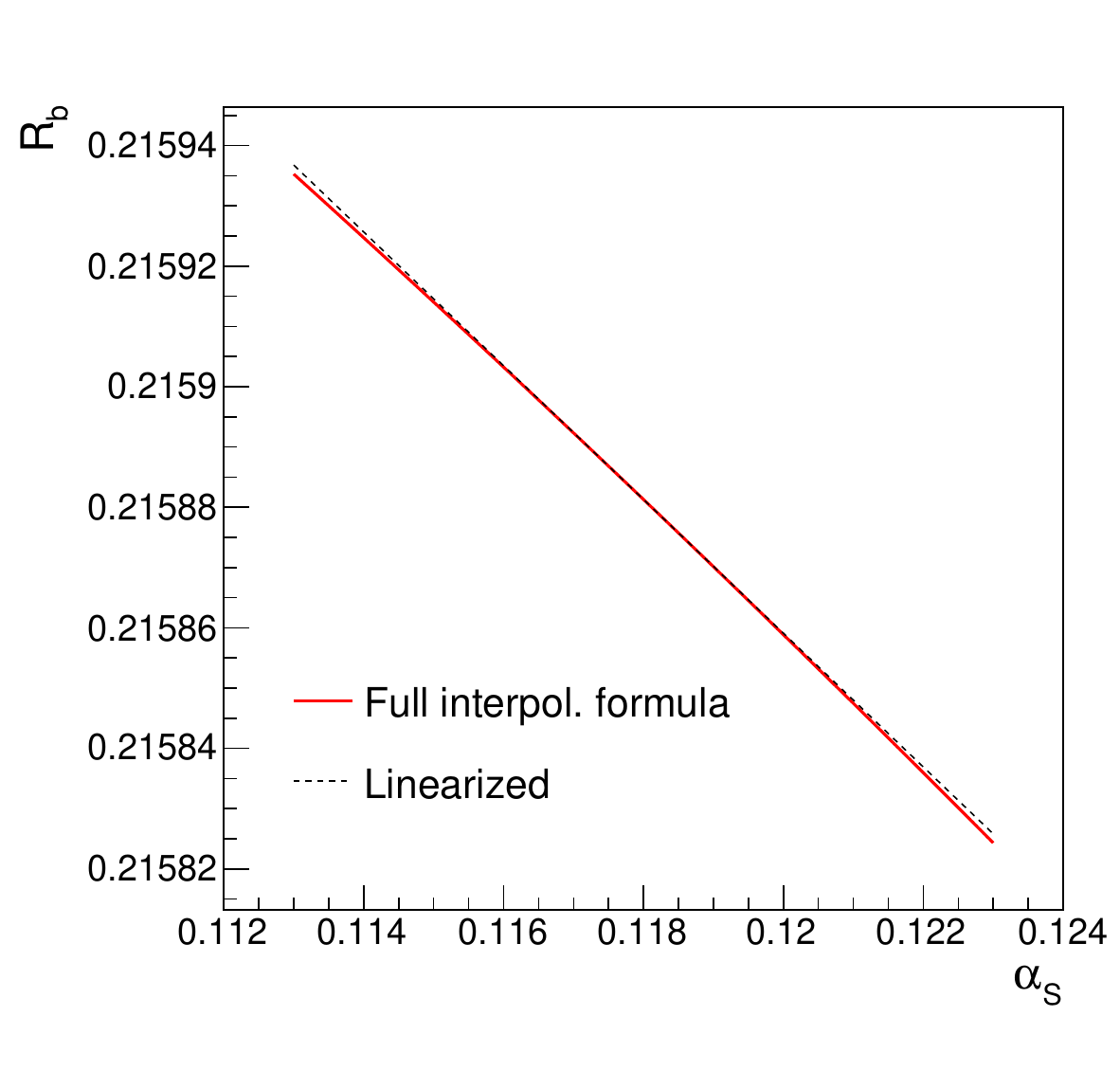}%
	\includegraphics[width=0.5\textwidth]{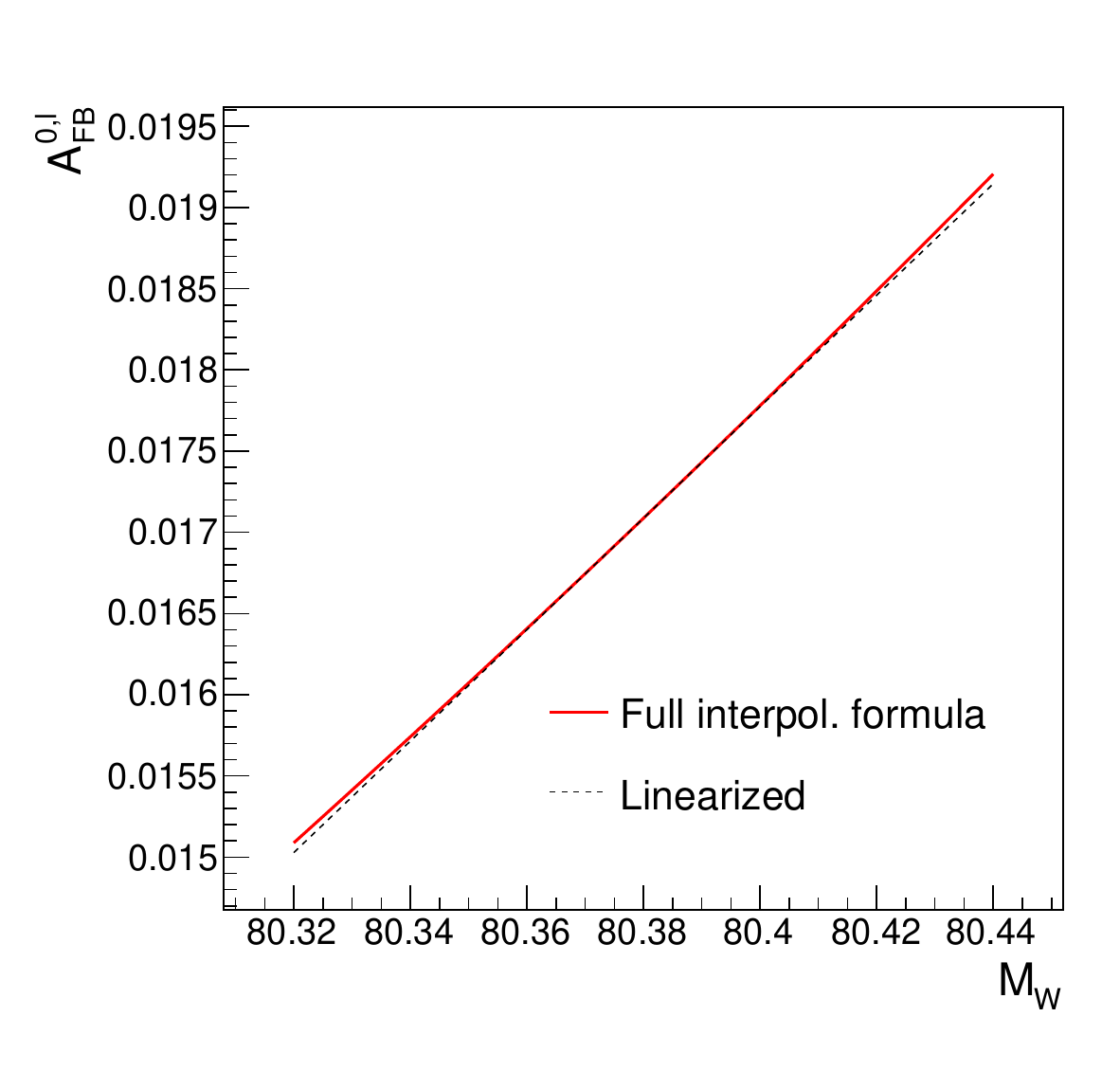}
	\caption{Examples of the dependence of EWPO predictions on experimental inputs. The plots show \Rb (left) and \AFBl (right) as functions of the inputs \alphas and \MW, respectively, using the full interpolation formulas from Ref.~\cite{Dubovyk:2019szj} and their linear approximation around the central prediction, as implemented in the code presented in this paper. The input quantities \alphas and \MW are varied by about five standard deviations around the values determined from direct measurements.}
	\label{fig:lin}
\end{figure}

As an alternative to the nuisance-parameter-free scheme, the likelihood of
Equation~\ref{eq:likelihood} can be formulated to include the SM input
observables, $\mathbf x^\textrm{in}$, in the observables $\mathbf x$ explicitly
considered in the multivariate Gaussian model. In that case, $\mathbf
	x_\textrm{pred}$ of Equation~\ref{eq:diff} is a function of both SM parameters
and Wilson coefficients, where this dependence is trivial for SM input
parameters: $\mathbf x^{\mathrm{in}}_\textrm{pred}=\mathbf x^\textrm{in}$. In
this alternative approach the random variable of the multivariate Gaussian is
$\mathbf x_\textrm{meas}$, not the difference between measurement and SM
prediction $\Delta \mathbf x^\textrm{SM}$. Given that the linear approximation
provides a sufficiently accurate model of uncertainties, a linear version of
the full interpolation formulas for SM predictions is implemented, also
utilizing the Jacobian matrix $\frac{\partial \mathbf
		x^\textrm{SM}_\textrm{pred}}{\partial \mathbf x^\textrm{in}}$:
\begin{equation}
	\mathbf x^\textrm{SM}_\textrm{pred}(\mathbf x_\textrm{in})=\mathbf x^\textrm{SM}_\textrm{pred}(\mathbf x^\textrm{in}_\textrm{meas})+\left(\mathbf x_\textrm{in}-\mathbf x^{\textrm{meas}}_\textrm{in}\right)\frac{\partial \mathbf x_\textrm{pred}^\textrm{SM}}{\partial \mathbf x_\textrm{in}}|_{\mathbf x_\textrm{in}=\mathbf x^{\textrm{meas}}_\textrm{in}}\,.
	\label{eq:lin}
\end{equation}
The linearized formula simplifies the SM fit and enables a stable, efficient combined fit of SM parameters and Wilson coefficients in the SMEFT.
The validity of the linear approximation is confirmed by comparison with the full interpolation formulas, with two examples illustrating relatively large (yet still negligible) non-linear contributions shown in Figure~\ref{fig:lin}.
Indeed, when the inputs to the interpolation formulas are varied within three standard deviations of their measured values, only the \alphas and \st dependence in certain predictions deviates by more than 2\% from the linear approximation.
The impact of \alphas variations on the SMEFT fit is minimal, and the non-linear dependence on \st is significant only when \st is treated as an input parameter.

As with parametric uncertainties, two options are provided for addressing
theory uncertainties. They can either be added directly to the covariance
matrix or be incorporated in the parametrization, in which case Gaussian
constraint terms are added for each nuisance parameter to the likelihood.
Theoretical uncertainties are accounted for in \MW (4 \MeV, from
Ref.~\cite{Awramik:2003rn}), \Rl, \Rc, \Rb, and \st (see
Ref.~\cite{Dubovyk:2019szj}). The uncertainties in \stb and \stc prediction are
negligible compared to the experimental precision of measurements. There is no
theory uncertainty for \MW or \st if used as input. Instead the replaced input
parameters \Deltaalpha or \Gmu have uncertainty of
$\sigma^\text{theo}_\text{old~inp.}=\frac{\partial x_\text{old~inp.}}{\partial x_\text{new~inp.}}\sigma^\text{theo}_\text{new~inp.}$, where
$\sigma^\text{theo}_\text{new~inp.}$ is the \MW or \st theory uncertainty in
the original \{$\alpha$,$M_Z$,$G_\mu$\} scheme calculation.

\subsection{The EWPD fit in the SM}
\label{sec:SMfit}

The linear parametrization of SM relations (Equation~\ref{eq:lin}) enables the
algebraic minimization of Equation~\ref{eq:likelihood}, thus solving the SM
fit. This fit is performed automatically when running the \ewpdlhc code. Free
parameters of the fit are \MZ, \MH, \mt, \alphas, and -- depending on the
electroweak input parameter scheme chosen -- two of \Deltaalpha, \MW, \Gmu,
\st, as well as nuisance parameters modelling theory uncertainties. The results
of fits using different input parameter sets show only marginal differences,
due to the differing role of the theoretical uncertainties on \MW and \st.

Fit results for the default configuration and using \Deltaalpha and \Gmu as
input, are presented in Table~\ref{tab:smfit}. The only significant deviations
of direct measurements from the global fit results are the well-known tensions
in \AFBb and \AlSLD measurements.

The results are in good agreement with established codes for the SM fit. For
instance, the latest \texttt{Gfitter} result~\cite{Haller:2022eyb} predicts
$\alphas = 0.1198 \pm 0.0029$, in good agreement with this result, $\alphas =
	0.1202 \pm 0.0028$. Half of the 2\,\MeV difference to the indirect
\texttt{Gfitter} $W$ mass prediction, $\MW = 80354 \pm 7~\MeV$, can be
attributed to a different value for \Deltaalpha. The \texttt{Gfitter} Higgs
mass $\MH = 100^{+25}_{-21}~\GeV$, has slightly smaller uncertainties than
presented here, mainly because non-linear effects in the \MH predictions become
important when the direct \MH constraint is ignored. The exact cause for the
one \GeV difference to the top mass prediction by \texttt{Gfitter}, $\mt=
	175.15^{+2.37}_{-2.39} \GeV$ could not be identified, part of it is related to
the small differences observed for \MW.

The indirect prediction of SM parameters (which have all been precisely
measured for more than a decade) is not the main use case of this tool, which
is designed to facilitate accurate SMEFT fits. The good agreement with more
sophisticated tools still demonstrates that it can also be useful for the
calculation of SM predictions and to obtain instant results of the electroweak
fit.

\begin{table}
	\small
	\centering
	\begin{tabular}{@{}lr@{}c@{}lr@{}c@{}lrr@{}c@{}l@{}}
		\toprule
		Observable                 & \multicolumn{3}{@{}c@{}}{Direct} & \multicolumn{3}{@{}c@{}}{Fit} & Pull        & \multicolumn{3}{@{}c@{}}{Indirect}                                                                        \\
		\midrule
		$\MH\,[\GeV]$              & $125.20$                         & $\,\pm\,$                     & $0.11$      & $125.20$                           & $\,\pm\,$ & $0.11$      & $-0.0$ & $106$     & $\,\pm\,$ & $27$      \\
		$\mt\,[\GeV]$              & $172.57$                         & $\,\pm\,$                     & $0.58$      & $172.68$                           & $\,\pm\,$ & $0.56$      & $0.2$  & $174.2$   & $\,\pm\,$ & $2.2$     \\
		$\alphas$                  & $0.11840$                        & $\,\pm\,$                     & $0.00080$   & $0.11854$                          & $\,\pm\,$ & $0.00077$   & $0.2$  & $0.1202$  & $\,\pm\,$ & $0.0028$  \\
		$\Deltaalpha$              & $0.05903$                        & $\,\pm\,$                     & $0.00010$   & $0.05902$                          & $\,\pm\,$ & $0.00010$   & $-0.1$ & $0.05886$ & $\,\pm\,$ & $0.00042$ \\
		$\Gmu\,[10^{-5}\GeV^{-2}]$ & $1.1663788$                      & $\,\pm\,$                     & $0.0000006$ & $1.1663788$                        & $\,\pm\,$ & $0.0000006$ & $0.0$  & $1.16658$ & $\,\pm\,$ & $0.00044$ \\
		$\MZ\,[\GeV]$              & $91.1875$                        & $\,\pm\,$                     & $0.0021$    & $91.1877$                          & $\,\pm\,$ & $0.0020$    & $0.1$  & $91.1960$ & $\,\pm\,$ & $0.0095$  \\
		\hline
		$\MW\,[\GeV]$              & $80.369$                         & $\,\pm\,$                     & $0.013$     & $80.358$                           & $\,\pm\,$ & $0.006$     & $-0.8$ & $80.356$  & $\,\pm\,$ & $0.006$   \\
		$\GammaZ\,[\GeV]$          & $2.4955$                         & $\,\pm\,$                     & $0.0023$    & $2.4947$                           & $\,\pm\,$ & $0.0006$    & $-0.3$ & $2.4947$  & $\,\pm\,$ & $0.0006$  \\
		$\Rl$                      & $20.767$                         & $\,\pm\,$                     & $0.025$     & $20.754$                           & $\,\pm\,$ & $0.007$     & $-0.5$ & $20.752$  & $\,\pm\,$ & $0.008$   \\
		$\Rc$                      & $0.1721$                         & $\,\pm\,$                     & $0.0030$    & $0.1722$                           & $\,\pm\,$ & $0.0001$    & $0.0$  & $0.1722$  & $\,\pm\,$ & $0.0001$  \\
		$\Rb$                      & $0.21629$                        & $\,\pm\,$                     & $0.00066$   & $0.21587$                          & $\,\pm\,$ & $0.00010$   & $-0.6$ & $0.21587$ & $\,\pm\,$ & $0.00010$ \\
		$\sigmahad\,[\pb]$         & $41480$                          & $\,\pm\,$                     & $32$        & $41488$                            & $\,\pm\,$ & $7$         & $0.2$  & $41489$   & $\,\pm\,$ & $7$       \\
		$\AlSLD$                   & $0.1513$                         & $\,\pm\,$                     & $0.0021$    & $0.1475$                           & $\,\pm\,$ & $0.0004$    & $-1.8$ & $0.1474$  & $\,\pm\,$ & $0.0005$  \\
		$\AlLEP$                   & $0.1465$                         & $\,\pm\,$                     & $0.0033$    & $0.1475$                           & $\,\pm\,$ & $0.0004$    & $0.3$  & $0.1476$  & $\,\pm\,$ & $0.0005$  \\
		$\AFBl$                    & $0.0171$                         & $\,\pm\,$                     & $0.0010$    & $0.0163$                           & $\,\pm\,$ & $0.0001$    & $-0.8$ & $0.0163$  & $\,\pm\,$ & $0.0001$  \\
		$\AFBb$                    & $0.0992$                         & $\,\pm\,$                     & $0.0016$    & $0.1031$                           & $\,\pm\,$ & $0.0003$    & $2.4$  & $0.1033$  & $\,\pm\,$ & $0.0003$  \\
		$\AFBc$                    & $0.0707$                         & $\,\pm\,$                     & $0.0035$    & $0.0737$                           & $\,\pm\,$ & $0.0002$    & $0.9$  & $0.0737$  & $\,\pm\,$ & $0.0002$  \\
		$\Ab$                      & $0.923$                          & $\,\pm\,$                     & $0.020$     & $0.935$                            & $\,\pm\,$ & $0.000$     & $0.6$  & $0.935$   & $\,\pm\,$ & $0.000$   \\
		$\Ac$                      & $0.670$                          & $\,\pm\,$                     & $0.027$     & $0.668$                            & $\,\pm\,$ & $0.000$     & $-0.1$ & $0.668$   & $\,\pm\,$ & $0.000$   \\
		$\GammaW\,[\GeV]$          & $2.085$                          & $\,\pm\,$                     & $0.042$     & $2.090$                            & $\,\pm\,$ & $0.000$     & $0.1$  & $2.090$   & $\,\pm\,$ & $0.000$   \\
		$\BrWhad$                  & $0.6741$                         & $\,\pm\,$                     & $0.0027$    & $0.6754$                           & $\,\pm\,$ & $0.0000$    & $0.5$  & $0.6754$  & $\,\pm\,$ & $0.0000$  \\
		\bottomrule
	\end{tabular}
	\caption{Input values and fit results for the SM fit of the first six observables in the table. Uncertainties on the remaining observables are obtained via error propagation. Pull values are calculated as the difference of fit result and direct measurement, divided by the uncertainty of the direct measurement. The column labeled ``indirect'' contains the result of a fit that does not include the direct measurement corresponding to each respective row.}
	\label{tab:smfit}
\end{table}

\subsection{SMEFT parametrization}
\label{sec:SMEFTpara}

Predictions in the SM are extended by a parametrization of the effect of
higher-dimensional operators, to obtain predictions in the SMEFT framework. The
SMEFT contributions to EWPOs can be described, at leading order, by 10 to 23
dimension-six Wilson coefficients, depending on symmetry assumptions. There are
no contributions of dimension-five or dimension-seven operators and higher
orders are strongly suppressed. At next-to-leading order 35 (more than 100)
operators contribute in the flavour universal (general) case, although the
contribution of most of the extra operators, contributing at loop-level, is
relatively small.

The predicted values of observables, $x_{\textrm{pred},i}$, can be decomposed
into the SM prediction, $x_{\textrm{pred},i}^\textrm{SM}$, and a SMEFT
correction, $\Delta_{\textrm{SMEFT},i}$, that depends on Wilson coefficients:
\begin{equation}
	x_{\textrm{pred},i}=  x_{\textrm{pred},i}^\textrm{SM} + \Delta_{\textrm{SMEFT},i} = x_{\textrm{pred},i}^\textrm{SM}+\sum_jA^{(6)}_{ij}\frac{c^{(6)}_j}{\Lambda^2} + \sum_{j,k}B^{(6)}_{ijk}\frac{c^{(6)}_jc^{(6)}_k}{\Lambda^4}+\sum_jA^{(8)}_{ij}\frac{c^{(8)}_j}{\Lambda^4}+\dots\,.
	\label{eq:smeftpara}
\end{equation}
Here, $A^{(6)}_{ij}$ and $B^{(6)}_{ijk}$ are a real-valued matrix and a tensor that parametrize the linear and quadratic dependence on dimension-six Wilson coefficients, while $A^{(8)}_{ij}$ parametrizes the linear dependence on dimension-eight operators.
The linear dependence on dimension-six Wilson coefficients, arising from the interference of amplitudes with dimension-six operator insertions with the SM, is expected to be the most important contribution, as it is the leading term in the $1/\Lambda$ expansion.

EWPD typically constrains deviations from the SM to be at most a few percent,
making it crucial to consider percent-level higher-order corrections to SM
predictions, as well as the precise input parameter dependence of $\mathbf
	x_\textrm{pred}^\textrm{SM}$. They introduce percent-level corrections to
the SM and thus $O(1)$ effects on Wilson coefficient constraints. In contrast,
percent-level corrections to $\Delta_{\textrm{SMEFT}}$ will affect results
typically at the same order of magnitude. Therefore, a leading order
parametrization of SMEFT effects is often sufficient, and the input parameter
dependence of $\Delta_{\textrm{SMEFT}}$ is neglected in \ewpdlhc, which uses
fixed parametrizations stored in \texttt{yaml} text files.

Baseline dimension-six parametrizations of the SMEFT correction have been
derived with the \texttt{SMEFTsim}~\cite{Brivio:2020onw} model at leading
order. First, linear and quadratic parametrizations of the partial widths for
all $W$ decays and polarized $Z$ decays are obtained by simulating the boson
decays in \texttt{MadGraph5\_aMC@NLO}~\cite{Alwall:2014hca,Alwall:2014bza},
with negligible numerical uncertainties. Fermions except for the top and bottom
quarks are considered massless, as is usually the assumption in LHC studies. In
a second step, parametrizations of EWPOs are calculated as sums, differences,
products or ratios of (polarized) partial widths and expanded to second order
in the Wilson coefficients. The Wilson-coefficient dependent shifts of \MW or
$\alpha$ are extracted directly from the UFO model. The linear dependence of
the $W$ width on Wilson coefficients affecting the $W$ mass is taken into
account following Ref.~\cite{Berthier:2016tkq}. Parametrizations are provided
for two input parameter schemes and for the $U(2)_q\times U(2)_u\times U(2)_d$
(referred to as \texttt{top} in \texttt{SMEFTsim}), $U(2)_q\times U(2)_u\times
	U(2)_d\times U(3)_\ell\times U(3)_e$ (\texttt{topU3l}), and $U(3)_q\times
	U(3)_u\times U(3)_d\times U(3)_\ell\times U(3)_e$ (\texttt{U35}), and general
SMEFT symmetry assumptions. The advantage of this approach is that it ensures
parametrizations are consistent with those used at the LHC, which either use
\texttt{SMEFTsim} or models that have been validated against
it~\cite{Durieux:2019lnv}. This method also includes contributions quadratic in
dimension-six Wilson coefficients, analogous to the approach used in LHC
analyses. The code allowing for the derivation of parametrizations is included
in \ewpdlhc, allowing users to re-derive them with different input values or,
with minor modifications, to use alternative EFT scenarios.

Two types of higher-order corrections to the SMEFT parametrizations are
included in \ewpdlhc as alternative parametrization that can also be used in
combination.

\begin{figure}
	\centering
	\includegraphics[width=0.99\textwidth]{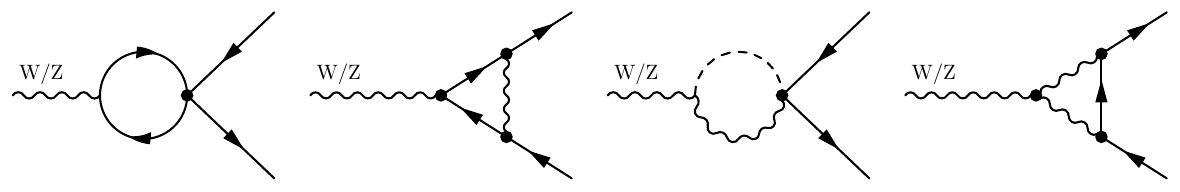}%
	\caption{Examples of next-to-leading order Feynman diagrams for $W$ and $Z$ boson decays in the SMEFT. Possible insertions of dimension-six operators are marked with dots.}
	\label{fig:diagrams}
\end{figure}

Next-to-leading-order QCD and EW perturbative
corrections~\cite{Dawson:2019clf,Dawson:2022bxd,Bellafronte:2023amz,Biekotter:2023xle,Biekotter:2023vbh}
modify tree-level parametrizations by a few percent and introduce new
dimension-six Wilson coefficients that only emerge at the loop level (example
Feynman diagrams are shown in Figure~\ref{fig:diagrams}). These corrections have
recently~\cite{Biekotter:2023xle,Biekotter:2023vbh} been computed in the
flavour-general case and for all five input parameter schemes listed in
Table~\ref{tab:schemes}. Analytic expressions, kindly provided by the authors
of Ref.~\cite{Biekotter:2023vbh}, are available as ancillary files accompanying
the corresponding arXiv submission. Numerical implementations of these
corrections have been incorporated in \ewpdlhc, using electroweak input
parameter values aligned with the \texttt{SMEFTsim} defaults. Expressions for a
broader set of derived observables and under various symmetry assumptions are
also made available in \ewpdlhc. The NLO parametrizations were cross-validated
against the numerical results in Ref.~\cite{Bellafronte:2023amz}. Observed
differences, at the 10\% level, are consistent with variations in input
parameter values, particularly the choice of $\alpha$.

Also available, in the \{$\alpha$,$M_Z$,$G_\mu$\} and \{$M_W$,$M_Z$,$G_\mu$\}
input parameter schemes and currently only for $Z$ pole observables, are the
complete $\Lambda^{-4}$~\cite{Corbett:2020bqv,Helset:2020yio,Corbett:2021eux}
corrections, which modify the dependence on dimension-six squared contributions
and introduce a new linear dependence on dimension-eight operators. They have
also been adapted to the flavour symmetries mentioned above.
\begin{table}
	\footnotesize
	\centering
	\begin{tabular}{@{}llR{2.5cm}R{2.5cm}R{2.5cm}R{2.5cm}@{}}
		\toprule
		$\delta\Al$    & Wilson coef.       & \texttt{SMEFTsim} & \texttt{EWPD2dim8} & \texttt{EWPDatLO} & \texttt{EWPDatNLO} \\
		\midrule
		Dim-6          & \cHWB              & 0.195             & 0.192              & 0.195             & 0.212              \\
		$O(\Lambda^2)$ & \cHDD              & 0.182             & 0.179              & 0.182             & 0.196              \\
		               & \cHlone            & 0.104             & 0.102              & 0.104             & 0.115              \\
		               & \cHlthree          & 0.104             & 0.102              & 0.104             & 0.122              \\
		               & \cHe               & 0.13              & 0.127              & 0.13              & 0.132              \\
		               & \cHqone            & $-$               & $-$                & $-$               & 0.013              \\
		               & \cHqthree          & $-$               & $-$                & $-$               & -0.008             \\
		               & \cHu               & $-$               & $-$                & $-$               & -0.017             \\
		               & \cW                & $-$               & $-$                & $-$               & -0.001             \\
		               & $c_{uB}$           & $-$               & $-$                & $-$               & -0.003             \\
		               & $c_{uW}$           & $-$               & $-$                & $-$               & 0.002              \\
		               & \clqone            & $-$               & $-$                & $-$               & 0.002              \\
		               & \clqthree          & $-$               & $-$                & $-$               & -0.003             \\
		               & \clu               & $-$               & $-$                & $-$               & -0.003             \\
		               & \cqe               & $-$               & $-$                & $-$               & 0.002              \\
		               & \ceu               & $-$               & $-$                & $-$               & -0.003             \\
		\midrule
		Dim-6          & \cHWB\cHWB         & -0.006            & -0.007             & $-$               & $-$                \\
		$O(\Lambda^4)$ & \cHWB\cHlone       & -0.019            & -0.018             & $-$               & $-$                \\
		               & \cHWB\cHlthree     & -0.007            & -0.018             & $-$               & $-$                \\
		               & \cHWB\cHe          & 0.003             & 0.003              & $-$               & $-$                \\
		               & \cHWB\cllone       & -0.006            & 0.0                & $-$               & $-$                \\
		               & \cHDD\cHWB         & -0.009            & -0.019             & $-$               & $-$                \\
		               & \cHDD\cHDD         & -0.003            & -0.006             & $-$               & $-$                \\
		               & \cHDD\cHlone       & -0.016            & -0.017             & $-$               & $-$                \\
		               & \cHDD\cHlthree     & -0.005            & -0.017             & $-$               & $-$                \\
		               & \cHDD\cHe          & 0.005             & 0.003              & $-$               & $-$                \\
		               & \cHDD\cllone       & -0.006            & 0.0                & $-$               & $-$                \\
		               & \cHlone\cHlone     & -0.008            & -0.008             & $-$               & $-$                \\
		               & \cHlone\cHlthree   & -0.01             & -0.016             & $-$               & $-$                \\
		               & \cHlone\cllone     & -0.003            & -0.0               & $-$               & $-$                \\
		               & \cHlthree\cHlthree & -0.002            & -0.008             & $-$               & $-$                \\
		               & \cHlthree\cllone   & -0.003            & -0.0               & $-$               & $-$                \\
		               & \cHe\cHlone        & -0.006            & -0.006             & $-$               & $-$                \\
		               & \cHe\cHlthree      & 0.002             & -0.006             & $-$               & $-$                \\
		               & \cHe\cHe           & 0.005             & 0.005              & $-$               & $-$                \\
		               & \cHe\cllone        & -0.004            & -0.0               & $-$               & $-$                \\
		               & \cHB\cHWB          & $-$               & 0.012              & $-$               & $-$                \\
		               & \cHW\cHWB          & $-$               & 0.012              & $-$               &                    \\\\
		\midrule
		Dim-8          & $c^{(8)}_{HWB}$    & $-$               & 0.006              & $-$               & $-$                \\
		$O(\Lambda^4)$ & $c^{(8)}_{HDD,2}$  & $-$               & 0.011              & $-$               & $-$                \\
		               & $c^{(8)}_{Hl}$     & $-$               & 0.003              & $-$               & $-$                \\
		               & $c^{(8)}_{Hl,2}$   & $-$               & 0.003              & $-$               & $-$                \\
		               & $c^{(8),(3)}_{Hl}$ & $-$               & 0.003              & $-$               & $-$                \\
		               & $c^{(8)}_{He}$     & $-$               & 0.004              & $-$               & $-$                \\
		\bottomrule
	\end{tabular}
	\caption{Linear shift in \Al for all dimension-six and dimension-eight Wilson coefficients contributing significantly, for $\Lambda=1\,\TeV$.
		The shifts are shown for \texttt{SMEFTsim}~\cite{Brivio:2020onw}, \texttt{EWPD2dim8}~\cite{Corbett:2021eux}, and \texttt{EWPDatNLO}~\cite{Biekotter:2023vbh} (both at LO and NLO). The comparison is made for the fully flavour symmetric case using the  \{$\MW$,$M_Z$,$G_\mu$\}  input parameter scheme.}
	\label{tab:paravalidAl}
\end{table}

The three parametrizations from \texttt{SMEFTsim}, Ref.~\cite{Corbett:2021eux}
(labeled \texttt{EWPD2dim8}), and those based on Ref.~\cite{Biekotter:2023vbh}
(labeled \texttt{EWPDatNLO} and \texttt{EWPDatLO} for the NLO and LO case) have
been compared to ensure consistency of the initial parametrizations and
correctness of the conversion to various flavour symmetries. The comparison of
the coefficients of the parametrizations for \Al is shown in
Table~\ref{tab:paravalidAl}, for the fully flavour symmetric case and the
\{$\MW$,$\MZ$,$G_\mu$\} input parameter scheme. The operator definitions and
notations from Ref.~\cite{Brivio:2020onw} are used for Warsaw
basis~\cite{Grzadkowski:2010es} Wilson coefficients throughout this paper while
the notation of Ref.~\cite{Corbett:2021eux} is used from dimension-eight Wilson
coefficients. At leading order, differences between the parametrizations are
less than 5\% for all observables and are mainly due to the $b$-quark mass,
which is non-zero only for the \texttt{SMEFTsim} parametrization, and the
limited numerical precision of the tables in Ref.~\cite{Corbett:2021eux}, which
were used for the \texttt{EWPD2dim8} parametrization.

\subsection{The EWPD fit in the SMEFT (at NLO)}
\label{sec:SMEFTfit}

In this section, the observables introduced in Section~\ref{sec:observables}
are analyzed using a parametrization that combines the SM predictions of
Section~\ref{sec:uncertainties} with the SMEFT parametrization in
Section~\ref{sec:SMEFTpara} to set limits on dimension-six Wilson coefficients.

This analysis assumes a $U(2)_q\times U(2)_u \times U(2)_d$ symmetry in the
quark sector, as there are no observables that allow to distinguish the first
two quark generations with high precision. A fully flavour-general fit would
result in several blind directions related to differences in light quark
couplings. Lepton flavour universality violation is allowed.

Unless noted
otherwise, the \{$M_W$,$M_Z$,$G_\mu$\} input parameter scheme is employed. The
SM value of $\alpha(Q^2\!=\!M_Z^2)$ is predicted by these input parameters, and
potential deviations from this value are constrained by including the
semi-experimental determination of \Deltaalpha as an observable in the fit.
Similarly, deviations from the SM prediction of \st are constrained by
measurements of the EWPOs \Ae, \Amu, and \Atau as well as the various
forward--backward asymmetries. The \texttt{EWPDatLO} and
\texttt{EWPDatNLO}~\cite{Biekotter:2023vbh} SMEFT parametrizations are employed
for this analysis.

While the \ewpdlhc tool contains parametrizations up to $\Lambda^{-4}$, this
analysis focuses on fits using only the $\Lambda^{-2}$ contributions, which
are produced instantaneously when running \ewpdlhc. Including $\Lambda^{-4}$
contributions requires a more complex numerical analysis due to the quadratic
dimension-six Wilson coefficient dependence. This is left to a future
publication.

\begin{figure}
	\includegraphics[width=0.33\textwidth]{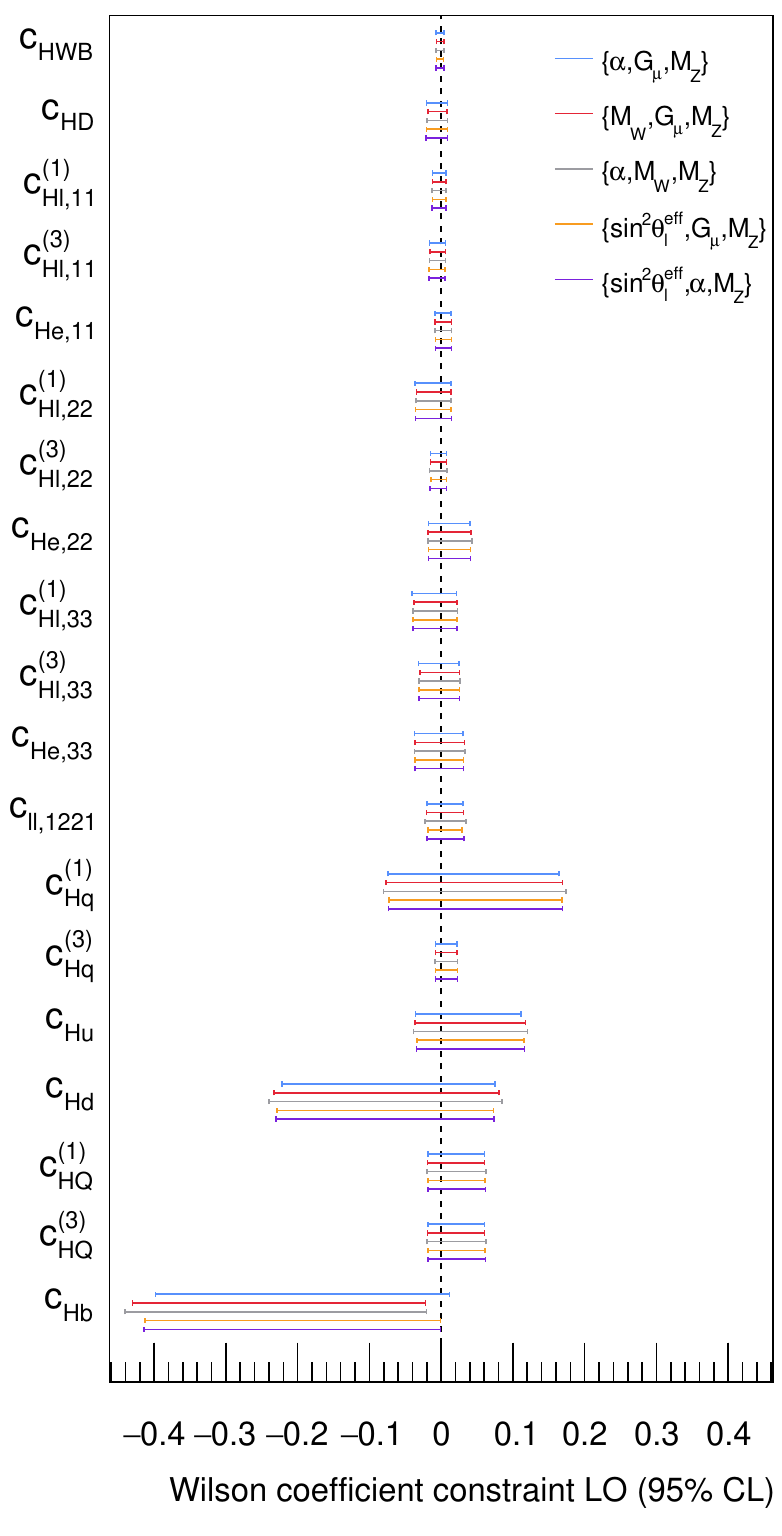}%
	\includegraphics[width=0.33\textwidth]{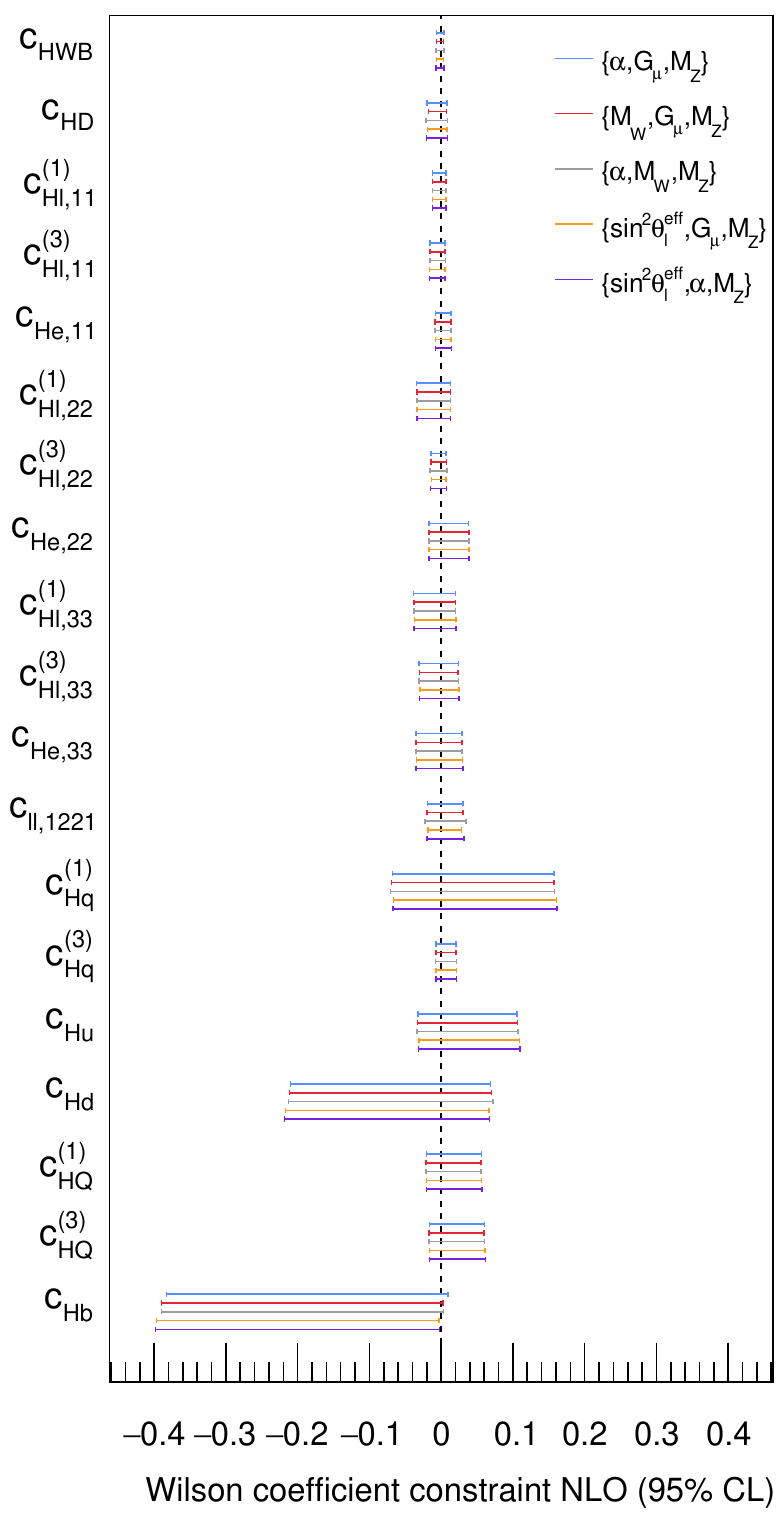}%
	\includegraphics[width=0.33\textwidth]{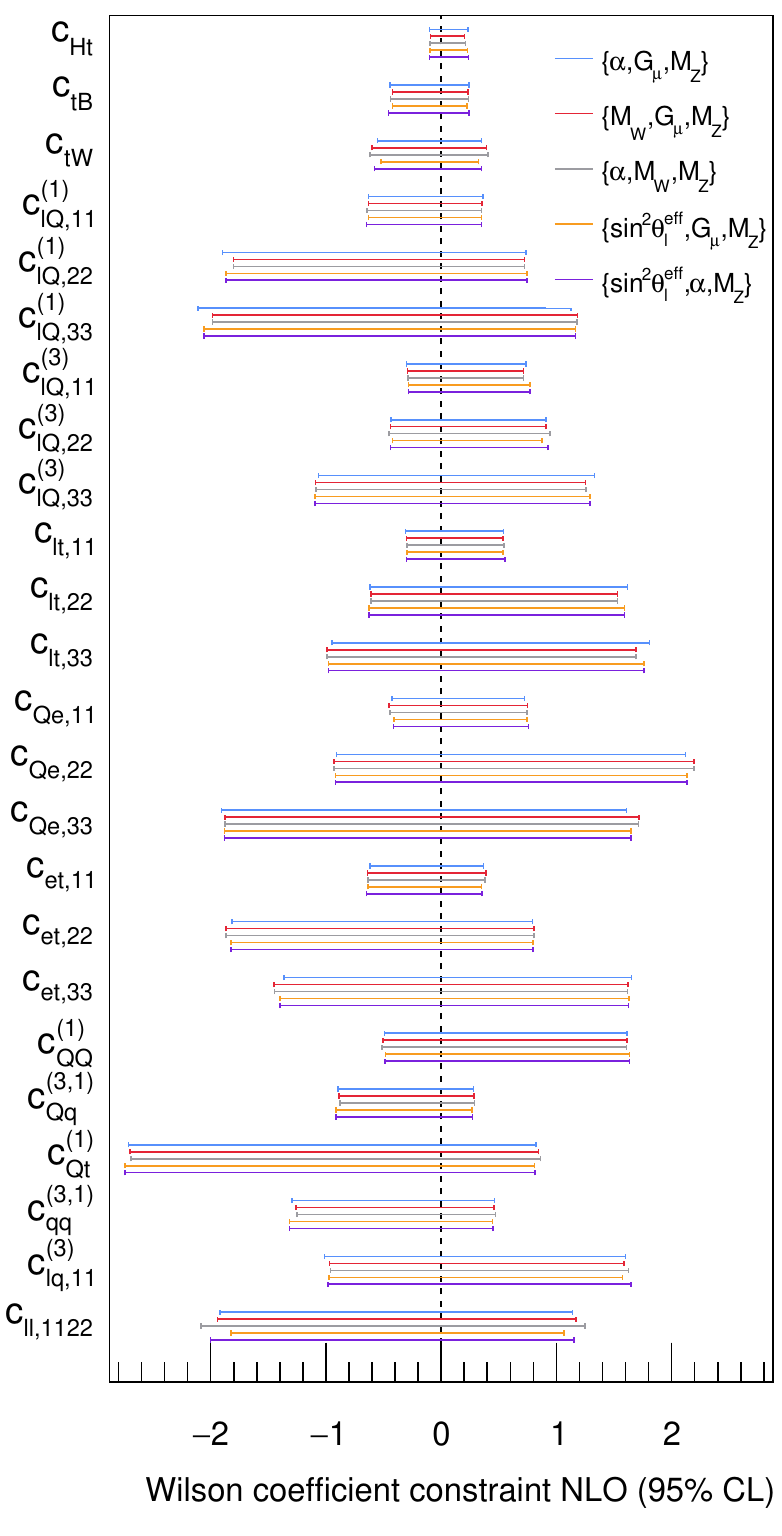}
	\caption{Comparison of one-at-a-time Wilson coefficient constraints in the SMEFT EWPD fit in five different input parameter schemes at $O(\Lambda^{-2})$. A $U(2)_q\times U(2)_u \times U(2)_d$ symmetry between the first two quark generations and $\Lambda=1\,\TeV$ is assumed. For the leftmost panel a LO SMEFT parametrization  is employed while a NLO parametrization is employed for the remaining constraints.}
	\label{fig:1Dconstraints}
\end{figure}

In total 30 observables are studied of which six are SM inputs.
Figure~\ref{fig:1Dconstraints} presents one-dimensional confidence intervals
for Wilson coefficients in all five input parameter schemes, derived by fitting
one parameter at a time. The first panel displays the limits obtained using the
LO SMEFT parametrization. In the second panel, the same 19 Wilson coefficients
are analyzed using the NLO SMEFT parametrization. NLO corrections have a
relatively small effect on results, typically shifting central values and
improving limits by approximately 5\%. The scheme dependence is already small
at LO -- because SM predictions include higher-order corrections -- and is
reduced further at NLO, particularly in schemes that use \MW as an input.
The
third panel displays the 24 most stringent constraints on Wilson coefficients
that appear only at loop level. Sensitivity exists in particular for operators
coupling to the top quark, owing to the larger impact of diagrams involving
heavy quark loops. While constraints on loop-level operators, except for \cHt,
are generally weaker than those for tree-level operators, they are often more
restrictive than limits derived from LHC measurements of top quark production.
For instance, in many cases they surpass the constraints from the comprehensive
SMEFT analysis of top-quark pair-production with additional leptons performed
by CMS~\cite{CMS:2023xyc}. However, the one-at-a-time EWPD constraints rely
heavily on a small set of precisely measured observables, leaving numerous
blind directions. These blind directions can only be addressed through
additional measurements at the LHC.

\begin{figure}
	\includegraphics[width=0.495\textwidth]{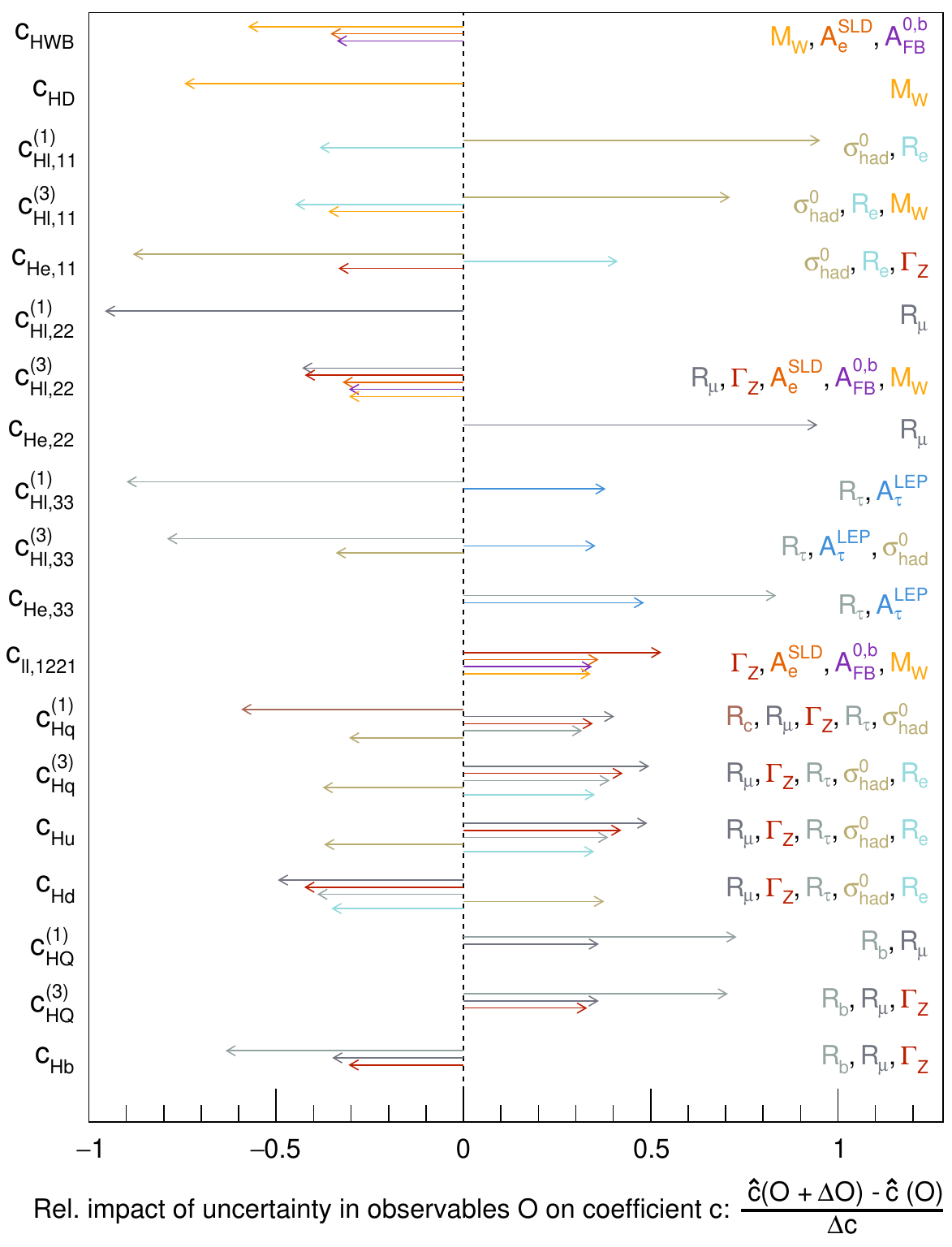}%
	\includegraphics[width=0.495\textwidth]{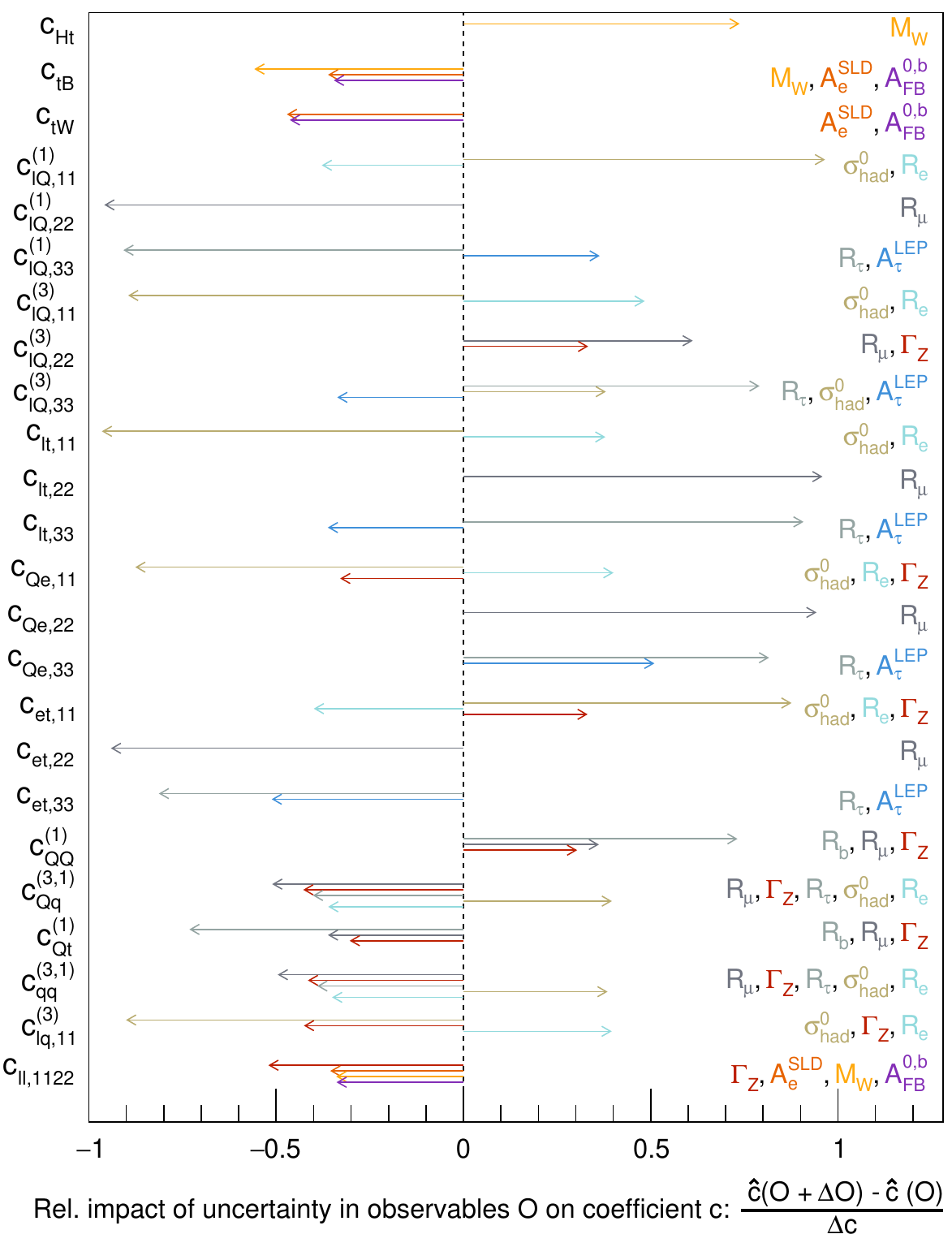}%
	\caption{Relative impact of individual observable uncertainties on the one-at-a-time determination of SMEFT Wilson coefficients.
For each observable, the global fit is repeated after increasing its central value by one standard deviation. The resulting change in the best-fit value of each Wilson coefficient is shown relative to its total uncertainty. Only observables that induce a shift exceeding 30\% are displayed for each coefficient. The analysis employs the NLO SMEFT parametrization, assuming a $U(2)_q \times U(2)_u \times U(2)_d$ flavour symmetry for the first two quark generations.}
	\label{fig:impacts}
\end{figure}

The impact of uncertainties in the measurement of electroweak precision observables
on the one-at-a-time constraints of Wilson coefficients (based on the NLO
parametrization) is illustrated in Figure~\ref{fig:impacts}.
The Wilson coefficients associated with bosonic operators, \cHWB and \cHDD, are
primarily constrained by precise measurements of \Gmu, $\alpha$, \MZ, \MW, and \st, and
are thus especially sensitive to the relatively large uncertainties in \MW and in the asymmetries \Ae and \AFB.
These constraints represent some of the most stringent bounds on new physics and could be
further tightened by improved measurements of \MW and \st at the LHC.
Wilson coefficients that modify only electron couplings are mainly constrained by
\sigmahad, while those affecting other fermion couplings are dominated by measurements
of ratios of $Z$ boson partial widths -- observables that are unlikely to see
substantial improvement at the LHC.

Multi-dimensional fits require extra care as unconstrained directions exist,
corresponding to linear combination of Wilson coefficients that do not affect
the studied observables. Both constrained and unconstrained directions are
identified, with the latter eliminated, through the following procedure. The
inverse of the covariance matrix of measurements, $V^{-1}$ (which, for this
purpose, includes theoretical and parametric uncertainties as outlined in
Section~\ref{sec:uncertainties}), is transformed using the linear
parametrization matrix $A^{(6)}$ (see Equation~\ref{eq:smeftpara}), to obtain
the inverse of the covariance matrix in the space of Wilson coefficients:
\begin{equation}
	V^{-1}_\text{SMEFT}  = (A^{(6)})^\mathsf{T} V^{-1} A^{(6)}\,.
\end{equation}
This matrix corresponds to the Fisher information matrix. Its eigenvectors are uncorrelated directions in the Wilson coefficient space and their corresponding eigenvalues, $\lambda$, are related to the expected uncertainty $\sigma$ of a constraint in direction of the eigenvector:
\begin{equation}
	\sigma = \frac{1}{\sqrt{\lambda}}\,.
\end{equation}

At LO there are 16 constrained directions with finite eigenvalues and three
unconstrained directions corresponding to zero eigenvalues. Two of the
unconstrained directions are Wilson coefficient combinations that cannot be
constrained in fermion-fermion to fermion-fermion scattering
alone~\cite{Brivio:2017bnu} and one corresponds to a combination of \cHQone and
\cHQthree that affects the top but not the bottom quark coupling to weak
bosons.

Constrained directions, along with the observables contributing most to each
constraint are listed in Table~\ref{tab:smeftfit1}. In the eigenvector basis,
where correlations are removed and unconstrained directions are eliminated, the
best-fit values are determined by solving a system of linear equations. The
pulls, defined as the difference of best-fit value and direct measurement,
divided by the uncertainty $\sigma$, are also shown in
Table~\ref{tab:smeftfit1}.

\begin{table}
	\small
	\centering
	\begin{tabular}{@{}llrr@{}}
		\toprule
		Constrained direction (LO parametrization)                                                                                         & Main contr.    & $\sigma$ & Pull   \\
		\midrule
		{\footnotesize$0.83c_{HWB}+0.31c_{Hl,22}^{(3)}+0.31c_{HD}+0.25c_{Hl,11}^{(3)}-0.15c_{ll,1221}-0.14c_{He,11}$}                      & \Deltaalpha    & 0.0022   & $-0.7$ \\
		{\footnotesize$0.68c_{Hl,11}^{(1)}+0.46c_{Hl,11}^{(3)}-0.44c_{He,11}-0.23c_{Hq}^{(3)}-0.17c_{Hl,22}^{(3)}-0.12c_{HWB}$}            & \sigmahad      & 0.0033   & $-0.6$ \\
		{\footnotesize$0.6c_{Hq}^{(3)}-0.48c_{Hl,22}^{(3)}-0.36c_{Hl,22}^{(1)}+0.27c_{He,22}+0.24c_{HWB}-0.2c_{He,11}$}                    & \Rmu, \GammaZ  & 0.006    & $0.7$  \\
		{\footnotesize$0.68c_{He,11}+0.42c_{HD}-0.25c_{Hl,22}^{(3)}+0.24c_{Hl,11}^{(3)}-0.23c_{Hl,22}^{(1)}+0.22c_{Hl,11}^{(1)}$}          & \AeSLD, \AFBb  & 0.0092   & $-0.2$ \\
		{\footnotesize$0.59c_{Hl,33}^{(3)}+0.55c_{Hl,33}^{(1)}-0.31c_{He,33}-0.31c_{He,11}-0.19c_{Hl,22}^{(3)}-0.17c_{Hl,11}^{(3)}$}       & \Rtau          & 0.0097   & $0.0$  \\
		{\footnotesize$0.51c_{Hl,22}^{(1)}-0.42c_{He,22}+0.36c_{Hl,11}^{(1)}+0.35c_{ll,1221}+0.31c_{Hq}^{(3)}-0.28c_{Hl,11}^{(3)}$}        & \GammaZ        & 0.012    & $-0.5$ \\
		{\footnotesize$0.63c_{HQ}^{(1)}+0.63c_{HQ}^{(3)}-0.38c_{Hq}^{(3)}-0.11c_{Hl,11}^{(3)}-0.1c_{Hb}-0.09c_{Hl,22}^{(3)}$}              & \Rb            & 0.016    & $0.4$  \\
		{\footnotesize$0.84c_{He,33}+0.38c_{HD}+0.23c_{Hl,33}^{(1)}+0.18c_{Hl,33}^{(3)}-0.14c_{He,11}-0.12c_{Hl,11}^{(3)}$}                & \AtauLEP       & 0.02     & $-0.5$ \\
		{\footnotesize$0.68c_{He,22}-0.37c_{Hl,11}^{(3)}+0.36c_{Hl,22}^{(3)}+0.25c_{HD}+0.24c_{Hl,11}^{(1)}+0.21c_{Hl,22}^{(1)}$}          & \AFBmu, \RWmue & 0.033    & $0.5$  \\
		{\footnotesize$0.64c_{Hl,22}^{(1)}-0.43c_{Hl,11}^{(1)}-0.39c_{Hl,22}^{(3)}+0.35c_{Hl,11}^{(3)}+0.3c_{He,22}-0.13c_{Hl,33}^{(1)}$}  & \RWmue         & 0.052    & $-0.3$ \\
		{\footnotesize$0.69c_{Hl,33}^{(1)}-0.49c_{Hl,33}^{(3)}-0.35c_{ll,1221}+0.22c_{Hq}^{(3)}-0.15c_{He,22}+0.13c_{Hl,22}^{(1)}$}        & \Rel           & 0.063    & $0.2$  \\
		{\footnotesize$0.97c_{Hq}^{(1)}+0.14c_{Hu}-0.1c_{Hq}^{(3)}-0.09c_{Hd}+0.07c_{Hb}-0.06c_{HD}$}                                      & \Rc            & 0.099    & $0.1$  \\
		{\footnotesize$0.85c_{Hu}+0.35c_{ll,1221}-0.17c_{Hq}^{(1)}+0.14c_{Hl,33}^{(1)}+0.13c_{Hb}-0.13c_{Hq}^{(3)}$}                       & \AFBc, \Ac     & 0.14     & $0.0$  \\
		{\footnotesize$0.77c_{ll,1221}-0.4c_{Hu}-0.26c_{Hl,33}^{(3)}-0.21c_{Hl,11}^{(1)}+0.2c_{Hl,33}^{(1)}-0.18c_{Hl,22}^{(1)}$}          & \RWtaumu       & 0.15     & $-0.5$ \\
		{\footnotesize$0.97c_{Hb}-0.12c_{Hu}+0.1c_{HD}$}                                                                                   & \Ab, \AFBb     & 0.23     & $-3.0$ \\
		{\footnotesize$0.99c_{Hd}+0.1c_{Hq}^{(1)}$}                                                                                        & \BrWmu, \BrWe  & 1.2      & $-0.3$ \\
		\midrule
		Constrained direction (NLO parametrization)                                                                                        & Main contr.    & $\sigma$ & Pull   \\
		\midrule
		{\footnotesize$0.83c_{HWB}+0.32c_{Hl,22}^{(3)}+0.31c_{HD}+0.25c_{Hl,11}^{(3)}-0.15c_{ll,1221}-0.12c_{He,11}$}                      & \Deltaalpha    & 0.0021   & $-0.7$ \\
		{\footnotesize$0.67c_{Hl,11}^{(1)}-0.47c_{He,11}+0.46c_{Hl,11}^{(3)}-0.25c_{Hq}^{(3)}-0.14c_{Hl,22}^{(3)}-0.11c_{HWB}$}            & \sigmahad      & 0.0032   & $-0.7$ \\
		{\footnotesize$0.6c_{Hq}^{(3)}-0.48c_{Hl,22}^{(3)}-0.35c_{Hl,22}^{(1)}+0.29c_{He,22}+0.24c_{HWB}-0.2c_{He,11}$}                    & \Rmu, \GammaZ  & 0.0057   & $0.7$  \\
		{\footnotesize$0.69c_{He,11}+0.42c_{HD}+0.29c_{Hl,11}^{(3)}-0.25c_{Hl,22}^{(1)}+0.23c_{Hl,11}^{(1)}-0.22c_{Hl,22}^{(3)}$}          & \AeSLD, \AFBb  & 0.0088   & $-0.3$ \\
		{\footnotesize$0.58c_{Hl,33}^{(3)}+0.55c_{Hl,33}^{(1)}-0.35c_{He,33}-0.24c_{He,11}-0.2c_{Hl,22}^{(3)}-0.17c_{Hl,22}^{(1)}$}        & \Rtau          & 0.0093   & $-0.0$ \\
		{\footnotesize$0.47c_{Hl,22}^{(1)}-0.44c_{He,22}+0.38c_{Hl,11}^{(1)}+0.33c_{ll,1221}+0.3c_{Hq}^{(3)}-0.24c_{Hl,11}^{(3)}$}         & \GammaZ        & 0.012    & $-0.6$ \\
		{\footnotesize$0.64c_{HQ}^{(1)}+0.62c_{HQ}^{(3)}-0.37c_{Hq}^{(3)}-0.11c_{Hl,11}^{(3)}-0.11c_{Hb}-0.09c_{Hl,22}^{(3)}$}             & \Rb            & 0.016    & $0.4$  \\
		{\footnotesize$0.82c_{He,33}+0.39c_{HD}+0.26c_{Hl,33}^{(1)}+0.22c_{Hl,33}^{(3)}-0.13c_{He,11}-0.12c_{Hl,11}^{(3)}$}                & \AtauLEP       & 0.019    & $-0.8$ \\
		{\footnotesize$0.54c_{He,22}+0.45c_{Hl,22}^{(3)}-0.44c_{Hl,11}^{(3)}+0.35c_{Hl,11}^{(1)}+0.24c_{HD}-0.18c_{Hl,33}^{(3)}$}          & \RWmue, \AFBmu & 0.037    & $0.5$  \\
		{\footnotesize$0.68c_{Hl,22}^{(1)}-0.35c_{Hl,11}^{(1)}+0.34c_{He,22}-0.29c_{Hl,22}^{(3)}+0.29c_{Hl,11}^{(3)}+0.18c_{Hl,33}^{(1)}$} & \RWmue         & 0.055    & $-0.2$ \\
		{\footnotesize$0.67c_{Hl,33}^{(1)}-0.45c_{Hl,33}^{(3)}-0.35c_{He,22}-0.33c_{ll,1221}+0.2c_{Hq}^{(3)}-0.17c_{Hl,22}^{(1)}$}         & \Rel           & 0.063    & $0.2$  \\
		{\footnotesize$0.99c_{Hq}^{(1)}-0.09c_{Hd}-0.09c_{Hq}^{(3)}-0.04c_{HD}+0.04c_{Hb}+0.03c_{Hl,11}^{(3)}$}                            & \Rc            & 0.098    & $0.0$  \\
		{\footnotesize$0.81c_{ll,1221}+0.28c_{Hu}-0.24c_{Hl,33}^{(3)}+0.24c_{Hl,33}^{(1)}+0.2c_{Hl,11}^{(3)}-0.19c_{Hl,11}^{(1)}$}         & \RWtaumu       & 0.15     & $-0.4$ \\
		{\footnotesize$0.91c_{Hu}-0.26c_{ll,1221}-0.14c_{HD}-0.12c_{Hq}^{(3)}+0.12c_{Hl,33}^{(3)}+0.1c_{Hl,11}^{(1)}$}                     & \Ac, \AFBc     & 0.16     & $0.4$  \\
		{\footnotesize$0.98c_{Hb}$}                                                                                                        & \Ab            & 0.27     & $-2.7$ \\
		{\footnotesize$0.97c_{Hd}+0.12c_{Qq}^{(3,1)}+0.11c_{Hq}^{(1)}$}                                                                    & \BrWmu, \BrWe  & 1.1      & $-0.3$ \\
		{\footnotesize$0.72c_{tW}+0.40c_{lQ,22}^{(3)}+0.38c_{lQ,33}^{(3)}+0.31c_{lQ,11}^{(3)}-0.12cW+0.12c_{HQ}^{(1)}$}                    & \GammaW        & 17       & 0.1    \\
		\bottomrule
	\end{tabular}
	\caption{Constrained directions in the SMEFT EWPD fit, assuming a $U(2)_q\times U(2)_u \times U(2)_d$ symmetry between the first two quark generations, for $\Lambda=1\,\TeV$ and at $O(\Lambda^{-2})$, using a LO (top) and NLO SMEFT parametrization (bottom).
		Linear combinations are normalized and only the five Wilson coefficients with the largest absolute value are shown, provided their absolute value is larger than 0.1.
		For each direction, the observables that contribute most to an uncorrelated $\chi^2$ for a shift in the eigenvector direction are indicated as ``Main contr.'', with a cut-off of at a 20\% fractional contribution.
		The uncertainty $\sigma$ corresponds to 68\% confidence level intervals.
		The pull is defined as the best-fit value of the Wilson coefficient direction, divided by the uncertainty $\sigma$.}
	\label{tab:smeftfit1}
\end{table}

The most tightly constrained direction corresponds to the difference between
the SM-predicted value of $\alpha(Q^2\!=\!M_Z^2)$ and the experimental value of
$\alpha(Q^2\!=\!0)$ in conjunction with the semi-experimental result on the
running to $Q^2=M_Z^2$. For a BSM physics model that introduces non-zero Wilson
coefficients in this direction with $O(1)$ couplings, such that
$\sum_i\left(c_{i}\right)^2=1$, this corresponds to a sensitivity to a mass
scale of about $\frac{1}{\sqrt{0.002}}\,\TeV\approx 20\,\TeV$.

Measurements of $W$ and $Z$ pole observables constrain 15 additional directions
at LO, with a precision ranging 0.003 (driven by the hadronic $Z$ pole cross
section measurement) to 1 (constrained only by the hadronic $W$ branching
fractions). Throughout this paper $\Lambda=1\,\TeV$ is assumed and results for
alternative scales $\Lambda^\prime$ can be obtained by multiplying with
$\left(\frac{\Lambda^\prime}{\TeV}\right)^2$.

The inclusion of additional operators at NLO disrupts some of the relationships
that hold between observables. As a result, more independent directions in
parameter space exist that can be constrained by EWPD. For example, already in
a flavour-universal scenario, the $Z\rightarrow b\bar b$ decay rate decouples
from the $Z\rightarrow d\bar d$ and $Z\rightarrow s\bar s$ rate, as it is more
strongly influenced by operators modifying quark couplings, due to the more
important role of heavy quark loops for this process. Additionally, left-handed
$Z$ couplings are related to $W$ couplings (see, e.g.,
Ref.~\cite{Breso-Pla:2021qoe}), enabling at LO the prediction of $W$ branching
fractions based on $Z$ boson branching fractions and asymmetries. At NLO, $Z$
decays are influenced by a broader set of four-fermion operators compared to
$W$ decays, breaking this relationship.

All constrained direction in the NLO parametrization are summarized in
Table~\ref{tab:smeftfit1}, too. These directions closely resemble those
obtained at LO, with minor changes in numerical coefficients and small
corrections from additional Wilson coefficients. Under the studied flavor
symmetry, which decouples third-generation quarks already at tree level, one
additional direction is constrained at NLO. This direction primarily
corresponds to modifications in the $c_{tW}$ and $c_{lQ}^{(3)}$ coefficients,
which contribute to NLO diagrams involving top and bottom quarks in loops, some
of which are shown in Figure~\ref{fig:diagrams}. The linear combination
primarily consists of Wilson coefficients that contribute only at loop level
and does not affect $Z$ decays but can be constrained through the $W$ boson
width. However, the sensitivity to this NLO contribution is poor, requiring
Wilson coefficients of magnitudes around 20 (for $\Lambda=1\,\TeV$) to produce
a one-standard-deviation shift in the observed value.

The overall agreement with the SM expectation of no significant deviations from
zero is excellent at NLO, corresponding to a $p$-value of 87\%, which is even
higher than the LO parametrization $p$-value of 76\%. The analysis does not
show substantial evidence of BSM physics. Indeed, there is only one direction
in Wilson coefficient space that deviate by more than a standard deviation from
zero, a three-standard-deviation excess in a direction predominantly affecting
the coupling of the $Z$ boson to right-handed bottom quarks. This deviation is
driven by the larger-than-expected values of \AFBb and \Ab measured at LEP and
SLD, respectively. A value of $\frac{c_{Hb}}{\Lambda^2}\approx\frac{1}{\TeV^2}$
would be necessary to match the measurement, which would imply BSM physics that
is either strongly coupled or at a relatively low mass scale.

Across all five input parameter schemes, the number of constrained directions
remains the same, with constraints and the composition of Wilson coefficients
differing only slightly. This consistency validates the NLO parametrizations in alternative schemes as well as
the uncertainty model, which is also crucial for scheme-independent results.
Despite the near scheme-independence of the likelihood, using a consistent
input parameter scheme remains critical when combining results with additional
data. Scheme differences could otherwise misalign blind directions, introducing
spurious constraints on certain combinations of Wilson coefficients.

To obtain the best-fit values and uncertainties for all observables, error
propagation is employed. Sensitive directions that are linear combinations of
both the SM input parameters and the Wilson coefficients are fit to the data.
Post-fit values for all observables are derived by propagating uncertainties
from these combinations. An exception is made for the weakest constrained
direction, which only exists at NLO. As it requires unlikely large Wilson
coefficient values (or small scales $\Lambda$) to affect the data, it is fixed
to zero. The results are summarized in Table~\ref{tab:smeftfit}. In most cases,
the fit values closely align with direct measurements due to the large number
of additional parameters, the Wilson coefficients, that independently modify
observables. There are two main exceptions to this. The first exception are
$A_f$ and $A_\textrm{FB}^{0,f}$, as well as $W$ boson (ratio of) branching
fractions measurements. These are straightforwardly related through
Equation~\ref{eq:AFBdef} and Equation~\ref{eq:RWdef}, respectively. The second
exception are the total widths of the $Z$ boson and $W$ boson, whose
relationship is more complex and will be discussed in the next section.

\begin{table}
	\small
	\centering
	\begin{tabular}{@{}lr@{}c@{}lr@{}c@{}lrr@{}c@{}l@{}}
		\toprule
		Observable                 & \multicolumn{3}{c}{Direct} & \multicolumn{3}{c}{Fit} & Pull         & \multicolumn{3}{c}{Indirect}                                                                       \\
		\midrule
		$\MH\,[\GeV]$              & $125.10$                   & $\,\pm\,$               & $0.11$       & $125.10$                     & $\,\pm\,$ & $0.11$       & $-0.0$ &          & $-$       &          \\
		$\mt\,[\GeV]$              & $172.57$                   & $\,\pm\,$               & $0.58$       & $172.57$                     & $\,\pm\,$ & $0.58$       & $0.0$  &          & $-$       &          \\
		$\alphas$                  & $0.11840$                  & $\,\pm\,$               & $0.00080$    & $0.11840$                    & $\,\pm\,$ & $0.00080$    & $0.0$  &          & $-$       &          \\
		$\MZ\,[\GeV]$              & $91.1876$                  & $\,\pm\,$               & $0.0021$     & $91.1874$                    & $\,\pm\,$ & $0.0021$     & $-0.1$ &          & $-$       &          \\
		$\Gmu\,[10^{-5}\GeV^{-2}]$ & $1.16637880$               & $\,\pm\,$               & $0.00000060$ & $1.16637880$                 & $\,\pm\,$ & $0.00000060$ & $-0.0$ &          & $-$       &          \\
		$\Deltaalpha$              & $0.059030$                 & $\,\pm\,$               & $0.000090$   & $0.059030$                   & $\,\pm\,$ & $0.000090$   & $-0.0$ &          & $-$       &          \\
		\midrule
		$\MW\,[\GeV]$              & $80.369$                   & $\,\pm\,$               & $0.013$      & $80.369$                     & $\,\pm\,$ & $0.013$      & $0.0$  &          & $-$       &          \\
		$\GammaZ\,[\GeV]$          & $2.4955$                   & $\,\pm\,$               & $0.0023$     & $2.4955$                     & $\,\pm\,$ & $0.0023$     & $0.0$  & $2.505$  & $\,\pm\,$ & $0.072$  \\
		$\Rel$                     & $20.804$                   & $\,\pm\,$               & $0.050$      & $20.786$                     & $\,\pm\,$ & $0.046$      & $-0.3$ & $20.71$  & $\,\pm\,$ & $0.52$   \\
		$\Rmu$                     & $20.784$                   & $\,\pm\,$               & $0.034$      & $20.784$                     & $\,\pm\,$ & $0.033$      & $0.0$  &          & $-$       &          \\
		$\Rtau$                    & $20.764$                   & $\,\pm\,$               & $0.045$      & $20.764$                     & $\,\pm\,$ & $0.045$      & $-0.0$ &          & $-$       &          \\
		$\sigmahad\,[\pb]$         & $41481$                    & $\,\pm\,$               & $32$         & $41480$                      & $\,\pm\,$ & $32$         & $-0.0$ & $41300$  & $\,\pm\,$ & $1000$   \\
		$\AeSLD$                   & $0.1516$                   & $\,\pm\,$               & $0.0021$     & $0.1494$                     & $\,\pm\,$ & $0.0017$     & $-1.0$ & $0.1459$ & $\,\pm\,$ & $0.0027$ \\
		$\AeLEP$                   & $0.1498$                   & $\,\pm\,$               & $0.0049$     & $0.1494$                     & $\,\pm\,$ & $0.0017$     & $-0.1$ & $0.1494$ & $\,\pm\,$ & $0.0018$ \\
		$\AmuSLD$                  & $0.142$                    & $\,\pm\,$               & $0.015$      & $0.147$                      & $\,\pm\,$ & $0.010$      & $0.3$  & $0.150$  & $\,\pm\,$ & $0.013$  \\
		$\AtauSLD$                 & $0.136$                    & $\,\pm\,$               & $0.015$      & $0.145$                      & $\,\pm\,$ & $0.004$      & $0.6$  & $0.145$  & $\,\pm\,$ & $0.004$  \\
		$\AtauLEP$                 & $0.1439$                   & $\,\pm\,$               & $0.0043$     & $0.1448$                     & $\,\pm\,$ & $0.0040$     & $0.2$  & $0.1510$ & $\,\pm\,$ & $0.011$  \\
		$\AFBe$                    & $0.0145$                   & $\,\pm\,$               & $0.0025$     & $0.0167$                     & $\,\pm\,$ & $0.0003$     & $0.9$  & $0.0167$ & $\,\pm\,$ & $0.0003$ \\
		$\AFBmu$                   & $0.0169$                   & $\,\pm\,$               & $0.0013$     & $0.0164$                     & $\,\pm\,$ & $0.0010$     & $-0.4$ & $0.0160$ & $\,\pm\,$ & $0.0015$ \\
		$\AFBtau$                  & $0.0188$                   & $\,\pm\,$               & $0.0017$     & $0.0162$                     & $\,\pm\,$ & $0.0004$     & $-1.5$ & $0.0161$ & $\,\pm\,$ & $0.0004$ \\
		$\Rc$                      & $0.1721$                   & $\,\pm\,$               & $0.0030$     & $0.1720$                     & $\,\pm\,$ & $0.0030$     & $-0.0$ &          & $-$       &          \\
		$\Rb$                      & $0.21629$                  & $\,\pm\,$               & $0.00066$    & $0.21630$                    & $\,\pm\,$ & $0.00066$    & $0.0$  &          & $-$       &          \\
		$\AFBc$                    & $0.0707$                   & $\,\pm\,$               & $0.0035$     & $0.0734$                     & $\,\pm\,$ & $0.0022$     & $0.8$  & $0.0746$ & $\,\pm\,$ & $0.0028$ \\
		$\Ab$                      & $0.923$                    & $\,\pm\,$               & $0.020$      & $0.897$                      & $\,\pm\,$ & $0.015$      & $-1.3$ & $0.871$  & $\,\pm\,$ & $0.021$  \\
		$\Ac$                      & $0.670$                    & $\,\pm\,$               & $0.027$      & $0.655$                      & $\,\pm\,$ & $0.022$      & $-0.6$ & $0.634$  & $\,\pm\,$ & $0.036$  \\
		$\AFBb$                    & $0.0992$                   & $\,\pm\,$               & $0.0016$     & $0.1009$                     & $\,\pm\,$ & $0.0014$     & $1.1$  & $0.1044$ & $\,\pm\,$ & $0.0024$ \\
		$\GammaW\,[\GeV]$          & $2.085$                    & $\,\pm\,$               & $0.042$      & $2.080$                      & $\,\pm\,$ & $0.017$      & $-0.1$ & $2.079$  & $\,\pm\,$ & $0.018$  \\
		$\BrWe$                    & $0.1071$                   & $\,\pm\,$               & $0.0016$     & $0.1084$                     & $\,\pm\,$ & $0.0010$     & $0.8$  & $0.1085$ & $\,\pm\,$ & $0.0012$ \\
		$\BrWmu$                   & $0.1063$                   & $\,\pm\,$               & $0.0015$     & $0.1087$                     & $\,\pm\,$ & $0.0009$     & $1.6$  & $0.1098$ & $\,\pm\,$ & $0.0012$ \\
		$\BrWtau$                  & $0.1138$                   & $\,\pm\,$               & $0.0021$     & $0.1088$                     & $\,\pm\,$ & $0.0011$     & $-2.4$ & $0.1063$ & $\,\pm\,$ & $0.0014$ \\
		$\RWmue$                   & $1.0034$                   & $\,\pm\,$               & $0.0063$     & $1.0031$                     & $\,\pm\,$ & $0.0058$     & $-0.1$ & $1.001$  & $\,\pm\,$ & $0.015$  \\
		$\RWtaue$                  & $0.994$                    & $\,\pm\,$               & $0.021$      & $1.004$                      & $\,\pm\,$ & $0.010$      & $0.5$  & $1.007$  & $\,\pm\,$ & $0.011$  \\
		$\RWtaumu$                 & $0.990$                    & $\,\pm\,$               & $0.011$      & $1.001$                      & $\,\pm\,$ & $0.009$      & $1.0$  & $1.025$  & $\,\pm\,$ & $0.016$  \\
		\bottomrule
	\end{tabular}
	\caption{Inputs to the EWPD SMEFT fit, alongside the fit results obtained using a NLO SMEFT parametrization.
		Pull values are calculated as the difference of fit result and direct measurement, divided by the uncertainty of the direct measurement. The column labeled ``indirect'' contains the result of a fit that does not include the direct measurement corresponding to each respective row. A dash indicates that the indirect prediction is impossible or possible only with very poor precision.}
	\label{tab:smeftfit}
\end{table}

\subsection{Comparison with existing likelihoods}
\label{sec:smefit}
Most existing EWPD likelihoods~\citep{Han:2004az,Pomarol:2013zra,Falkowski:2014tna,Efrati:2015eaa,Berthier:2015gja,deBlas:2017wmn,daSilvaAlmeida:2018iqo,Biekotter:2018ohn,Aebischer:2018iyb,Ellis:2018gqa,Falkowski:2019hvp,Ellis:2020unq,Corbett:2021eux,Bellafronte:2023amz} are constructed using SMEFT parametrizations based on the \{$\alpha$, $M_Z$, $G_\mu$\} input parameter scheme. However, this scheme is not compatible with the \{$M_W$, $M_Z$, $G_\mu$\} scheme increasingly favoured in LHC analyses~\cite{Brivio:2021yjb}. To support a consistent combination of EWPD and LHC data, \ewpdlhc provides SMEFT parametrizations in five alternative input schemes, summarized in Table~\ref{tab:schemes}. It also incorporates the latest electroweak precision measurements and fully accounts for correlated uncertainties, an essential feature when considering alternative scheme choices.

\begin{figure}
\centering
\includegraphics[width=0.5\textwidth]{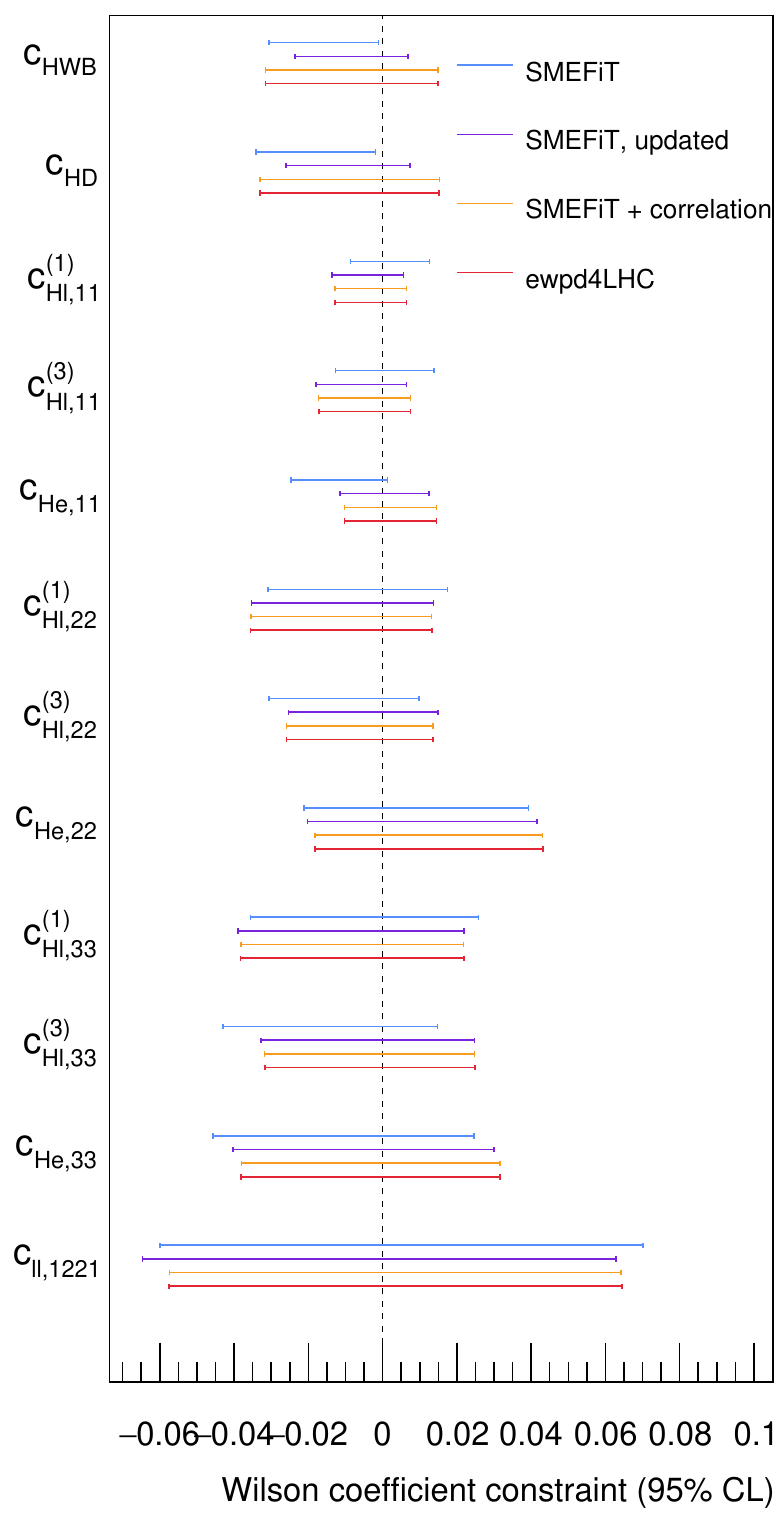}%
\caption{Comparison of one-at-a-time constraints on Wilson coefficients from EWPD, obtained using \ewpdlhc and different configurations of \textsc{SMEFiT}~\cite{Giani:2023gfq,Celada:2024mcf}. The default \textsc{SMEFiT} results are shown alongside those using updated experimental inputs and SM predictions, consistent with \ewpdlhc. A third configuration includes the correlation model from \ewpdlhc. The SMEFT parametrization in \textsc{SMEFiT} remains unchanged. Only operators affecting lepton couplings are included, and $\Lambda=1\,\TeV$ is assumed.}
\label{fig:SMEFiT}
\end{figure}

A likelihood based on the \{$M_W$, $M_Z$, $G_\mu$\} scheme is also implemented in \textsc{SMEFiT}~\cite{Giani:2023gfq,Celada:2024mcf}, where it has been validated against the likelihood presented in Ref.~\cite{Brivio:2017bnu}. Figure~\ref{fig:SMEFiT} compares the resulting one-at-a-time constraints on Wilson coefficients affecting only lepton couplings, as obtained from \ewpdlhc and different configurations of \textsc{SMEFiT}. The default \textsc{SMEFiT} results differ from \ewpdlhc constraints by up to one standard deviation, which can be traced to outdated input values for measurements and SM predictions. When the latest inputs and the correlation model from \ewpdlhc are incorporated into \textsc{SMEFiT}, constraints are weakened by up to 50\%. After both updates, the two tools yield excellent agreement.

\section{Impact of recent LHC measurements on the EWPD likelihood}
\label{sec:impact}
In 2024, the ATLAS and CMS collaborations have published three measurements that have provided important updates on weak boson properties.
A measurement of the $W$ mass and width~\cite{ATLAS:2024erm}, the lepton universality of $W$ branching fractions~\cite{ATLAS:2024tlf}, and the effective leptonic weak mixing angle~\cite{CMS:2024ony}.\footnote{While finalizing this paper,
	a precise measurement of the $W$ mass~\cite{CMS:2024nau} was presented by the CMS collaboration.
	It is not considered here as it is a preliminary result and because it assumes, in its current iteration, the SM $W$ width.}
This section examines how these measurements influence the EWPD likelihood in the SMEFT.

\subsection{ATLAS measurement of $W$ boson mass and width}
\label{sec:MWGammaW}
\begin{figure}
	\centering
	\includegraphics[width=0.6\textwidth]{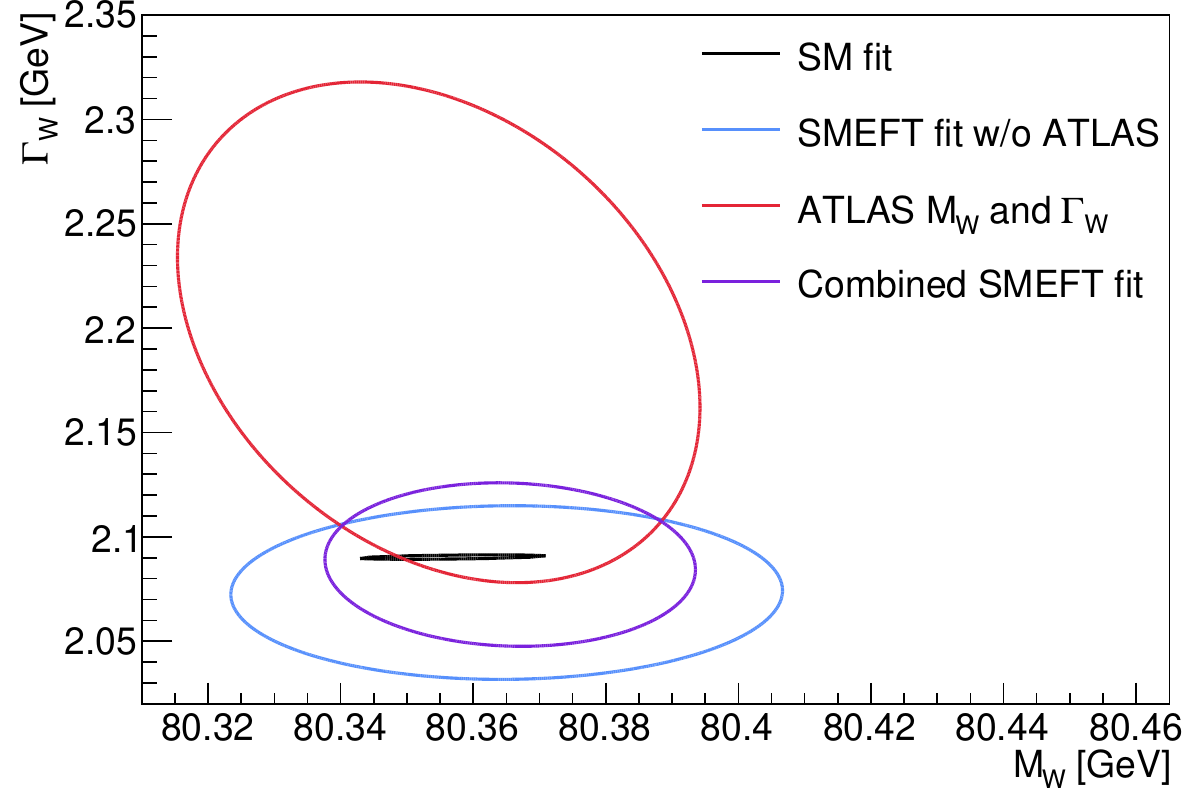}%
	\caption{The direct ATLAS measurement of $M_W$ and $\Gamma_W$ compared to the SM fit result as well as the SMEFT fit results at $O(\Lambda^{-2})$, without assuming flavour universality. The SMEFT fits are performed with and without the ATLAS measurement, to demonstrate its impact.
		Only variations of Wilson coefficients that have a significant effect for values of $\frac{c_i}{\Lambda^2}\lesssim\frac{1}{\TeV^2}$ are considered.
		The value of \MW in the SMEFT fit corresponds to the combination of direct measurements, as indirect constraints are weak while both direct and indirect constraints exist for \GammaW.}
	\label{fig:W}
\end{figure}

The simultaneous measurement of the $W$ boson mass and width by ATLAS
has an important impact on the EWPD likelihood.
In Figure ~\ref{fig:W}, the ATLAS measurement is compared with predictions from the SM and SMEFT fits, using the framework described in Section~\ref{sec:likelihood}.

Without ATLAS data, the \MW used as input to the fit is a combined value from
measurements from LEP~\cite{ALEPH:2013dgf}, D0~\cite{D0:2013jba}, and
LHCb~\cite{LHCb:2021bjt}, as the PDG average already contains an ATLAS
measurement of \MW~\cite{ATLAS:2017rzl} that is superseded by
Ref.~\cite{ATLAS:2024erm} while the CDF measurement is excluded due to
incompatibility~\cite{LHC-TeVMWWorkingGroup:2023zkn}. The combination uses the
CT18 set of parton distribution functions (PDFs), in line with the procedure of
the LHC-TeV \MW working group~\cite{LHC-TeVMWWorkingGroup:2023zkn}, resulting
in combined value of $\MW^\textrm{direct,\,no\,ATLAS}=80.365\pm17\,\GeV$.

In the SM, both \MW and \GammaW can be predicted with higher precision than
direct measurements achieve. Consequently, incorporating the ATLAS measurement
has small effect on \MW (that is not shown in Figure~\ref{fig:W}) while its
impact is negligible for \GammaW.

However, in the SMEFT framework, $M_W$ cannot be predicted indirectly with high
precision. The SM value of \MW is determined, at tree level, by $\alpha$,
$M_Z$, and $G_\mu$, but in the SMEFT \MW also depends on multiple Wilson
coefficients. This dependence arises from field redefinitions necessary in the
presence of certain dimension-six operators and from modifications to the muon
decay rate used to determine $G_\mu$, which receives contributions from
operators affecting $W$ couplings to leptons and an operator introducing a
four-fermion interaction~\cite{Brivio:2021yjb}. As a result, the SMEFT fit
without the ATLAS data yields an \MW value identical to the LEP+D0+LHCb
combination of direct measurements.

The ATLAS measurement, being more precise than the combination of other direct
measurements (as long as the CDF and CMS results are excluded), substantially
enhances the precision of \MW in the SMEFT. In input parameter schemes that
exclude \MW, this directly constrains the combination of Wilson coefficient
that influences \MW, providing one of the leading constraints on BSM physics.
In the schemes including \MW, this improvement results in more precise SM
predictions, equally improving SMEFT constraints.

In contrast, \GammaW in the SMEFT can be predicted, at least at LO, with higher
precision indirectly than is possible with direct measurements (albeit much
lower than the precision of the SM prediction).\footnote{It can be shown
	straightforwardly demonstrated this is sensible, using the known dimension-six
	Wilson coefficient dependence of only a few observables. Specifically, in the
	\{$M_W$,$M_Z$,$G_\mu$\} scheme, the relative deviation in the leptonic $W$
	branching fraction due to dimension-six Wilson coefficients is a linear
	combination (with strictly positive coefficients) of the relative deviation
	from the SM expectation of $\alpha$, the invisible $Z$ width and the partial
	width of the $Z$ boson for decays into left-handed leptons (the decay rate into
	left-handed leptons can be inferred from the partial $Z$ boson widths, in
	combination with the asymmetry parameters $A_\ell$). All three quantities are
	measured with a precision of 0.5\% or better. Consequently, the partial
	$W\rightarrow\ell\nu$ width can be predicted with even higher precision. By
	dividing this result by $\mathcal B(W\rightarrow\textrm{leptons})$ yields
	\GammaW, where the uncertainty of this \GammaW prediction is primarily
	determined by the 0.8\% uncertainty in the measurement of the leptonic
	branching fraction.} This prediction relies on two central SMEFT assumption:
that the SM gauge symmetries are valid -- relating $W$ and $Z$ couplings,
allowing for the translation of the more precise $Z$ coupling constraints to
$W$ coupling constraints -- and that there are no new light states into which
SM particles, including the $W$ boson, could decay. Although the ATLAS
measurements of \GammaW is nearly as precise as the combination of previous
direct measurements, its impact on the global SMEFT fit is modest because the
indirect SMEFT prediction for \GammaW has a precision of 18 \MeV precision,
compared to the 49\,\MeV precision of the ATLAS measurement.

If NLO corrections to SMEFT parametrizations are taken into account, it is
generally no longer possible to predict \GammaW indirectly. However, as
discussed in Section~\ref{sec:SMEFTfit}, observable modifications to \GammaW
would require extremely large Wilson coefficient values for operators appearing
exclusively at loop level, particularly $c_{tW}$ and $c_{lQ}^{(3)}$, which are
largely excluded by measurements of $t\bar{t}W$ production. This non-trivial
relationship between the $W$ boson width and the $t\bar{t}W$ production
cross-section further underscores the importance of combining EWPD and LHC
data. For Figure~\ref{fig:W}, the weakly constrained linear combination
involving $c_{tW}$ and $c_{lQ}^{(3)}$ has been fixed to zero, consistent with
the approach described in the previous section.

The publication of the correlation of the $W$ boson mass and width measurements
is of particular importance for the SMEFT analysis. The SMEFT fit prefers a
lower value of \GammaW than ATLAS, which, due to the negative correlation
between \MW and \GammaW, leads to a larger \MW value. In fact, the global SMEFT
fit result for \MW is 6\,\MeV larger than it would be if zero correlation were assumed
for the ATLAS \MW and \GammaW measurements. Furthermore, the uncertainty in \MW
in the SMEFT fit is larger than it would be if the SM width were assumed in its
extraction, but smaller than the uncertainty that would result from a fit that
is agnostic of \GammaW.

Intriguingly, the ATLAS measurement of \GammaW deviates by about two standard
deviations from the SM prediction. However, this deviation cannot be attributed
to BSM physics compatible with SMEFT assumptions, as the SMEFT fit, even with
the ATLAS data included, aligns closely with the SM value of \GammaW.

\subsection{ATLAS measurement of lepton flavour universality in $W$ decays}
\label{sec:WBR}
Another precise measurement of $W$ boson properties recently published by ATLAS is the precise study of lepton flavour universality in $W$ decays~\cite{ATLAS:2024tlf}.
The ratio of $W$ branching fractions $R^{\mu/e}_W$ is determined with an exceptional precision of 0.45\%.
This ratio,  which compares the $W$ boson decays into muons and neutrinos against electrons and neutrinos, is sensitive to anomalous couplings of the $W$ boson to leptons.
In the framework of the SMEFT it constrains exactly two Warsaw basis Wilson coefficients (at LO -- the small NLO contributions are discussed at the end of this section):
\begin{equation}
	R^{\mu/e}_W=\frac{\mathcal{B}(W\rightarrow \mu\nu)}{\mathcal{B}(W\rightarrow e\nu)}\approx1+2\delta g^{W\mu}-2\delta g^{We}=1+\frac{2v^2}{\Lambda^2}\left(c^{(3)}_{H\ell,22}-c^{(3)}_{H\ell,11}\right)\,,
\end{equation}
where,  $g^{W\ell}$ is the dimensionless coupling of the $W$ to a charged lepton $\ell$ and its corresponding neutrino, and $\delta g^{W\ell}$ denotes its deviation from the SM expectation.
The approximation $\Gamma(W\rightarrow\ell\nu)=\Gamma_{\textrm{SM}}(W\rightarrow\ell\nu)(1+\frac{\delta g_{W}}{g_{W}})^2\approx\Gamma_{\textrm{SM}}(W\rightarrow\ell\nu)(1+2\frac{\delta g^{W\ell}}{g^{W\ell}})$ is used.
In contrast to the total width, partial widths for individual lepton flavours cannot be inferred from $Z$ boson measurements. This would require flavour-specific measurements of $Z$ boson decays to neutrinos, when only the inclusive invisible width can be observed.

To mitigate uncertainties in lepton identification, ATLAS does not directly
measure $R^{\mu/e}_W$ but instead fits the parameter
\begin{equation}
	R^{\mu/e}_{WZ}=  R^{\mu/e}_{WZ}/\sqrt{  R^{\mu/e}_{Z}}\,,
\end{equation}
where $R^{\mu/e}_Z$ represents the ratio of $Z$ boson branching fractions into muon and electron pairs.
This ratio is combined with the $R^{\mu/e}_Z$ result from LEP+SLD, which has its own SMEFT parameter dependence:
\begin{align}
	R^{\mu/e}_Z & =\frac{\mathcal{B}(Z\rightarrow \mu^+\mu^-)}{\mathcal{B}(Z\rightarrow e^+e^-)}\approx1-4\delta g^{Z\mu}_A+4\delta g^{Ze}_A\nonumber \\ &=1+\frac{2v^2}{\Lambda^2}\left(c^{(3)}_{H\ell,22}-c^{(3)}_{H\ell,11}+c^{(1)}_{H\ell,22}-c^{(1)}_{H\ell,11}-c_{He,22}+c_{He,11}\right)\,,
\end{align}
where the dependence $\Gamma(Z\rightarrow\ell\ell)\propto (g^{Z\ell}_V)^2+(g^{Z\ell}_A)^2$ of the $Z$ boson partial width on the vector coupling $g^{Z\ell}_V\approx0$ and axial vector couplings $g^{Z\ell}_A\approx-0.5$ to leptons is used.

The precise ATLAS measurement constrains the following linear combinations of
couplings and Wilson coefficients:
\begin{align}
	R^{\mu/e}_{WZ} & =\frac{\mathcal{B}(W\rightarrow \mu\nu)}{\mathcal{B}(W\rightarrow e\nu)}/\sqrt{\frac{\mathcal{B}(Z\rightarrow \mu\mu)}{\mathcal{B}(Z\rightarrow ee)}}\approx1+2\delta g^{W\mu}-2\delta g^{We}+2\delta g^{Z\mu}_A-2\delta g^{Ze}_A\nonumber \\ &=1+\frac{v^2}{\Lambda^2}\left(c^{(3)}_{H\ell,22}-c^{(3)}_{H\ell,11}-c^{(1)}_{H\ell,22}+c^{(1)}_{H\ell,11}+c_{He,22}-c_{He,11}\right)\,.
\end{align}
The Wilson-coefficient dependence  is similar to that of $R^{\mu/e}_Z$, but it is a factor of two weaker, with the signs of $c^{(1)}_{H\ell,22}$ and $c^{(1)}_{H\ell,11}$ flipped.
Hence the observables constrains an independent direction in Wilson coefficient space while leaving a third direction in the space of LFU violating Wilson coefficients unconstrained.

Without the inclusion of the ATLAS measurement, the global SMEFT fit yields a
value of $R^{\mu/e}_{WZ} = 1.0027 \pm 0.0059$, which is slightly less precise
than the constraint of $R^{\mu/e}_{W}$ (as shown in Table~\ref{tab:smeftfit}).
This indicates that a $R^{\mu/e}_{WZ}$ measurement actually has a more
significant impact on the SMEFT fit than a measurement of $R^{\mu/e}_{W}$ with
the same precision. When the ATLAS $R^{\mu/e}_{WZ}$ measurement is included,
the global fit result -- and consequently the constraint on the corresponding
direction in Wilson coefficient space -- is improved to $R^{\mu/e}_{WZ} =
	1.0003 \pm 0.0034$.

For completeness, the LO Wilson coefficient dependence of $e$--$\mu$
universality violating anomalous $W$ and $Z$ couplings is given below, based on
Ref.~\cite{Dawson:2019clf} and using $g^{Z\ell}_{V/A}=g^{Z\ell}_{R}\pm
	g^{Z\ell}_{L}$:
\begin{align}
	\delta g^{W\mu}-\delta g^{We}     & = \frac{v^2}{\Lambda^2}\left(c^{(3)}_{H\ell,22}-c^{(3)}_{H\ell,11}\right)\,,                                                                             \\
	\delta g^{Z\mu}_V-\delta g^{Ze}_V & = \frac{v^2}{2\Lambda^2}\left(-c^{(3)}_{H\ell,22}+c^{(3)}_{H\ell,11} -c^{(1)}_{H\ell,22}+c^{(1)}_{H\ell,11} -c_{He,22}+c_{He,11}\right)\label{eq:dgV}\,, \\
	\delta g^{Z\mu}_A-\delta g^{Ze}_A & = \frac{v^2}{2\Lambda^2}\left(c^{(3)}_{H\ell,22}-c^{(3)}_{H\ell,11} +c^{(1)}_{H\ell,22}-c^{(1)}_{H\ell,11} -c_{He,22}+c_{He,11}\right)\label{eq:dgA}\,.
\end{align}\\

\subsection{CMS measurement of the effective leptonic weak mixing angle}
\label{sec:st}
The CMS collaboration recently presented a new measurement of the  effective leptonic weak mixing angle, \st~\cite{CMS:2024ony}.
Unlike the measurements of $W$ mass and width, which focus on fairly model-independent characteristics of the reconstructed distributions (see also the discussion in Ref.~\cite{Bjorn:2016zlr}), and the $W$ branching fraction measurements, which constrains  inclusive rates, the \st measurement does not lend itself to straightforward interpretation within the SMEFT framework.
This is because the measurement relies on a mass-dependent analysis of the forward--backward asymmetry in Drell--Yan events, which in the SMEFT context involves a non-trivial dependence not only on lepton but also on quark couplings and four-fermion operators.
Moreover, the measurements in the electron and muon channels have different Wilson coefficient dependencies, preventing their combination in the SMEFT.

However, the CMS measurement can be reinterpreted as one of the most precise
test of lepton-flavour universality. It is primarily sensitive to vector
couplings of the $Z$ boson, thus constraining the third Wilson combinations
combination left unconstrained by the $R^{\mu/e}_Z$ and $R^{\mu/e}_{WZ}$
measurements.

For the LFU interpretation, the ratio of \st measurements in muon and electron
channel is denoted $R^{\mu/e}_{\sin^2\theta^\ell_{\textrm{eff}}}$. The exact
value implied by the CMS measurement is difficult to determine without a
detailed correlation model for the systematic uncertainties. An estimate based
on leptons reconstructed in the central part of the detector (see Table~4 of
Ref.~\cite{CMS:2024ony}), which offers similar coverage for electrons and
muons, yields a value of
$R^{\mu/e}_{\sin^2\theta^\ell_{\textrm{eff}}}=0.9987\pm0.0012~(\textrm{stat.})\pm0.0010~(\textrm{syst.})$.
This estimate assumes that theoretical and PDF uncertainties cancel due to the
similarity in phase space between electron and muon channel measurements. While
some theory uncertainties might affect only one lepton species and PDF
uncertainties may not be fully correlated due to differences in lepton
acceptance and identification efficiency, these effects are expected to be
small. For sensitivity estimation, this approach is considered conservative, as
a more rigorous analysis, including electrons reconstructed in the forward part
of the detector, could further improve the precision. However, a critical
caveat involves the central value of this estimate. During the CMS
determination of the \st results in electron and muon channel, nuisance
parameters, such as those related to the PDFs, will likely be pulled to
different central values, which can differently affect \st in each channel,
whereas consistent parameters should be used in the derivation of
$R^{\mu/e}_{\sin^2\theta^\ell_{\textrm{eff}}}$.

The uncertainty in the extraction of
$R^{\mu/e}_{\sin^2\theta^\ell_{\textrm{eff}}}$ is dominated by statistical
uncertainties and can be straightforwardly improved with future measurements
utilizing larger datasets. In fact, a large part of the systematic uncertainty
in the CMS measurement also arises from statistical limitations due to the
finite size of Monte Carlo simulated samples. In a dedicated LFU analysis,
these uncertainties could be mitigated by using identical parton level events
(i.e. events before simulating final state photon radiation or detector and
reconstruction effects that differentiate electrons and muons) for analyzing
electron and muon channels, or by reweighting simulated distributions to match
channels at parton level, resulting in correlated Monte Carlo statistical
uncertainties that cancel in the ratio
$R^{\mu/e}_{\sin^2\theta^\ell_{\textrm{eff}}}$.

Using the relationship
\begin{equation}
	\frac{g^{Zf}_{V}}{g^{Zf}_{A}}=1-4|Q_f|\stf
	\label{eq:st}
\end{equation}
(see e.g. Ref.~\cite{ALEPH:2005ab}) as well as $\frac{g^{Z\ell}_V}{g^{Z\ell}_A}\ll1$, and $g^{Z\ell}_A\approx-0.5$ one finds:
\begin{align}
	\label{eq:approxRst}
	R^{\mu/e}_{\sin^2\theta^\ell_{\textrm{eff}}} & =\frac{\sin^2\theta^\mu_{\textrm{eff}}}{\sin^2\theta^e_{\textrm{eff}}}\approx\frac{1-\frac{g^{Z\mu}_V}{g^{Z\mu}_A}}{1-\frac{g^{Ze}_V}{g^{Ze}_A}}\approx1-\frac{g^{Z\mu}_V}{g^{Z\mu}_A}+\frac{g^{Ze}_V}{g^{Ze}_A}\approx1+2\left( g^{Z\mu}_V- g^{Ze}_V\right) \\
	                                             & =1+\frac{v^2}{\Lambda^2}\left(-c^{(3)}_{H\ell,22}+c^{(3)}_{H\ell,11}-c^{(1)}_{H\ell,22}+c^{(1)}_{H\ell,11}-c_{He,22}+c_{He,11}\right)\,.
\end{align}
The measurement of the observable evidently constrains a direction in Wilson coefficient space that is linearly independent of the constraints from $R^{\mu/e}_Z$ and $R^{\mu/e}_{WZ}$.

With the experimental result of
$R^{\mu/e}_{\sin^2\theta^\ell_{\textrm{eff}}}=0.9987\pm0.0016$ it is thus
possible to determine the difference in the vector couplings of muons and
electrons with high precision. A more accurate expansion of
Equation~\ref{eq:approxRst} that takes into account a non-zero SM value of
$g^{Z\mu}_V$ yields
\begin{align}
	g^{Z\mu}_V-  g^{Ze}_V=\left(g^{Z\ell,\textrm{SM}}_A-g^{Z\ell,\textrm{SM}}_V\right)\left(1-R^{\mu/e}_{\sin^2\theta^\ell_{\textrm{eff}}}\right)=0.46\left(R^{\mu/e}_{\sin^2\theta^\ell_{\textrm{eff}}}-1\right)
\end{align}
from which one obtains
\begin{equation}
	g^{Z\mu}_V- g^{Ze}_V = (-6 \pm 7)\times10^{-4}\,.
\end{equation}
This is over three times more precise than the current difference in PDG values for  $g^{Z\mu}_V$ and $g^{Ze}_V$, which is $(15 \pm 23)\times10^{-4}$.
Combining the electron vector coupling measurement from LEP+SLD with the above coupling-difference constraint also yields an improved determination of the muon coupling to the $Z$ boson:
\begin{equation}
	g_V^{Z\mu}= (-382\pm4 - 6 \pm 7)\times10^{-4}= (-388 \pm 8)\times10^{-4}\,.
\end{equation}
This result improves on the precision of the PDG value by a factor of three.
As it depends only weakly on theoretical uncertainties and is statistically limited it can be further improved with a larger dataset or by an ATLAS measurement.
The main caveats in this derivation are the assumptions that four-fermion operators or anomalous quark couplings have no significant impact, which is likely a reasonable approximation, and that PDF pulls match between the electron and muon channels in Ref.~\cite{CMS:2024ony}, which cannot be validated externally and requires a dedicated fit of  $R^{\mu/e}_{\sin^2\theta^\ell_{\textrm{eff}}}$  by the CMS collaboration.

\subsection{Comparison of lepton flavour universality tests}
\begin{figure}
	\includegraphics[width=0.33\textwidth]{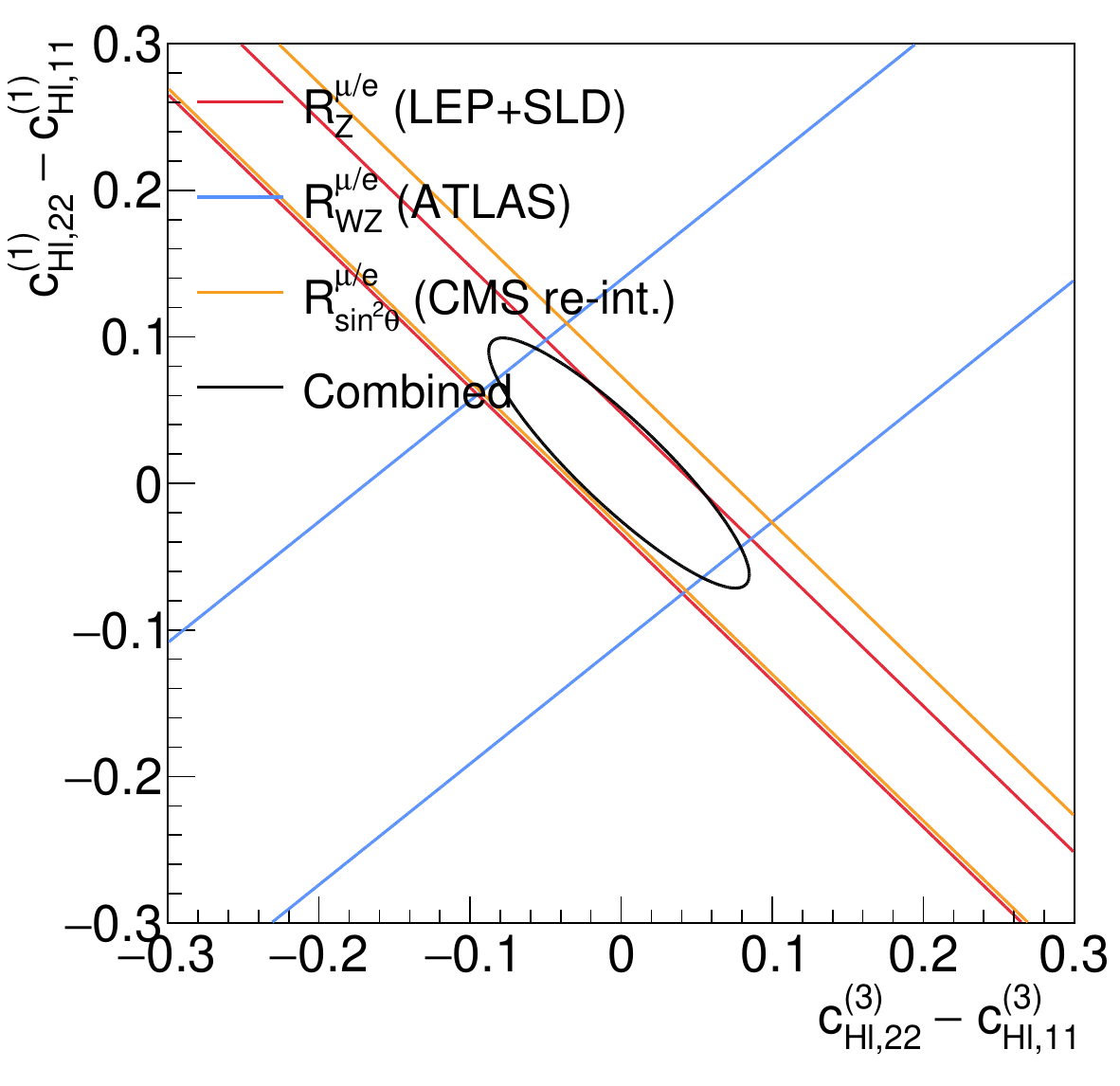}%
	\includegraphics[width=0.33\textwidth]{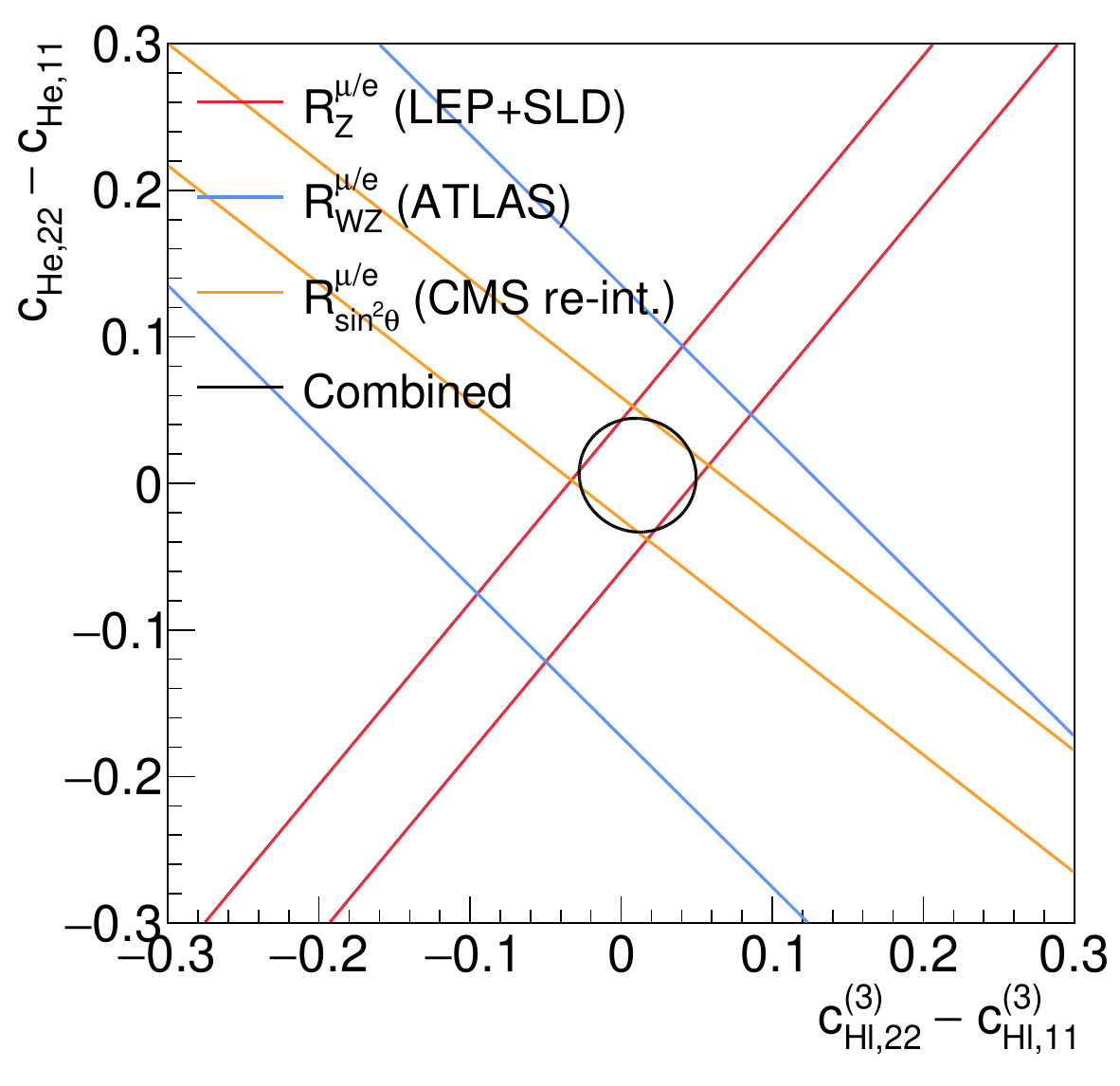}%
	\includegraphics[width=0.33\textwidth]{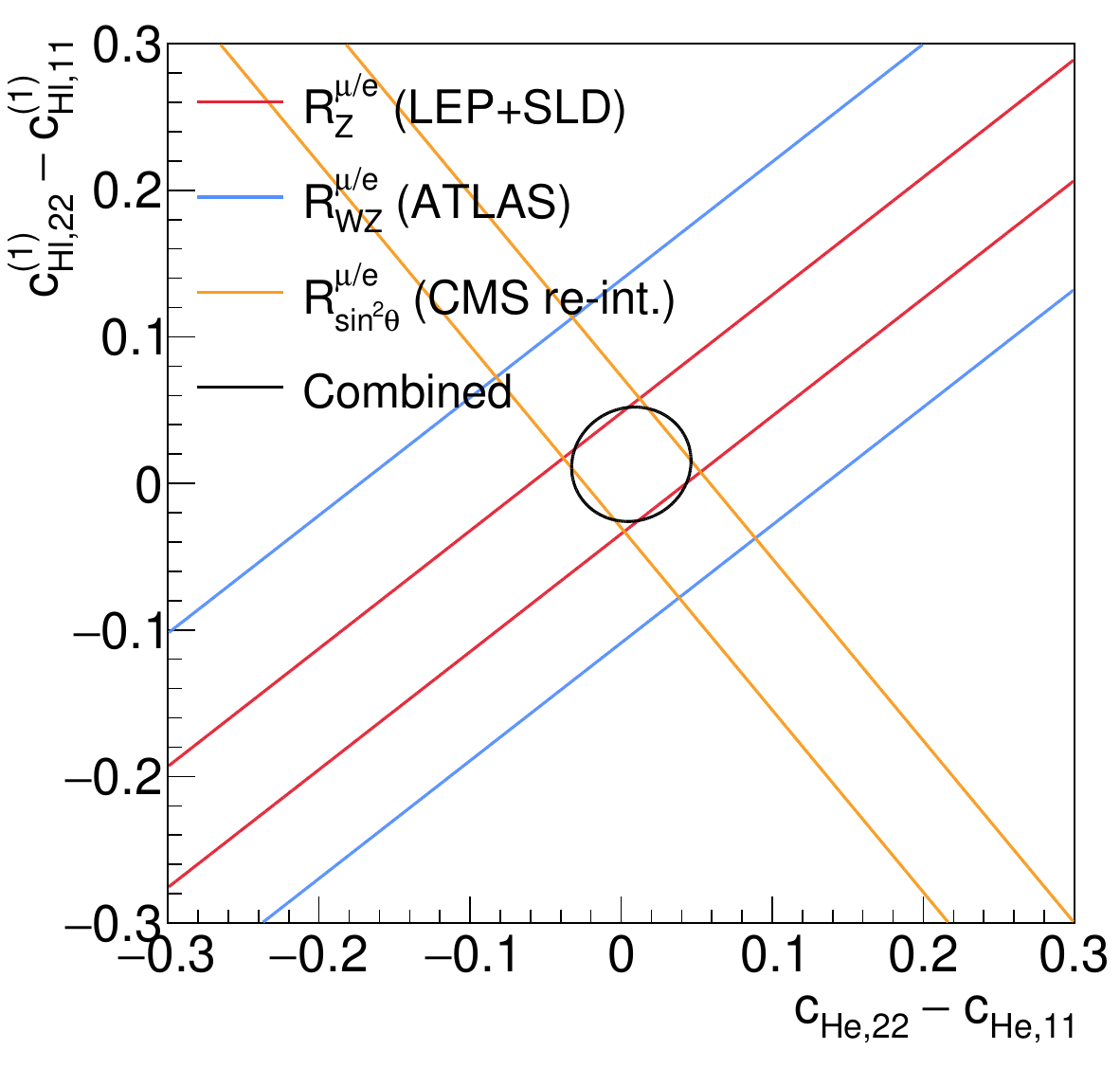}
	\caption{Constraints on the three Wilson coefficient combinations introducing leading-order $e$--$\mu$ lepton-flavour universality violating effects in weak boson couplings, at  95\% confidence level, for $\Lambda=1\,\TeV$. The Wilson coefficient combination not shown in each of the plots is fixed to zero. The impact of the LEP+SLD measurement of  $R^{\mu/e}_{Z}$~\cite{ALEPH:2005ab}, the ATLAS measurement of  $R^{\mu/e}_{WZ}$~\cite{ATLAS:2024tlf}, and the reinterpretation of the CMS weak mixing angle measurement~\cite{CMS:2024ony} as a measurement of $R^{\mu/e}_{\sin^2\theta^\ell_{\textrm{eff}}}$ is compared.
		For the individual ratio measurements, only one degree of freedom is constrained. Hence, the corresponding confidence intervals are based on a one-dimensional $\chi^2$ distribution. Confidence intervals for the combination correspond to a two-dimensional $\chi^2$ distribution.}
	\label{fig:lfu}
\end{figure}

In \ewpdlhc the exact NLO Wilson coefficient dependence of $R^{\mu/e}_{WZ}$,
$R^{\mu/e}_{Z}$, and $R^{\mu/e}_{\sin^2\theta^\ell_{\textrm{eff}}}$ on the
three types of Wilson coefficients is implemented, which is used to derive
confidence levels for Wilson coefficients in Figure~\ref{fig:lfu}. It is clear
from the three plots that each of the three observables is almost perfectly
suited to constrain a direction in parameter space that is unconstrained by the
other two observables.

The constraints based on $R^{\mu/e}_{Z}$ and
$R^{\mu/e}_{\sin^2\theta^\ell_{\textrm{eff}}}$, of which only the latter can be
improved in the near future, are more precise than those based on
$R^{\mu/e}_{WZ}$. A dedicated measurement of
$R^{\mu/e}_{\sin^2\theta^\ell_{\textrm{eff}}}$ by either ATLAS or CMS thus
clearly has the potential to become the most precise test of lepton flavour
universality.

At NLO, additional Wilson coefficients contribute, although their impact is at
least a factor of 40 smaller compared to the Wilson coefficients contributing
at tree level. The most interesting and numerically largest contributions arise
due to four-fermion operators coupling top quark and leptons, \clQone , \clt,
\clQthree, \cet, and \cQe. While \clQone and \clt as well as \cet and \cQe
cannot be distinguished by the three observables, the remaining three
directions are constrained by the three types of measurements, similar to the
LO Wilson coefficients.

\section{Interpretation of the ATLAS Drell--Yan triple-differential cross-section measurement and impact on the global EWPD fit}
\label{sec:Z3D}
In the previous section the measurement of $R^{\mu/e}_{\sin^2\theta^\ell_{\textrm{eff}}}$ was identified as the potentially most precise test of the lepton flavour universality of weak boson couplings.
However, certain approximations were necessary to extract couplings values from the CMS measurement of \st.
For instance, it was assumed that some systematic uncertainties as well as quark-coupling-modifying operators affect electron and muon channels identically and thus do not interfere in the interpretation of $R^{\mu/e}_{\sin^2\theta^\ell_{\textrm{eff}}}$.
Moreover, the influence of four-fermion operators directly coupling quarks to leptons was ignored.
In this section, a more accurate analysis of the forward--backward asymmetry will be performed that accounts for all relevant SMEFT operators and employs a detailed correlation model for systematic uncertainties. This approach will demonstrate that such an analysis can not only test flavour universality but also constrain anomalous quark couplings.

The ATLAS triple-differential cross-section measurement of Drell--Yan
production~\cite{ATLAS:2017rue}, based on LHC Run~1 data at 8\,\TeV, will be
analyzed for this purpose. This model-independent and granular cross-section
measurement, published with a detailed breakdown of systematic uncertainties in
each analysis channel, allows for an accurate interpretation in the SMEFT.

A SMEFT analysis of $Z$ production at the LHC was previously performed in Ref.~\cite{Breso-Pla:2021qoe},
with a focus on quark couplings only and a greatly simplified uncertainty model.

\subsection{Analysis setup}
In Drell--Yan production of a lepton pair in hadron collision, a
forward--backward asymmetry exists for the direction of the outgoing leptons.
Depending on the invariant mass of the lepton pair, the outgoing negatively
charged lepton is emitted either more or less frequently in the direction of
the incoming quark (as opposed to anti-quark). This asymmetry is caused by the
parity-violating couplings of the $Z$ boson. Near the $Z$ pole, the asymmetric
contribution to the Drell--Yan cross-section is, at leading order, proportional
to the product of the $Z$ boson axial vector couplings and vector couplings to
the colliding quarks and leptons, $g^{Zq}_Vg^{Zq}_Ag^{Z\ell}_Vg^{Z\ell}_A$ (see
for example Equation~3 of Ref.~\cite{CMS:2011utm}). The pole asymmetry is
particularly sensitive to variations in the vector coupling of leptons,
$g^{Z\ell}_V$, as the coupling is close to zero in the SM. Away from the $Z$
pole, the asymmetry is primarily caused by the interference between $Z$ boson
and $\gamma^*$ mediated amplitudes, which is proportional to the $Z$ boson
axial vector couplings to the two fermion types and their photon couplings. The
sensitivity of the forward--backward asymmetry to lepton couplings enables
precise measurements of the effective leptonic weak mixing angle, \st, which is
related to fermion couplings according to Equation~\ref{eq:st}.

ATLAS measured~\cite{ATLAS:2017rue} the Drell--Yan cross-section as a function
of three observables: the invariant mass of the lepton pair, $m_{\ell\ell}$,
the absolute dilepton rapidity, $|y_{\ell\ell}|$, and the angular variable
$\cos\theta^{*}$. In the dilepton centre-of-mass frame, $\theta^{*}$ is defined
as the angle of the outgoing negatively charged lepton to the longitudinal
boost direction of the dilepton system. The longitudinal boost directions
typically corresponds to the incoming valence quark direction, so
$\cos\theta^{*}$ exhibits the forward--backward asymmetry described above. The
asymmetry is more pronounced for large values of $|y_{\ell\ell}|$, where the
boost direction is more likely aligned with the valence quark direction.

The ATLAS measurement was performed in three analysis channels: with a pair of
muons reconstructed in the central part of the detector, a pair of electrons
reconstructed in the central detector, and a pair of electrons with one
reconstructed in the central and the other in the forward part of the detector.
While the measurement of forward electrons is experimentally challenging, it
offers sensitivity to dilepton pairs highly boosted along the beam direction,
for which the asymmetry is less likely to be diluted by a misassignment of the
quark direction.

Over 1000 data points were reported by ATLAS and are available, together with
the impact of various sources of uncertainties, on the HEPData webpage.\footnote{\url{https://www.hepdata.net/record/77492}}
To reduce the complexity of the
presented interpretation, some data points were merged. Instead of analyzing
the full $\cos\theta^{*}$ dependence of the cross-section, only the
forward--backward asymmetry $A_{FB}$ is considered. It is calculated in each
$m_{\ell\ell}\times|y_{\ell\ell}|$ bin as the cross-section of
$\cos\theta^{*}>0$ events minus the cross-section of $\cos\theta^{*}<0$ events,
divided by the total cross-section in that bin. While the integration over
$\cos\theta^{*}$ bins reduces the statistical power of the analysis, it
simplifies the analysis considerably. Rapidity bins are combined such that the
resulting data points correspond to a bin width of at least 0.8 in
$|y_{\ell\ell}|$, which is sufficient to capture rapidity increase of $A_{FB}$.
The full granularity of the $m_{\ell\ell}$ binning is retained, as SMEFT
operators exhibit strong mass dependence. The lowest mass bin, measured only in
the central--central lepton channels, is omitted due to its limited sensitivity
to \st or dimension-six operators. This procedure reduces the number of
analyzed data points to 51: 15 in the $e^+e^-$ central--forward channel and 18
in the central--central channels.

More than 300 unique sources of systematics affect the measurement,
with the dominant ones affecting lepton identification and momentum measurement.
The effect of each source on the merged analysis bins is determined and used to
create a $51\times51$ covariance matrix for the subsequent statistical
analysis.

A SM prediction of Drell--Yan production is generated at next-to-leading order
QCD in the \textsc{Powheg} BOX framework~\cite{Alioli:2010xd,Barze:2012tt} with
the \textsc{NNPDF3.1nnlo} set of parton distribution
functions~\cite{NNPDF:2017mvq}. Higher-order corrections to \st are taken into
account by setting the weak mixing angle to 0.23130, the SM value predicted
from the measurements of \{$M_W$,$M_Z$,$G_\mu$\} using the formulas discussed
in Section~\ref{sec:predictions}. The linear dependence on \MW is included as a
parametric uncertainty in the analysis, while the smaller variations due to
other SM input quantities are neglected.

The dominant uncertainty in the SM prediction originates from PDFs, stemming
from uncertainties in the relative contributions of up-quark and down-quark
initial states to Drell–Yan production and their predicted momentum fractions.
Due to different SM couplings, up-quark and down-quark final states yield different
asymmetries while other quark flavours create no asymmetry as quarks and anti
quarks carry, on average, equal momentum. Additionally, the distribution of the
momentum fraction carried by up and down (anti) quarks affects the number of
events in which the anti quark carries a larger momentum fraction than the
valence quark, diluting the observable asymmetry. PDF uncertainties are
propagated using the Hessian variations of the PDF set and incorporated into
the statistical analysis using nuisance parameters with Gaussian constraints.
The fit thus simultaneously extracts parameters of interest and updated
PDFs from the data, albeit in an approximation that is only valid if pulls
and nuisance parameter constraints remain small~\cite{Paukkunen:2014zia}.

A further subtlety arises from the fact that the \textsc{NNPDF3.1nnlo} set used for predictions is derived under the assumption of SM couplings. In principle, PDFs can exhibit non-negligible dependence on SMEFT Wilson coefficients~\citep{Carrazza:2019sec,Hammou:2023heg,Gao:2022srd,Costantini:2024xae}. For instance, significant deviations in quark couplings to the $Z$ boson would alter the interpretation of neutral-current measurements employed in PDF determinations.
However, existing studies suggest that such effects generally have only a modest impact on SMEFT fits~\cite{Hammou:2023heg,Gao:2022srd,Costantini:2024xae}.
Moreover, many measurements important for the PDF constraints relevant to this analysis are expected to be largely insensitive to SMEFT effects.
For example, the interpretation of $W$ charge asymmetry in terms of up-quark and down-quark PDFs remains valid because the SMEFT modifies $W^+$ and $W^-$ couplings to up- and down-quarks in a correlated manner,  and the flavour symmetry assumption adopted in this analysis limits variations in second-generation quark couplings.
A comprehensive understanding of PDF-SMEFT interplay would require an extension of the analysis in Ref.~\cite{Costantini:2024xae} to include a more complete set of SMEFT effects, for example for single boson production and deep inelastic scattering data, followed by a combined global fit of PDFs and SMEFT Wilson coefficients. However, such an undertaking is beyond the scope of this work.

The values of \AFB extracted from the measurement, as well as the SM prediction
and their respective uncertainties, were validated against the measurement and
predictions published by ATLAS~\cite{ATLAS:2017rue}, using a matching PDF and
value for \st. Excellent agreement for the measured value was found, for which
however only a combination of electron and muon channels is published. Good
agreement was also found for the theoretical predictions, despite the absence
of NNLO QCD and NLO EW k-factors, which are applied only for the ATLAS
prediction.

The linear impact of dimension-six operators on \AFB is calculated at leading
order and in the \{$M_W$,$M_Z$,$G_\mu$\} input parameter scheme using
\texttt{MadGraph5\_aMC@NLO} with the \texttt{SMEFTsim}~3.0 model. As in the
EWPD fit presented in the first part of the paper, a $U(2)_q\times U(2)_u\times
	U(2)_d$ asymmetry is assumed in the generation of SMEFT effects. Variations of
up-quark and down-quark couplings are thus always accompanied by the same
variation of charm-quark and strange-quark couplings. As the latter are
symmetric in $\cos\theta^{*}$, \AFB is diluted and the sensitivity to light
quark couplings is slightly weaker than in a fully flavour general approach.

\begin{figure}
	\includegraphics[width=\textwidth]{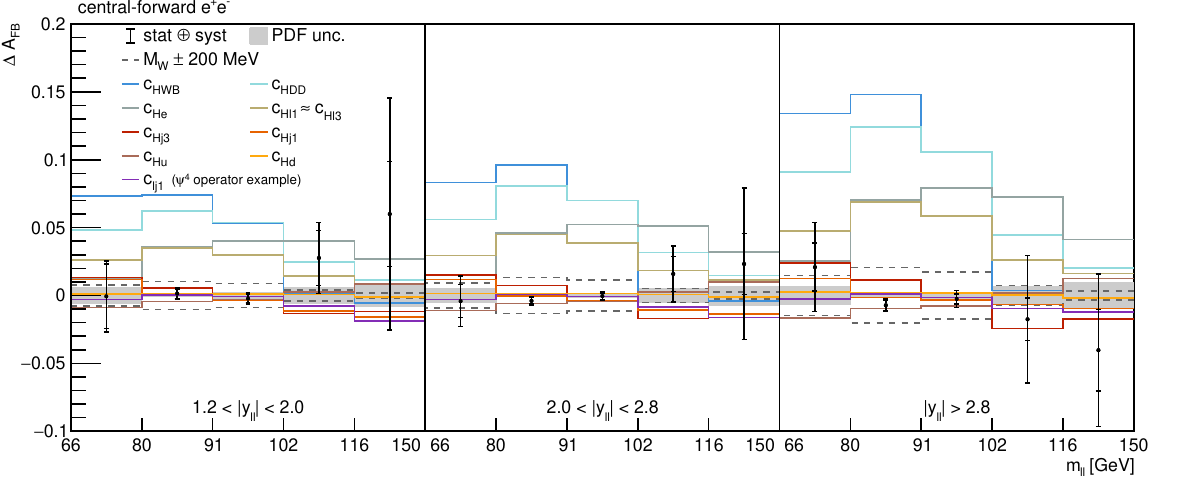}\\
	\includegraphics[width=\textwidth]{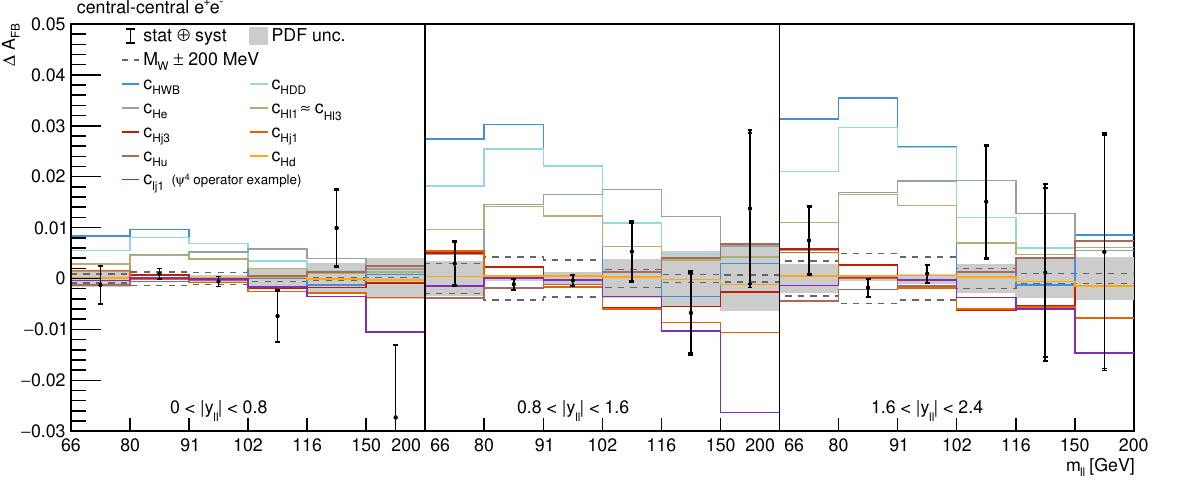}\\
	\includegraphics[width=\textwidth]{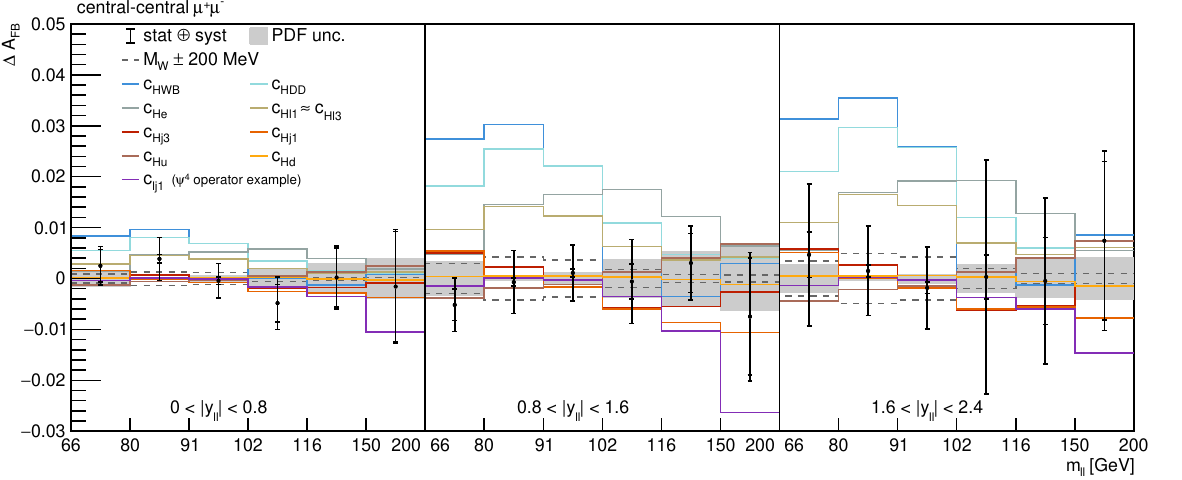}
	\caption{Difference between the measured~\cite{ATLAS:2017rue}  forward--backward asymmetry \AFB in Drell--Yan production in 8\,\TeV proton--proton collisions and the SM prediction.
		The linearized shifts in \AFB due to systematic uncertainties and dimension-six Wilson coefficients, for $\frac{c_i}{\Lambda^2}=\frac{1}{\TeV^2}$, are also shown.
		The panels display, from top to bottom, the central--forward $e^+e^-$, central--central $e^+e^-$, and central--central $\mu^+\mu^-$ channels.
		The inner error bars of data points represent the statistical uncertainty while the outer error bars represent the total uncertainty of the measurement.
		The impact of the Wilson coefficient \cHlone closely matches the impact of \cHlthree. Similarly, the effect of the various four-fermion operators is comparable, with only \cljone shown as a representative example.}
	\label{fig:Z3D}
\end{figure}

Figure~\ref{fig:Z3D} shows the difference between the measured \AFB and the SM
prediction, denoted as $\Delta\AFB$. The figure also displays the deviations
from the nominal values introduced by systematic uncertainties and
dimension-six operators. The data is in good agreement with the SM prediction.
Systematic uncertainties are large, in particular in the central--forward
$e^+e^-$ channel due to experimental challenges in forward-electron identification. In the
$\mu^+\mu^-$ channel, uncertainties are large for more forward events, mainly
due to the sagitta bias, which results from potential detector misalignment and
introduces a charge-dependent muon momentum uncertainty. PDF uncertainties are
minimal in the bin below the $Z$ pole and increase on either side. Variations
in \MW affect the SM prediction of the weak mixing angle, leading to shifts in
the predicted asymmetry.

For a fixed value of \MW, the Wilson coefficients \cHWB and \cHDD modify the
couplings of all fermions to both the $Z$ boson and the photon, thereby
influencing \AFB. The coefficient \cHe affects only couplings of the $Z$ boson
to right-handed leptons while \cHlone and \cHlthree affect the coupling of the
$Z$ boson to charged left-handed leptons. The effect of \cHlone and \cHlthree
on $Z$ boson couplings to charged leptons is identical but only the latter
affects $\alpha$, leading to a small difference in their influence on \AFB. On
the $Z$ pole, the sensitivity is maximal for a simultaneous increase of
left-handed and right-handed lepton couplings, which corresponds to a
modification of the vector coupling, while the sensitivity to changes in
opposite directions, which corresponds to a modification of the lepton axial
vector coupling, is relatively small. The coefficients \cHu and \cHd modify the
couplings of right-handed up and down quarks to the $Z$ boson, respectively,
while \cHqone and \cHqthree affect left-handed quark couplings. The greater
sensitivity to modifications in lepton couplings compared to quark couplings is
evident. Seven four-fermion operators (\cljone, \cljthree, \clu, \cld, \ceu,
\ced, and \cje) couple up and down quarks to leptons of various chirality. For
clarity, only one of these operators is shown as an example in the figure. The
impact of the four-fermion operators on \AFB increases strongly with the
invariant mass and is suppressed near the $Z$ pole. All operators except \cHWB
and \cHDD carry fermion indices that are not explicitly shown, with operators
with subscript $_{11}$ ($_{22}$) affecting electron (muon) channel only.

\subsection{Extraction of the effective leptonic weak mixing angle}
\label{sec:sttest}
Before discussing the SMEFT interpretation of the data, this section presents an extraction of the effective leptonic weak mixing angle using the setup described above.
The ATLAS collaboration has previously measured \st using the 8\,\TeV dataset~\cite{ATLAS:2018gqq}, which serves as a benchmark of the methodology of this analysis.
However, the ATLAS analysis used a different approach, where multiple angular coefficients of the Drell--Yan cross-section were simultaneously extracted in a fit.
The value of the angular coefficient $A_4$, which quantifies the $\cos\theta^{*}$ dependence of the cross-section, was used to determine \st.
Differences in event selection criteria, $m_{\ell\ell}$ binning, slightly different calibrations, and different SM predictions also prevent a direct comparison with the value of \st obtained in this analysis.

A modification of \st introduces a shift in \AFB that is approximately linear.
Consequently, it can be determined by algebraically solving the $\chi^2$ fit,
analogous to Equation~\ref{eq:likelihood}. For this purpose, measurement
uncertainties are included in the $51\times51$ covariance matrix, while PDF
uncertainties are treated as nuisance parameters.

\begin{table}
	\centering
	\begin{tabular}{@{}llr@{}}
		\toprule
		\st $[10^{-5}]$               & ATLAS $A_4$              & Interpretation of ATLAS Z3D \\
		\midrule
		Central--forward $e^+e^-$     & $23166\pm29\pm 23\pm 22$ & $23189\pm36\pm29\pm22$      \\
		Central--central $e^+e^-$     & $23148\pm48\pm31\pm37$   & $23169\pm58\pm26\pm38$      \\
		Central--central $\mu^+\mu^+$ & $23123\pm40\pm27 \pm 35$ & $23099\pm41\pm40\pm35$      \\
		Combined                      & $23140\pm21\pm 16\pm24$  & $23167\pm24\pm20\pm24$      \\
		\bottomrule
	\end{tabular}
	\caption{Comparison of the value of \st (in units of $10^{-5}$) extracted by ATLAS in Ref.~\cite{ATLAS:2018gqq} in an analysis based on the measurement of the angular coefficient $A_4$ and the extraction based on the Drell-Yan triple-differential cross-section measurement (Z3D)~\cite{ATLAS:2017rue} performed in this paper. Uncertainties correspond, in order, to statistical, systematic, and PDF uncertainties.}
	\label{tab:st}
\end{table}

Results based on the reinterpretation of the triple-differential cross-sections
measurement (Z3D) in all three channels and their combination are presented in
Table~\ref{tab:st}, alongside the results of the ATLAS measurement based on
$A_4$. Central values in each channel differ by less than the size of the
statistical uncertainty alone, an appropriate level of compatibility
considering the differences in methodologies.

The discrepancy between electron and muon channel is larger in this
interpretation, driven by the $80\,\GeV < m_{ll} < 91\,\GeV$ bins, which have
the smallest uncertainties and are most strongly impacted by variations of \st.
These bins show opposite deviations from the SM prediction in the two channels.
A similar effect would be less apparent in the $A_4$ measurement, which jointly
analyzes these events with the neighbouring bins, which tend to deviate in the
opposite direction.

Statistical uncertainties of the Z3D interpretation are slightly larger, in
particular in the electron channels. Two effects, which seem to cancel each
other out in the muon channels, could contribute to these differences. On the
one hand, the Z3D measurement has looser selection requirements. On the other
hand, the fit of the angular coefficient shape of the $A_4$ measurement is a
more powerful use of the data compared to analyzing \AFB only. Systematic
uncertainties of the Z3D analysis are slightly larger in the muon channel,
while they are compatible with the $A_4$ results in the electron channel. These
differences could originate from different calibrations used in the ATLAS
measurements or the different utilization of the data, e.g., the difference in
mass binning. Despite the different PDF sets used, the presented analysis has
nearly identical PDF uncertainties.

Overall, the agreement between the \st interpretation presented here and the
dedicated ATLAS measurement is strong, both in terms of central values and
uncertainties, with some expected differences. The consistency suggests that
the analysis framework can be expected to yield reasonable results in the SMEFT
interpretation ahead.

\subsection{General SMEFT constraints}
The fit of all 30 dimension-six Wilson coefficients affecting the Z3D
measurement is, like the \st interpretation, performed algebraically. The nine
most stringently constrained combinations of Wilson coefficients -- obtained
using the methodology discussed in Section~\ref{sec:SMEFTfit} -- are listed in
Table~\ref{tab:smeftfitZ3D}. Constraints on additional independent directions
are of least a factor of two weaker but still considered in the following
analysis.

\begin{table}
	\centering
	\begin{tabular}{@{}llr@{}}
		\toprule
		Constrained direction (leading contributions)                                                                                                             & $\sigma$ & Pull   \\
		\midrule
		{\footnotesize$-0.66c_{HWB}-\!0.58c_{HD}-\!0.29c_{He,11}-\!0.26c_{Hl,11}^{(3)}-\!0.26c_{Hl,11}^{(1)}$}                                                    & 0.0098   & $1.5$  \\
		{\footnotesize$-0.48c_{He,22}-\!0.42c_{Hl,22}^{(3)}-\!0.41c_{Hl,22}^{(1)}+\!0.39c_{He,11}+\!0.33c_{Hl,11}^{(3)}+\!0.33c_{Hl,11}^{(1)}-\!0.17c_{HWB}$}     & 0.021    & $-1.1$ \\
		{\footnotesize$+0.6c_{He,11}-\!0.46c_{HWB}-\!0.36c_{Hq}^{(3)}+\!0.25c^{(3)}_{lj,11}+\!0.24c_{HD}+\!0.21c_{He,22}+\!0.18c_{Hl,22}^{(3)}$}                  & 0.16     & $1.3$  \\
		{\footnotesize$+0.59c^{(1)}_{lj,11}-\!0.53c_{le,11}-\!0.47c_{lu,11}+\!0.3c_{eu,11}+\!0.15c_{ld,11}-\!0.14c^{(3)}_{lj,11}$}                                & 0.26     & $-0.5$ \\
		{\footnotesize$-0.59c_{He,22}+\!0.37c_{Hq}^{(1)}-\!0.29c_{Hl,11}^{(3)}-\!0.29c_{Hl,11}^{(1)}+\!0.25c^{(3)}_{lj,11}+\!0.25c_{He,11}+\!0.2c_{Hl,22}^{(3)}$} & 0.4      & $-0.1$ \\
		{\footnotesize$-0.61c_{Hq}^{(1)}+\!0.48c_{Hu}+\!0.34c^{(3)}_{lj,11}+\!0.25c_{eu,11}-\!0.25c_{He,22}+\!0.21c^{(1)}_{lj,11}+\!0.17c_{HWB}$}                 & 0.58     & $-0.9$ \\
		{\footnotesize$+0.53c_{lu,11}+\!0.44c^{(3)}_{lj,11}+\!0.41c^{(1)}_{lj,11}+\!0.39c_{Hq}^{(1)}+\!0.3c_{eu,11}-\!0.2c_{ld,11}-\!0.17c_{Hu}$}                 & 1.1      & $-1.1$ \\
		{\footnotesize$+0.61c_{Hu}+\!0.41c_{Hq}^{(1)}-\!0.25c^{(3)}_{lj,11}+\!0.24c_{lu,11}-\!0.23c_{eu,11}-\!0.23c_{Hq}^{(3)}+\!0.21c_{ed,11}$}                  & 1.8      & $-0.3$ \\
		{\footnotesize$-0.47c^{(1)}_{lj,11}-\!0.36c_{ed,11}-\!0.36c_{lu,11}+\!0.34c_{eu,11}+\!0.33c_{Hq}^{(1)}+\!0.29c_{Hu}+\!0.26c^{(3)}_{lj,11}$}               & 2.5      & $-0.8$ \\
		\bottomrule
	\end{tabular}
	\caption{Uncorrelated linear combinations of Wilson coefficients constrained by the Drell--Yan triple-differential cross-section measurement, for $\Lambda=1\,\TeV$.
		The linear combinations are normalized and at most six Wilson coefficients with the largest absolute value are shown, if the absolute value is larger than 0.1.
		The uncertainty $\sigma$ corresponds to 68\% confidence level intervals.
		The pull is defined as the best-fit value of the Wilson coefficient direction, determined by a global minimization of the $\chi^2$, divided by the uncertainty $\sigma$.}
	\label{tab:smeftfitZ3D}
\end{table}

The most constrained direction corresponds to variations in both electron and
muon couplings, akin to a change in \st. A value one-and-a-half standard
deviations larger than the SM is preferred for this Wilson coefficient
combination. This might seem at odds with the \st measurement of~\ref{tab:st},
which is closer to the SM prediction. However, in the SMEFT fit quark coupling
and four fermion operators also affect the interpretation, pulling \AFB in the
opposite direction.

The second most tightly constrained direction aligns well with the Wilson
coefficient dependence of $R^{\mu/e}_{\sin^2\theta^\ell_{\textrm{eff}}}$ in
Equation~\ref{eq:approxRst}, confirming the power of this measurement to constrain LFU
violation. The remaining directions, with constraints nearly an order of
magnitude weaker, modify a mix of lepton couplings, quark couplings, and
four-fermion interactions.

A good approximation of the full likelihood corresponding to the Z3D interpretation can be
reconstructed from Table~\ref{tab:smeftfitZ3D}. To better understand the
improvements brought by the Z3D interpretation to the set of EWPD observables,
it is useful to propagate the uncertainties in Wilson coefficient to obtain
uncertainties in couplings and EWPOs. In the following sections, the implied
constraints on lepton and quark couplings will be discussed separately.

\subsection{Constraints on lepton couplings}

\begin{figure}
	\includegraphics[width=0.5\textwidth]{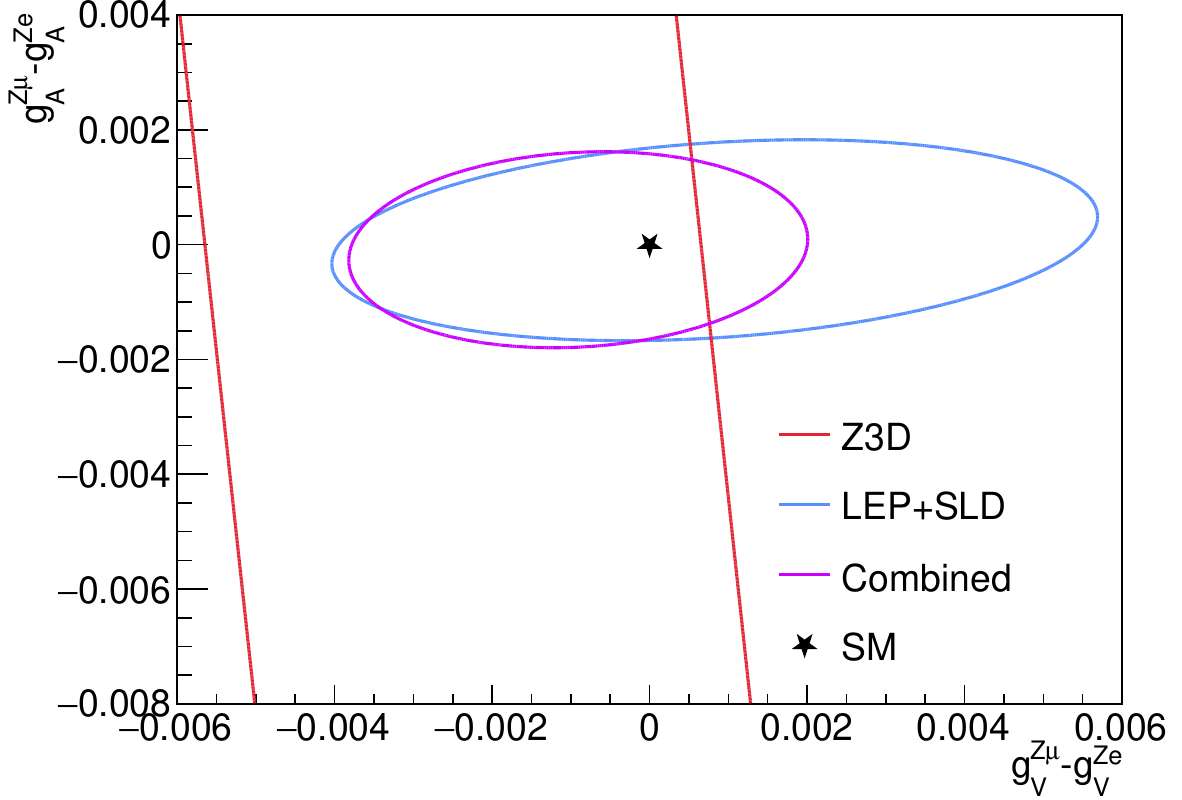}
	\includegraphics[width=0.5\textwidth]{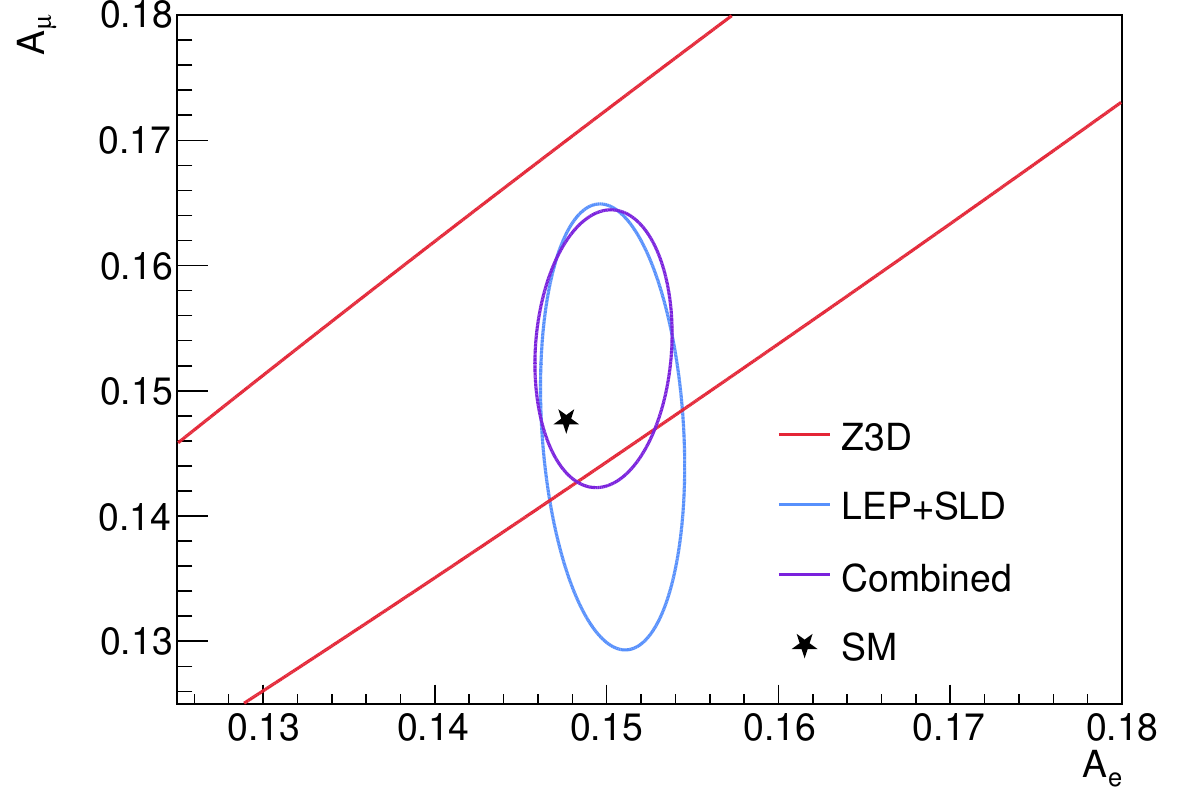}
	\caption{Constraints on the difference in muon  and electron vector and axial vector couplings (left) and the asymmetry parameters \Amu and \Ae (right), at 95\% CL. They are derived from a SMEFT fit involving all relevant Wilson coefficients.
		Constraints labeled ``Z3D'' correspond to the re-analysis of the  Drell--Yan triple-differential cross-section measurement, while LEP+SLD refers to the EWPD fit discussed in Section~\ref{sec:likelihood}.     For the coupling measurement, only one degree of freedom is constrained by the Z3D measurement and the corresponding confidence intervals are based on a one-dimensional $\chi^2$ distribution.
	}
	\label{fig:leptons}
\end{figure}

Constraints on the differences of the $Z$ boson couplings of electrons and
muons are derived from the general fit results using Gaussian error propagation
and the formulas in Equation~\ref{eq:dgV} and Equation~\ref{eq:dgA}. The
constraints obtained are illustrated in the left-hand plot of
Figure~\ref{fig:leptons}. The \AFB measurement can indeed constrain lepton
couplings, even when allowing all relevant dimension-six operators to vary. As
discussed in Section~\ref{sec:st}, \AFB is in particularly sensitive to the
difference in vector couplings, $g^{Z\mu}_V- g^{Ze}_V$, with minimal
sensitivity to axial vector couplings. A combination with the LEP+SLD fit, also
shown in Figure~\ref{fig:leptons}, leverages the strength of each measurement:
the Z3D analysis provides better constraints on vector couplings, while at LEP
and SLD partial width measurements pin down axial vector couplings.

The difference in vector couplings $g^{Z\mu}_V- g^{Ze}_V$, assuming SM axial
vector couplings, is compared for different scenarios in Table~\ref{tab:lfu}.
The presence of four-fermion operators, which can also violate LFU, leads to
only a slight increase in uncertainty, as these operators are constrained in
high-mass bins. However, as they are not constrained to zero, they introduce a
significant shift of the central value.

\begin{table}
	\centering
	\begin{tabular}{@{}llr@{}}
		\toprule
		Channels analyzed                    & Assumption                & $g^{Z\mu}_V- g^{Ze}_V~[10^{-4}]$ \\
		\midrule
		$\mu^+\mu^-$CC+$e^+e^-$CC+$e^+e^-$CF & $-$                       & $-25 \pm 16$                     \\
		$\mu^+\mu^-$CC+$e^+e^-$CC+$e^+e^-$CF & No four-fermion operators & $-18 \pm 16$                     \\
		$\mu^+\mu^-$CC+$e^+e^-$CC            & $-$                       & $-29 \pm 18$                     \\
		$\mu^+\mu^-$CC+$e^+e^-$CC            & No four-fermion operators & $-16 \pm 17$                     \\
		\bottomrule
	\end{tabular}
	\caption{Comparison of 68\% CL vector coupling constraints $g^{Z\mu}_V- g^{Ze}_V$ from the SMEFT fit to the ATLAS Drell--Yan triple-differential cross-section measurement in different scenarios. CC (CF) refers to central--central (central--forward) lepton channels. For these results $g^{Z\mu}_A- g^{Ze}_A$ is fixed to zero while all other Wilson coefficient combinations are allowed to vary.
	}
	\label{tab:lfu}
\end{table}

It is instructive to compare the measurement using central--central lepton
channels only with an estimate based on
$R^{\mu/e}_{\sin^2\theta^\ell_{\textrm{eff}}}$. The latter yields, based on
Table~\ref{tab:st}, $g^{Z\mu}_V- g^{Ze}_V=(-13 \pm 17)\times10^{-4}$, which is
already close to the SMEFT fit without four-fermion operators presented in
Table~\ref{tab:lfu}. When PDFs are fixed to their central values, the estimate
based on $R^{\mu/e}_{\sin^2\theta^\ell_{\textrm{eff}}}$ matches exactly the
SMEFT fit result. This suggests that the difference arises from different pulls
of the PDFs in the individual channels, highlighting the importance of
extracting $R^{\mu/e}_{\sin^2\theta^\ell_{\textrm{eff}}}$ in a simultaneous fit
of all channels. The close agreement of the
$R^{\mu/e}_{\sin^2\theta^\ell_{\textrm{eff}}}$ estimate and the full SMEFT fit
confirms that the reinterpretation of \st as a ratio measurement holds fairly
general validity in the SMEFT. The shift introduced by four-fermion operators,
which the $R^{\mu/e}_{\sin^2\theta^\ell_{\textrm{eff}}}$ estimate cannot
account for, is a challenge for this approximation. However, such large
four-fermion operator contributions would impact the \AFB of Drell--Yan
production much more significantly at masses higher than those studied in this
analysis. The absence of reported excesses at high mass suggests that any
modifications are indeed small.

Finally, the SMEFT analysis also improves the precision of the asymmetry
parameter $A_\mu$ with respect to the EWPD fit in Section~\ref{sec:SMEFTfit}.
The constraints on $A_\mu$ and $A_e$ from the SMEFT fit are shown in the
right-hand plot of Figure~\ref{fig:leptons}. Due to unknown quark couplings,
which simultaneously shift \AFB in both electron and muon channels, the
constraints on $A_\mu$ and $A_e$ are highly correlated. As shown in the figure,
combining the SMEFT results with the LEP+SLD likelihood yields an improved
result for \Amu, benefiting from the tight constraints on \Ae provided by LEP
and SLD.

\subsection{Constraints on quark couplings}
The \AFB measurement can also constrain quark couplings within a general SMEFT
fit, using lepton coupling constraints from LEP and SLD in a global
combination. While the sensitivity to quark couplings is weaker than that to
lepton couplings, as the latter benefit from the accidentally small SM value of
$g_V^\ell$, it is of great importance. Firstly, \AFB at a hadron colliders
offers unique sensitivity to up quarks and down quarks. This sensitivity as
highlighted in Ref.~\cite{Breso-Pla:2021qoe} and estimated under simplified
assumptions using the ATLAS $A_4$ measurement~\cite{ATLAS:2018gqq}. Couplings
to light quark are otherwise weakly constrained from precision measurements, as
only charm quarks and bottom quarks can be reliably identified experimentally.
Secondly, the quark couplings that have been precisely measured in the past,
i.e., those of charm quarks and bottom quarks, are in tension with the SM. In
particular, deviations in the bottom quark asymmetries \Ab and \AFBb result in
a three standard deviation discrepancy from the SM in the SMEFT fit, as shown
in Table~\ref{tab:smeftfit1}.

Visualizing constraints on light quark couplings is challenging because \AFB at
the LHC only reflects the combined effects of up and down quark couplings. The
asymmetry is thus affected by four degrees of freedom: the vector and axial
vector couplings of the two quark flavours. To simplify the interpretation, the
$U(2)_q\times U(2)_u\times U(2)_d$ symmetry assumption is employed, which
requires identical couplings for up quarks and charm quarks as well as down
quarks and strange quarks. In a combined analysis with LEP+SLD data this
constrains the $\Gamma_{Z\rightarrow u\bar u}$ and $\Gamma_{Z\rightarrow d\bar
		d}$ partial widths, linking them to the $\Gamma_{Z\rightarrow c\bar c}$ and
$\Gamma_{Z\rightarrow s\bar s}$ partial widths. This assumption essentially
precludes scenarios where $\Gamma_{Z\rightarrow u\bar u}$ and
$\Gamma_{Z\rightarrow c\bar c}$ differ from the SM while their sum does not, a
scenario compatible with EWPD but unlikely.

\begin{figure}
	\centering
	\includegraphics[width=0.6\textwidth]{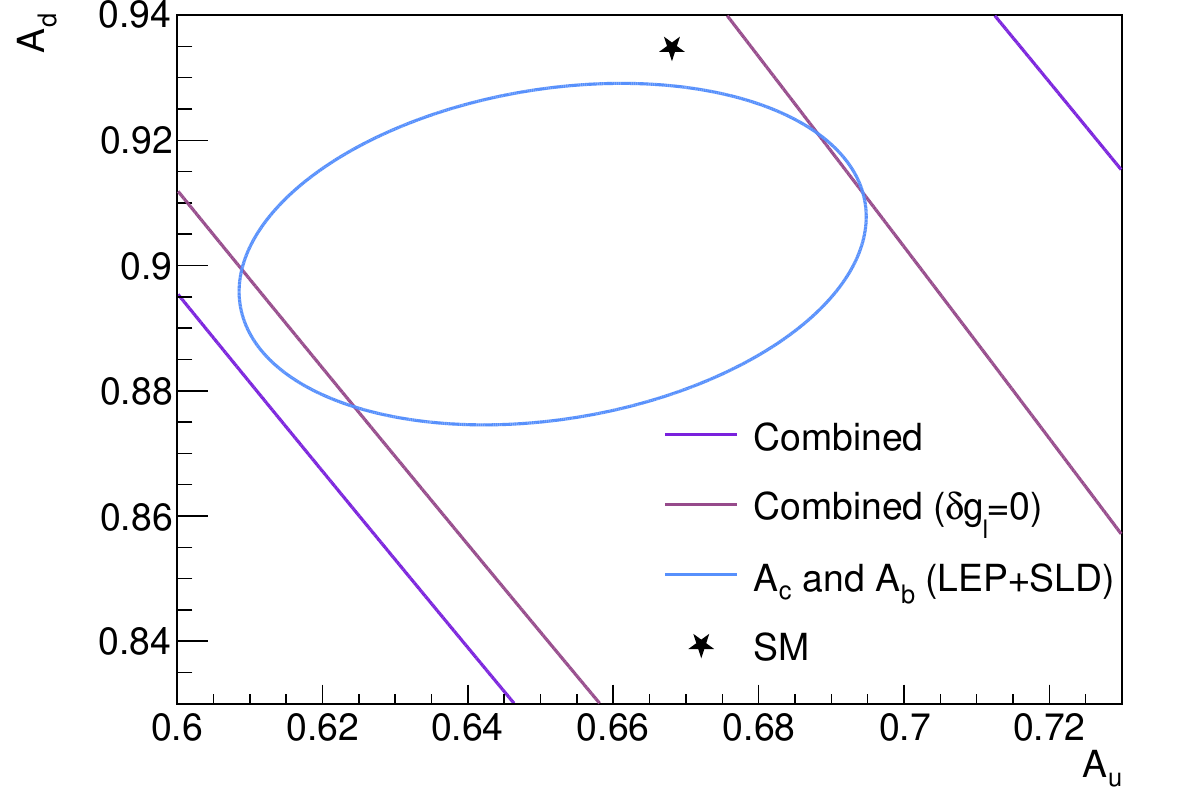}
	\caption{Constraints on the quark coupling asymmetry parameters $A_u$ and $A_d$, at 95\% CL. Derived from a combined SMEFT fit of EWPD and the ATLAS Drell--Yan triple-differential cross-section measurement, where all relevant dimension-six Wilson coefficients are allowed to vary. For the result labeled $\delta g_\ell=0$, lepton couplings are fixed to their SM value.
		For comparison, the quark asymmetry parameters $A_c$ and $A_b$ from the EWPD fit of Section~\ref{sec:likelihood} are also shown, with $A_c$ and $A_u$ ($A_b$ and $A_d$) sharing the coordinate axes.
		For the light quark  measurement, only one degree of freedom is constrained. Hence, the corresponding confidence intervals are based on a one-dimensional $\chi^2$ distribution.
		The indicated SM value is valid for both sets of parameters.}
	\label{fig:Aq}
\end{figure}

With all $Z$ boson partial width constrained by LEP and SLD data, only two light quark coupling
combinations affecting $A_u$ and $A_d$ but not altering partial widths are
unconstrained. One combination is tightly constrained by the Z3D measurement,
as shown in Figure~\ref{fig:Aq}.
The result is limited by lepton coupling uncertainties.
Assuming no deviations from the SM expectation in lepton couplings, the $A_u$ and $A_d$ constraints of Figure~\ref{fig:Aq} improve to a sensitivity that is comparable to that of $A_c$ and $A_b$ measurements.

Thus, future improvements in the precision of light quark constraints from \AFB
measurement could provide insights whether quark couplings differ in general
from the SM, potentially providing new insights into a longstanding discrepancy
in the electroweak fit.

\section*{Conclusion}

This paper presented a computer code designed to calculate the likelihood of
electroweak precision data within the framework of the Standard Model Effective
Field Theory. The code integrates state-of-the-art experimental and theoretical
results, including next-to-leading order effects in both perturbative and SMEFT
expansions. It offers five electroweak input parameter schemes and
a more thorough treatment of uncertainties than previous
approaches, fully accounting for the input parameter dependence of EWPO
predictions. The assumptions and conventions align with those used in LHC
analyses, making this code, which produces text file and \texttt{Roofit}
output, ideally suited for more extensive fits that incorporate LHC data.
For the first time, EWPD fits were performed and compared in five different
input parameter schemes, at leading and next-to-leading order.
The impact of next-to-leading order corrections and the scheme choice was found
to be small.

The impact of recent LHC measurements on the EWPD likelihood was analyzed. The
analysis demonstrated that the ATLAS measurements of the $W$ boson
mass~\cite{ATLAS:2024erm} and the lepton flavor universality in $W$ branching
fractions~\cite{ATLAS:2024tlf} provide constraints in the SMEFT that are more
precise than those obtained by fitting all prior data (if the CDF measurement
of the $W$ mass is discarded). While the same is not true for the \GammaW
measurement, which leads despite its precision to only a marginal improvement
in the SMEFT fit, the significance of a joint \MW-\GammaW measurement was
highlighted. Furthermore, it was shown that the CMS effective leptonic weak
mixing angle measurement~\cite{CMS:2024ony} can be used to construct one of the
most precise test of LFU in weak boson couplings to date, by measuring the
ratio $R^{\mu/e}_{\sin^2\theta^\ell_{\textrm{eff}}}$. This was further explored
through a SMEFT interpretation of an LHC Run~1 ATLAS Drell--Yan
triple-differential cross-section measurement~\cite{ATLAS:2017rue}. The
interpretation not only validated the LFU test and improved constraints on muon
couplings by nearly a factor of two but also simultaneously constrained the
quark coupling asymmetry parameters $A_u$ and $A_d$, providing an important
cross-check of $A_c$ and $A_b$ measurements from LEP and SLD.

A precise measurement of $R^{\mu/e}_{\sin^2\theta^\ell_{\textrm{eff}}}$ by the
experimental collaborations is strongly encouraged, as it has the potential to
be the most powerful test of lepton universality of weak boson couplings. The
publication of CMS or ATLAS measurements of either \AFB or the angular
coefficient $A_4$, as a function of rapidity and mass, separately for both
electron and muon channels, would allow revisiting the presented SMEFT
interpretation using Run~2 data. A detailed breakdown of experimental
uncertainties and the publication of Standard Model predictions would be
extremely valuable for this endeavor while extending the measured mass range
beyond the $Z$ pole would help constraining four-fermion operators. Ideally,
such analyses should be undertaken by the experimental collaborations, which
have the best understanding of experimental limitations and can optimize their
measurements for the purpose of a SMEFT interpretation.

\section*{Acknowledgements}
I would like to express my gratitude to Anke Biekötter, Ken Mimasu, and Matthias Schott for valuable discussions and comments on the manuscript.
Anke Biekötter I would also like to thank for providing NLO SMEFT parametrizations in multiple input parameter schemes and for numerical comparisons.
I am especially thankful to Matthias Schott for his continuous support throughout this project. I also extend my thanks to Mike Trott for his initial introduction to EWPD in the SMEFT.
Suppot of the BMBF (Federal Ministry of Education and Research) is gratefully acknowledged.
\bibliography{bib}

\providecommand{\href}[2]{#2}\begingroup\raggedright\begin{thebibliography}{10}

\bibitem{Buchmuller:1985jz}
W.~Buchmuller and D.~Wyler, \emph{{Effective Lagrangian Analysis of New
  Interactions and Flavor Conservation}},
  \href{https://doi.org/10.1016/0550-3213(86)90262-2}{\emph{Nucl. Phys. B}
  {\bfseries 268} (1986) 621}.

\bibitem{Grzadkowski:2010es}
B.~Grzadkowski, M.~Iskrzynski, M.~Misiak and J.~Rosiek, \emph{{Dimension-Six
  Terms in the Standard Model Lagrangian}},
  \href{https://doi.org/10.1007/JHEP10(2010)085}{\emph{JHEP} {\bfseries 10}
  (2010) 085} [\href{https://arxiv.org/abs/1008.4884}{{\ttfamily 1008.4884}}].

\bibitem{Brivio:2017vri}
I.~Brivio and M.~Trott, \emph{{The Standard Model as an Effective Field
  Theory}}, \href{https://doi.org/10.1016/j.physrep.2018.11.002}{\emph{Phys.
  Rept.} {\bfseries 793} (2019) 1}
  [\href{https://arxiv.org/abs/1706.08945}{{\ttfamily 1706.08945}}].

\bibitem{ALEPH:2005ab}
{\scshape ALEPH, DELPHI, L3, OPAL, SLD, LEP Electroweak Working Group, SLD
  Electroweak Group, SLD Heavy Flavour Group} collaboration, \emph{{Precision
  electroweak measurements on the $Z$ resonance}},
  \href{https://doi.org/10.1016/j.physrep.2005.12.006}{\emph{Phys. Rept.}
  {\bfseries 427} (2006) 257}
  [\href{https://arxiv.org/abs/hep-ex/0509008}{{\ttfamily hep-ex/0509008}}].

\bibitem{ATLAS:2024erm}
{\scshape ATLAS} collaboration, \emph{{Measurement of the W-boson mass and
  width with the ATLAS detector using proton-proton collisions at $\sqrt{s}$ =
  7 TeV}},  \href{https://arxiv.org/abs/2403.15085}{{\ttfamily 2403.15085}}.

\bibitem{ATLAS:2024tlf}
{\scshape ATLAS} collaboration, \emph{{Precise test of lepton flavour
  universality in $W$-boson decays into muons and electrons in $pp$ collisions
  at $\sqrt{s}=13\,\text {T}\text {e}\hspace{-1.00006pt}\text {V} $ with the
  ATLAS detector}},
  \href{https://doi.org/10.1140/epjc/s10052-024-13070-4}{\emph{Eur. Phys. J. C}
  {\bfseries 84} (2024) 993}
  [\href{https://arxiv.org/abs/2403.02133}{{\ttfamily 2403.02133}}].

\bibitem{CMS:2024ony}
{\scshape CMS} collaboration, \emph{{Measurement of the Drell--Yan
  forward-backward asymmetry and of the effective leptonic weak mixing angle in
  proton-proton collisions at $\sqrt{s}$ = 13 TeV}},
  \href{https://arxiv.org/abs/2408.07622}{{\ttfamily 2408.07622}}.

\bibitem{ATLAS:2017rue}
{\scshape ATLAS} collaboration, \emph{{Measurement of the Drell-Yan
  triple-differential cross section in $pp$ collisions at $\sqrt{s} = 8$ TeV}},
  \href{https://doi.org/10.1007/JHEP12(2017)059}{\emph{JHEP} {\bfseries 12}
  (2017) 059} [\href{https://arxiv.org/abs/1710.05167}{{\ttfamily
  1710.05167}}].

\bibitem{Han:2004az}
Z.~Han and W.~Skiba, \emph{{Effective theory analysis of precision electroweak
  data}}, \href{https://doi.org/10.1103/PhysRevD.71.075009}{\emph{Phys. Rev. D}
  {\bfseries 71} (2005) 075009}
  [\href{https://arxiv.org/abs/hep-ph/0412166}{{\ttfamily hep-ph/0412166}}].

\bibitem{Pomarol:2013zra}
A.~Pomarol and F.~Riva, \emph{{Towards the Ultimate SM Fit to Close in on Higgs
  Physics}}, \href{https://doi.org/10.1007/JHEP01(2014)151}{\emph{JHEP}
  {\bfseries 01} (2014) 151} [\href{https://arxiv.org/abs/1308.2803}{{\ttfamily
  1308.2803}}].

\bibitem{Falkowski:2014tna}
A.~Falkowski and F.~Riva, \emph{{Model-independent precision constraints on
  dimension-6 operators}},
  \href{https://doi.org/10.1007/JHEP02(2015)039}{\emph{JHEP} {\bfseries 02}
  (2015) 039} [\href{https://arxiv.org/abs/1411.0669}{{\ttfamily 1411.0669}}].

\bibitem{Efrati:2015eaa}
A.~Efrati, A.~Falkowski and Y.~Soreq, \emph{{Electroweak constraints on
  flavorful effective theories}},
  \href{https://doi.org/10.1007/JHEP07(2015)018}{\emph{JHEP} {\bfseries 07}
  (2015) 018} [\href{https://arxiv.org/abs/1503.07872}{{\ttfamily
  1503.07872}}].

\bibitem{Berthier:2015gja}
L.~Berthier and M.~Trott, \emph{{Consistent constraints on the Standard Model
  Effective Field Theory}},
  \href{https://doi.org/10.1007/JHEP02(2016)069}{\emph{JHEP} {\bfseries 02}
  (2016) 069} [\href{https://arxiv.org/abs/1508.05060}{{\ttfamily
  1508.05060}}].

\bibitem{deBlas:2017wmn}
J.~de~Blas, M.~Ciuchini, E.~Franco, S.~Mishima, M.~Pierini, L.~Reina et~al.,
  \emph{{The Global Electroweak and Higgs Fits in the LHC era}},
  \href{https://doi.org/10.22323/1.314.0467}{\emph{PoS} {\bfseries EPS-HEP2017}
  (2017) 467} [\href{https://arxiv.org/abs/1710.05402}{{\ttfamily
  1710.05402}}].

\bibitem{daSilvaAlmeida:2018iqo}
E.~da~Silva~Almeida, A.~Alves, N.~Rosa~Agostinho, O.J.P.~\'Eboli and
  M.C.~Gonzalez-Garcia, \emph{{Electroweak Sector Under Scrutiny: A Combined
  Analysis of LHC and Electroweak Precision Data}},
  \href{https://doi.org/10.1103/PhysRevD.99.033001}{\emph{Phys. Rev. D}
  {\bfseries 99} (2019) 033001}
  [\href{https://arxiv.org/abs/1812.01009}{{\ttfamily 1812.01009}}].

\bibitem{Biekotter:2018ohn}
A.~Biek\"otter, T.~Corbett and T.~Plehn, \emph{{The Gauge-Higgs Legacy of the
  LHC Run II}},
  \href{https://doi.org/10.21468/SciPostPhys.6.6.064}{\emph{SciPost Phys.}
  {\bfseries 6} (2019) 064} [\href{https://arxiv.org/abs/1812.07587}{{\ttfamily
  1812.07587}}].

\bibitem{Aebischer:2018iyb}
J.~Aebischer, J.~Kumar, P.~Stangl and D.M.~Straub, \emph{{A Global Likelihood
  for Precision Constraints and Flavour Anomalies}},
  \href{https://doi.org/10.1140/epjc/s10052-019-6977-z}{\emph{Eur. Phys. J. C}
  {\bfseries 79} (2019) 509}
  [\href{https://arxiv.org/abs/1810.07698}{{\ttfamily 1810.07698}}].

\bibitem{Ellis:2018gqa}
J.~Ellis, C.W.~Murphy, V.~Sanz and T.~You, \emph{{Updated Global SMEFT Fit to
  Higgs, Diboson and Electroweak Data}},
  \href{https://doi.org/10.1007/JHEP06(2018)146}{\emph{JHEP} {\bfseries 06}
  (2018) 146} [\href{https://arxiv.org/abs/1803.03252}{{\ttfamily
  1803.03252}}].

\bibitem{Falkowski:2019hvp}
A.~Falkowski and D.~Straub, \emph{{Flavourful SMEFT likelihood for Higgs and
  electroweak data}},
  \href{https://doi.org/10.1007/JHEP04(2020)066}{\emph{JHEP} {\bfseries 04}
  (2020) 066} [\href{https://arxiv.org/abs/1911.07866}{{\ttfamily
  1911.07866}}].

\bibitem{Ellis:2020unq}
J.~Ellis, M.~Madigan, K.~Mimasu, V.~Sanz and T.~You, \emph{{Top, Higgs, Diboson
  and Electroweak Fit to the Standard Model Effective Field Theory}},
  \href{https://doi.org/10.1007/JHEP04(2021)279}{\emph{JHEP} {\bfseries 04}
  (2021) 279} [\href{https://arxiv.org/abs/2012.02779}{{\ttfamily
  2012.02779}}].

\bibitem{Corbett:2021eux}
T.~Corbett, A.~Helset, A.~Martin and M.~Trott, \emph{{EWPD in the SMEFT to
  dimension eight}}, \href{https://doi.org/10.1007/JHEP06(2021)076}{\emph{JHEP}
  {\bfseries 06} (2021) 076}
  [\href{https://arxiv.org/abs/2102.02819}{{\ttfamily 2102.02819}}].

\bibitem{Bellafronte:2023amz}
L.~Bellafronte, S.~Dawson and P.P.~Giardino, \emph{{The importance of flavor in
  SMEFT Electroweak Precision Fits}},
  \href{https://doi.org/10.1007/JHEP05(2023)208}{\emph{JHEP} {\bfseries 05}
  (2023) 208} [\href{https://arxiv.org/abs/2304.00029}{{\ttfamily
  2304.00029}}].

\bibitem{Celada:2024mcf}
E.~Celada, T.~Giani, J.~ter Hoeve, L.~Mantani, J.~Rojo, A.N.~Rossia et~al.,
  \emph{{Mapping the SMEFT at high-energy colliders: from LEP and the (HL-)LHC
  to the FCC-ee}}, \href{https://doi.org/10.1007/JHEP09(2024)091}{\emph{JHEP}
  {\bfseries 09} (2024) 091}
  [\href{https://arxiv.org/abs/2404.12809}{{\ttfamily 2404.12809}}].

\bibitem{ATLAS:2022xyx}
{\scshape ATLAS} collaboration, \emph{{Combined effective field theory
  interpretation of Higgs boson and weak boson production and decay with ATLAS
  data and electroweak precision observables}},  {ATLAS PUB Note}
  \href{https://cds.cern.ch/record/2816369/}{ATL-PHYS-PUB-2022-037} (2022).

\bibitem{CMS:2024kvw}
{\scshape CMS} collaboration, \emph{{Combined effective field theory
  interpretation of Higgs boson, electroweak vector boson, top quark, and
  multi-jet measurements}},  {CMS Physics Analysis Summary}
  \href{https://cds.cern.ch/record/2911229/}{CMS-PAS-SMP-24-003} (2024).

\bibitem{Verkerke:2003ir}
W.~Verkerke and D.P.~Kirkby, \emph{{The RooFit toolkit for data modeling}},
  \href{https://arxiv.org/abs/physics/0306116}{{\ttfamily physics/0306116}}.

\bibitem{Brivio:2021yjb}
I.~Brivio, S.~Dawson, J.~de~Blas, G.~Durieux, P.~Savard, A.~Denner et~al.,
  \emph{{Electroweak input parameters}},
  \href{https://arxiv.org/abs/2111.12515}{{\ttfamily 2111.12515}}.

\bibitem{Brivio:2017btx}
I.~Brivio, Y.~Jiang and M.~Trott, \emph{{The SMEFTsim package, theory and
  tools}}, \href{https://doi.org/10.1007/JHEP12(2017)070}{\emph{JHEP}
  {\bfseries 12} (2017) 070}
  [\href{https://arxiv.org/abs/1709.06492}{{\ttfamily 1709.06492}}].

\bibitem{Brivio:2020onw}
I.~Brivio, \emph{{SMEFTsim 3.0 \textemdash{} a practical guide}},
  \href{https://doi.org/10.1007/JHEP04(2021)073}{\emph{JHEP} {\bfseries 04}
  (2021) 073} [\href{https://arxiv.org/abs/2012.11343}{{\ttfamily
  2012.11343}}].

\bibitem{Dawson:2019clf}
S.~Dawson and P.P.~Giardino, \emph{{Electroweak and QCD corrections to $Z$ and
  $W$ pole observables in the standard model EFT}},
  \href{https://doi.org/10.1103/PhysRevD.101.013001}{\emph{Phys. Rev. D}
  {\bfseries 101} (2020) 013001}
  [\href{https://arxiv.org/abs/1909.02000}{{\ttfamily 1909.02000}}].

\bibitem{Dawson:2022bxd}
S.~Dawson and P.P.~Giardino, \emph{{Flavorful electroweak precision observables
  in the Standard Model effective field theory}},
  \href{https://doi.org/10.1103/PhysRevD.105.073006}{\emph{Phys. Rev. D}
  {\bfseries 105} (2022) 073006}
  [\href{https://arxiv.org/abs/2201.09887}{{\ttfamily 2201.09887}}].

\bibitem{Biekotter:2023xle}
A.~Biek\"otter, B.D.~Pecjak, D.J.~Scott and T.~Smith, \emph{{Electroweak input
  schemes and universal corrections in SMEFT}},
  \href{https://doi.org/10.1007/JHEP07(2023)115}{\emph{JHEP} {\bfseries 07}
  (2023) 115} [\href{https://arxiv.org/abs/2305.03763}{{\ttfamily
  2305.03763}}].

\bibitem{Biekotter:2023vbh}
A.~Biek\"otter, B.D.~Pecjak and T.~Smith, \emph{{Using the effective weak
  mixing angle as an input parameter in SMEFT}},
  \href{https://doi.org/10.1007/JHEP04(2024)073}{\emph{JHEP} {\bfseries 04}
  (2024) 073} [\href{https://arxiv.org/abs/2312.08446}{{\ttfamily
  2312.08446}}].

\bibitem{Helset:2020yio}
A.~Helset, A.~Martin and M.~Trott, \emph{{The Geometric Standard Model
  Effective Field Theory}},
  \href{https://doi.org/10.1007/JHEP03(2020)163}{\emph{JHEP} {\bfseries 03}
  (2020) 163} [\href{https://arxiv.org/abs/2001.01453}{{\ttfamily
  2001.01453}}].

\bibitem{Liu:2022vgo}
Y.~Liu, Y.~Wang, C.~Zhang, L.~Zhang and J.~Gu, \emph{{Probing top-quark
  operators with precision electroweak measurements*}},
  \href{https://doi.org/10.1088/1674-1137/ac82e1}{\emph{Chin. Phys. C}
  {\bfseries 46} (2022) 113105}
  [\href{https://arxiv.org/abs/2205.05655}{{\ttfamily 2205.05655}}].

\bibitem{Aguilar-Saavedra:2018ksv}
D.~Barducci et~al., \emph{{Interpreting top-quark LHC measurements in the
  standard-model effective field theory}},
  \href{https://arxiv.org/abs/1802.07237}{{\ttfamily 1802.07237}}.

\bibitem{Janot:2019oyi}
P.~Janot and S.~Jadach, \emph{{Improved Bhabha cross section at LEP and the
  number of light neutrino species}},
  \href{https://doi.org/10.1016/j.physletb.2020.135319}{\emph{Phys. Lett. B}
  {\bfseries 803} (2020) 135319}
  [\href{https://arxiv.org/abs/1912.02067}{{\ttfamily 1912.02067}}].

\bibitem{ParticleDataGroup:2024cfk}
{\scshape Particle Data Group} collaboration, \emph{{Review of particle
  physics}}, \href{https://doi.org/10.1103/PhysRevD.110.030001}{\emph{Phys.
  Rev. D} {\bfseries 110} (2024) 030001}.

\bibitem{ATLAS:2016nqi}
{\scshape ATLAS} collaboration, \emph{{Precision measurement and interpretation
  of inclusive $W^+$ , $W^-$ and $Z/\gamma ^*$ production cross sections with
  the ATLAS detector}},
  \href{https://doi.org/10.1140/epjc/s10052-017-4911-9}{\emph{Eur. Phys. J. C}
  {\bfseries 77} (2017) 367}
  [\href{https://arxiv.org/abs/1612.03016}{{\ttfamily 1612.03016}}].

\bibitem{ATLAS:2020xea}
{\scshape ATLAS} collaboration, \emph{{Test of the universality of $\tau$ and
  $\mu$ lepton couplings in $W$-boson decays with the ATLAS detector}},
  \href{https://doi.org/10.1038/s41567-021-01236-w}{\emph{Nature Phys.}
  {\bfseries 17} (2021) 813}
  [\href{https://arxiv.org/abs/2007.14040}{{\ttfamily 2007.14040}}].

\bibitem{CMS:2022mhs}
{\scshape CMS} collaboration, \emph{{Precision measurement of the W boson decay
  branching fractions in proton-proton collisions at $\sqrt{s}$ = 13 TeV}},
  \href{https://doi.org/10.1103/PhysRevD.105.072008}{\emph{Phys. Rev. D}
  {\bfseries 105} (2022) 072008}
  [\href{https://arxiv.org/abs/2201.07861}{{\ttfamily 2201.07861}}].

\bibitem{LHCb:2016zpq}
{\scshape LHCb} collaboration, \emph{{Measurement of forward $W\to e\nu$
  production in $pp$ collisions at $\sqrt{s}=8\,$TeV}},
  \href{https://doi.org/10.1007/JHEP10(2016)030}{\emph{JHEP} {\bfseries 10}
  (2016) 030} [\href{https://arxiv.org/abs/1608.01484}{{\ttfamily
  1608.01484}}].

\bibitem{CDF:2022hxs}
{\scshape CDF} collaboration, \emph{{High-precision measurement of the $W$
  boson mass with the CDF II detector}},
  \href{https://doi.org/10.1126/science.abk1781}{\emph{Science} {\bfseries 376}
  (2022) 170}.

\bibitem{CMS:2024nau}
{\scshape CMS} collaboration, \emph{{Measurement of the W boson mass in
  proton-proton collisions at $\sqrt s$ = 13 TeV}},  {CMS Physics Analysis
  Summary} \href{https://cds.cern.ch/record/2910372}{CMS-PAS-SMP-23-002}
  (2024).

\bibitem{Davier:2019can}
M.~Davier, A.~Hoecker, B.~Malaescu and Z.~Zhang, \emph{{A new evaluation of the
  hadronic vacuum polarisation contributions to the muon anomalous magnetic
  moment and to $\alpha(m_Z^2)$}},
  \href{https://doi.org/10.1140/epjc/s10052-020-7792-2}{\emph{Eur. Phys. J. C}
  {\bfseries 80} (2020) 241}
  [\href{https://arxiv.org/abs/1908.00921}{{\ttfamily 1908.00921}}].

\bibitem{Steinhauser:1998rq}
M.~Steinhauser, \emph{{Leptonic contribution to the effective electromagnetic
  coupling constant up to three loops}},
  \href{https://doi.org/10.1016/S0370-2693(98)00503-6}{\emph{Phys. Lett. B}
  {\bfseries 429} (1998) 158}
  [\href{https://arxiv.org/abs/hep-ph/9803313}{{\ttfamily hep-ph/9803313}}].

\bibitem{Celada:2024cxw}
E.~Celada, G.~Durieux, K.~Mimasu and E.~Vryonidou, \emph{{Triboson production
  in the SMEFT}},  \href{https://arxiv.org/abs/2407.09600}{{\ttfamily
  2407.09600}}.

\bibitem{Brivio:2017bnu}
I.~Brivio and M.~Trott, \emph{{Scheming in the SMEFT... and a
  reparameterization invariance!}},
  \href{https://doi.org/10.1007/JHEP07(2017)148}{\emph{JHEP} {\bfseries 07}
  (2017) 148} [\href{https://arxiv.org/abs/1701.06424}{{\ttfamily
  1701.06424}}].

\bibitem{FlavourLatticeAveragingGroupFLAG:2021npn}
{\scshape Flavour Lattice Averaging Group (FLAG)} collaboration, \emph{{FLAG
  Review 2021}},
  \href{https://doi.org/10.1140/epjc/s10052-022-10536-1}{\emph{Eur. Phys. J. C}
  {\bfseries 82} (2022) 869}
  [\href{https://arxiv.org/abs/2111.09849}{{\ttfamily 2111.09849}}].

\bibitem{Trott:2023jrw}
M.~Trott, \emph{{$\alpha_s$ as an input parameter in the SMEFT}},
  \href{https://arxiv.org/abs/2306.14784}{{\ttfamily 2306.14784}}.

\bibitem{Hoang:2020iah}
A.H.~Hoang, \emph{{What is the Top Quark Mass?}},
  \href{https://doi.org/10.1146/annurev-nucl-101918-023530}{\emph{Ann. Rev.
  Nucl. Part. Sci.} {\bfseries 70} (2020) 225}
  [\href{https://arxiv.org/abs/2004.12915}{{\ttfamily 2004.12915}}].

\bibitem{Dubovyk:2019szj}
I.~Dubovyk, A.~Freitas, J.~Gluza, T.~Riemann and J.~Usovitsch,
  \emph{{Electroweak pseudo-observables and Z-boson form factors at two-loop
  accuracy}}, \href{https://doi.org/10.1007/JHEP08(2019)113}{\emph{JHEP}
  {\bfseries 08} (2019) 113}
  [\href{https://arxiv.org/abs/1906.08815}{{\ttfamily 1906.08815}}].

\bibitem{Awramik:2006uz}
M.~Awramik, M.~Czakon and A.~Freitas, \emph{{Electroweak two-loop corrections
  to the effective weak mixing angle}},
  \href{https://doi.org/10.1088/1126-6708/2006/11/048}{\emph{JHEP} {\bfseries
  11} (2006) 048} [\href{https://arxiv.org/abs/hep-ph/0608099}{{\ttfamily
  hep-ph/0608099}}].

\bibitem{Awramik:2003rn}
M.~Awramik, M.~Czakon, A.~Freitas and G.~Weiglein, \emph{{Precise prediction
  for the W boson mass in the standard model}},
  \href{https://doi.org/10.1103/PhysRevD.69.053006}{\emph{Phys. Rev. D}
  {\bfseries 69} (2004) 053006}
  [\href{https://arxiv.org/abs/hep-ph/0311148}{{\ttfamily hep-ph/0311148}}].

\bibitem{Cho:2011rk}
G.-C.~Cho, K.~Hagiwara, Y.~Matsumoto and D.~Nomura, \emph{{The MSSM confronts
  the precision electroweak data and the muon g-2}},
  \href{https://doi.org/10.1007/JHEP11(2011)068}{\emph{JHEP} {\bfseries 11}
  (2011) 068} [\href{https://arxiv.org/abs/1104.1769}{{\ttfamily 1104.1769}}].

\bibitem{Haller:2022eyb}
J.~Haller, A.~Hoecker, R.~Kogler, K.~M\"onig and J.~Stelzer, \emph{{Status of
  the global electroweak fit with Gfitter in the light of new precision
  measurements}}, \href{https://doi.org/10.22323/1.414.0897}{\emph{PoS}
  {\bfseries ICHEP2022} (2022) 897}
  [\href{https://arxiv.org/abs/2211.07665}{{\ttfamily 2211.07665}}].

\bibitem{Alwall:2014hca}
J.~Alwall, R.~Frederix, S.~Frixione, V.~Hirschi, F.~Maltoni, O.~Mattelaer
  et~al., \emph{{The automated computation of tree-level and next-to-leading
  order differential cross sections, and their matching to parton shower
  simulations}}, \href{https://doi.org/10.1007/JHEP07(2014)079}{\emph{JHEP}
  {\bfseries 07} (2014) 079} [\href{https://arxiv.org/abs/1405.0301}{{\ttfamily
  1405.0301}}].

\bibitem{Alwall:2014bza}
J.~Alwall, C.~Duhr, B.~Fuks, O.~Mattelaer, D.G.~\"Ozt\"urk and C.-H.~Shen,
  \emph{{Computing decay rates for new physics theories with FeynRules and
  MadGraph 5\_aMC@NLO}},
  \href{https://doi.org/10.1016/j.cpc.2015.08.031}{\emph{Comput. Phys. Commun.}
  {\bfseries 197} (2015) 312}
  [\href{https://arxiv.org/abs/1402.1178}{{\ttfamily 1402.1178}}].

\bibitem{Berthier:2016tkq}
L.~Berthier, M.~Bj\o{}rn and M.~Trott, \emph{{Incorporating doubly resonant
  $W^\pm$ data in a global fit of SMEFT parameters to lift flat directions}},
  \href{https://doi.org/10.1007/JHEP09(2016)157}{\emph{JHEP} {\bfseries 09}
  (2016) 157} [\href{https://arxiv.org/abs/1606.06693}{{\ttfamily
  1606.06693}}].

\bibitem{Durieux:2019lnv}
F.~Maltoni et~al., \emph{{Proposal for the validation of Monte Carlo
  implementations of the standard model effective field theory}},
  \href{https://arxiv.org/abs/1906.12310}{{\ttfamily 1906.12310}}.

\bibitem{Corbett:2020bqv}
T.~Corbett, \emph{{The Feynman rules for the SMEFT in the background field
  gauge}}, \href{https://doi.org/10.1007/JHEP03(2021)001}{\emph{JHEP}
  {\bfseries 03} (2021) 001}
  [\href{https://arxiv.org/abs/2010.15852}{{\ttfamily 2010.15852}}].

\bibitem{CMS:2023xyc}
{\scshape CMS} collaboration, \emph{{Search for physics beyond the standard
  model in top quark production with additional leptons in the context of
  effective field theory}},
  \href{https://doi.org/10.1007/JHEP12(2023)068}{\emph{JHEP} {\bfseries 12}
  (2023) 068} [\href{https://arxiv.org/abs/2307.15761}{{\ttfamily
  2307.15761}}].

\bibitem{Breso-Pla:2021qoe}
V.~Bres\'o-Pla, A.~Falkowski and M.~Gonz\'alez-Alonso, \emph{{A$_{FB}$ in the
  SMEFT: precision Z physics at the LHC}},
  \href{https://doi.org/10.1007/JHEP08(2021)021}{\emph{JHEP} {\bfseries 08}
  (2021) 021} [\href{https://arxiv.org/abs/2103.12074}{{\ttfamily
  2103.12074}}].

\bibitem{Giani:2023gfq}
T.~Giani, G.~Magni and J.~Rojo, \emph{{SMEFiT: a flexible toolbox for global
  interpretations of particle physics data with effective field theories}},
  \href{https://doi.org/10.1140/epjc/s10052-023-11534-7}{\emph{Eur. Phys. J. C}
  {\bfseries 83} (2023) 393}
  [\href{https://arxiv.org/abs/2302.06660}{{\ttfamily 2302.06660}}].

\bibitem{ALEPH:2013dgf}
{\scshape ALEPH, DELPHI, L3, OPAL, LEP Electroweak} collaboration,
  \emph{{Electroweak Measurements in Electron-Positron Collisions at
  W-Boson-Pair Energies at LEP}},
  \href{https://doi.org/10.1016/j.physrep.2013.07.004}{\emph{Phys. Rept.}
  {\bfseries 532} (2013) 119}
  [\href{https://arxiv.org/abs/1302.3415}{{\ttfamily 1302.3415}}].

\bibitem{D0:2013jba}
{\scshape D0} collaboration, \emph{{Measurement of the $W$ boson mass with the
  D0 detector}}, \href{https://doi.org/10.1103/PhysRevD.89.012005}{\emph{Phys.
  Rev. D} {\bfseries 89} (2014) 012005}
  [\href{https://arxiv.org/abs/1310.8628}{{\ttfamily 1310.8628}}].

\bibitem{LHCb:2021bjt}
{\scshape LHCb} collaboration, \emph{{Measurement of the W boson mass}},
  \href{https://doi.org/10.1007/JHEP01(2022)036}{\emph{JHEP} {\bfseries 01}
  (2022) 036} [\href{https://arxiv.org/abs/2109.01113}{{\ttfamily
  2109.01113}}].

\bibitem{ATLAS:2017rzl}
{\scshape ATLAS} collaboration, \emph{{Measurement of the $W$-boson mass in pp
  collisions at $\sqrt{s}=7$ TeV with the ATLAS detector}},
  \href{https://doi.org/10.1140/epjc/s10052-017-5475-4}{\emph{Eur. Phys. J. C}
  {\bfseries 78} (2018) 110}
  [\href{https://arxiv.org/abs/1701.07240}{{\ttfamily 1701.07240}}].

\bibitem{LHC-TeVMWWorkingGroup:2023zkn}
{\scshape LHC-TeV~MW~Working~Group} collaboration, \emph{{Compatibility and
  combination of world W-boson mass measurements}},
  \href{https://doi.org/10.1140/epjc/s10052-024-12532-z}{\emph{Eur. Phys. J. C}
  {\bfseries 84} (2024) 451}
  [\href{https://arxiv.org/abs/2308.09417}{{\ttfamily 2308.09417}}].

\bibitem{Bjorn:2016zlr}
M.~Bj\o{}rn and M.~Trott, \emph{{Interpreting $W$ mass measurements in the
  SMEFT}}, \href{https://doi.org/10.1016/j.physletb.2016.10.003}{\emph{Phys.
  Lett. B} {\bfseries 762} (2016) 426}
  [\href{https://arxiv.org/abs/1606.06502}{{\ttfamily 1606.06502}}].

\bibitem{CMS:2011utm}
{\scshape CMS} collaboration, \emph{{Measurement of the weak mixing angle with
  the Drell-Yan process in proton-proton collisions at the LHC}},
  \href{https://doi.org/10.1103/PhysRevD.84.112002}{\emph{Phys. Rev. D}
  {\bfseries 84} (2011) 112002}
  [\href{https://arxiv.org/abs/1110.2682}{{\ttfamily 1110.2682}}].

\bibitem{Alioli:2010xd}
S.~Alioli, P.~Nason, C.~Oleari and E.~Re, \emph{{A general framework for
  implementing NLO calculations in shower Monte Carlo programs: the POWHEG
  BOX}}, \href{https://doi.org/10.1007/JHEP06(2010)043}{\emph{JHEP} {\bfseries
  06} (2010) 043} [\href{https://arxiv.org/abs/1002.2581}{{\ttfamily
  1002.2581}}].

\bibitem{Barze:2012tt}
L.~Barze, G.~Montagna, P.~Nason, O.~Nicrosini and F.~Piccinini,
  \emph{{Implementation of electroweak corrections in the POWHEG BOX: single W
  production}}, \href{https://doi.org/10.1007/JHEP04(2012)037}{\emph{JHEP}
  {\bfseries 04} (2012) 037} [\href{https://arxiv.org/abs/1202.0465}{{\ttfamily
  1202.0465}}].

\bibitem{NNPDF:2017mvq}
{\scshape NNPDF} collaboration, \emph{{Parton distributions from high-precision
  collider data}},
  \href{https://doi.org/10.1140/epjc/s10052-017-5199-5}{\emph{Eur. Phys. J. C}
  {\bfseries 77} (2017) 663}
  [\href{https://arxiv.org/abs/1706.00428}{{\ttfamily 1706.00428}}].

\bibitem{Paukkunen:2014zia}
H.~Paukkunen and P.~Zurita, \emph{{PDF reweighting in the Hessian matrix
  approach}}, \href{https://doi.org/10.1007/JHEP12(2014)100}{\emph{JHEP}
  {\bfseries 12} (2014) 100} [\href{https://arxiv.org/abs/1402.6623}{{\ttfamily
  1402.6623}}].

\bibitem{Carrazza:2019sec}
S.~Carrazza, C.~Degrande, S.~Iranipour, J.~Rojo and M.~Ubiali, \emph{{Can New
  Physics hide inside the proton?}},
  \href{https://doi.org/10.1103/PhysRevLett.123.132001}{\emph{Phys. Rev. Lett.}
  {\bfseries 123} (2019) 132001}
  [\href{https://arxiv.org/abs/1905.05215}{{\ttfamily 1905.05215}}].

\bibitem{Hammou:2023heg}
E.~Hammou, Z.~Kassabov, M.~Madigan, M.L.~Mangano, L.~Mantani, J.~Moore et~al.,
  \emph{{Hide and seek: how PDFs can conceal new physics}},
  \href{https://doi.org/10.1007/JHEP11(2023)090}{\emph{JHEP} {\bfseries 11}
  (2023) 090} [\href{https://arxiv.org/abs/2307.10370}{{\ttfamily
  2307.10370}}].

\bibitem{Gao:2022srd}
J.~Gao, M.~Gao, T.J.~Hobbs, D.~Liu and X.~Shen, \emph{{Simultaneous CTEQ-TEA
  extraction of PDFs and SMEFT parameters from jet and $ t\overline{t} $
  data}}, \href{https://doi.org/10.1007/JHEP05(2023)003}{\emph{JHEP} {\bfseries
  05} (2023) 003} [\href{https://arxiv.org/abs/2211.01094}{{\ttfamily
  2211.01094}}].

\bibitem{Costantini:2024xae}
{\scshape PBSP} collaboration, \emph{{SIMUnet: an open-source tool for
  simultaneous global fits of EFT Wilson coefficients and PDFs}},
  \href{https://doi.org/10.1140/epjc/s10052-024-13079-9}{\emph{Eur. Phys. J. C}
  {\bfseries 84} (2024) 805}
  [\href{https://arxiv.org/abs/2402.03308}{{\ttfamily 2402.03308}}].

\bibitem{ATLAS:2018gqq}
{\scshape ATLAS} collaboration, \emph{{Measurement of the effective leptonic
  weak mixing angle using electron and muon pairs from $Z$-boson decay in the
  ATLAS experiment at $\sqrt s = 8$ TeV}},  {ATLAS CONF Note}
  \href{{https://cds.cern.ch/record/2630340/}}{ATLAS-CONF-2018-037} (2018).

\bibitem{ATLAS:2020jwz}
{\scshape ATLAS} collaboration, \emph{{Measurement of the associated production
  of a Higgs boson decaying into $b$-quarks with a vector boson at high
  transverse momentum in $pp$ collisions at $\sqrt{s} = 13$ TeV with the ATLAS
  detector}}, \href{https://doi.org/10.1016/j.physletb.2021.136204}{\emph{Phys.
  Lett. B} {\bfseries 816} (2021) 136204}
  [\href{https://arxiv.org/abs/2008.02508}{{\ttfamily 2008.02508}}].

\end{thebibliography}\endgroup
\newpage
\appendix
\section{How to run the code}
\label{sec:howto}

The code is available for download at:\\
\href{https://github.com/ewpd4lhc/ewpd4lhc}{https://github.com/ewpd4lhc/ewpd4lhc}

It requires \texttt{python3} with the \texttt{numpy} and \texttt{yaml} modules.
A \texttt{ROOT} installation with \texttt{python} bindings is required for \texttt{Roofit} output and workspace manipulation.
On CERN LXPLUS, no preparation is required as all of these requirements are met.

In the configuration file \texttt{config/ewpd4lhc.cfg} one can modify:
\begin{itemize}
	\item The list of observables included in the likelihood.
	\item The input scheme for calculations (see Table~\ref{tab:schemes}).
	\item The treatment of theoretical and parametric uncertainties (either as part of the covariance of the multivariate Gaussian or as nuisance parameters, see Section~\ref{sec:uncertainties})
	\item The SMEFT symmetry assumption (see also Table~\ref{tab:schemes}).
	\item Optionally, a subset of Wilson coefficients to be included.
	\item The source of dimension-six linear, dimension-six squared, and dimension-eight parametrizations (see also Table~\ref{tab:schemes}).
\end{itemize}
Measurement data, which may be adapted by the user, is stored as yaml files in the \texttt{data} folder with multiple alternative SMEFT parametrizations stored in the same location.

A \texttt{yaml} output file describing the SMEFT EWPD likelihood can be created by running the main executable:
\begin{verbatim}
./ewpd4lhc.py
\end{verbatim}
where by default \texttt{input/ewpo.cfg} is taken as input and the SMEFT likelihood is stored in a textfile name \texttt{ewpd\_out.yml}.
Alternative predefined configuration files can be found in the same folder.
The output describes the multivariate Gaussian model:
The predicted values of all observables, the total uncertainty -- possibly including theory and parametrization as sources of uncertainties -- and correlation, the Wilson coefficient dependence, and possibly the dependence of SM predictions on input parameters as well as theory nuisance parameters.
During the execution of the tool, SM and SMEFT fits are performed and the results printed.

Optional arguments can be listed with:
\begin{verbatim}
./ewpd4lhc.py --help
\end{verbatim}
For example, a ROOT workspace can be created either directly:
\begin{verbatim}
./ewpd4lhc.py --root_output ROOTFILENAME.root
\end{verbatim}
Or in a seperate step from the output textfile (allowing the user to modify or build more complex likelihoods) with:
\begin{verbatim}
./ewpd4lhc.py --output YAMLFILE.yml
./yaml2root.py --input YAMLFILE.yml --output ROOTFILENAME.root
\end{verbatim}

A simple script for fitting the \texttt{ROOT} output is part of the code, too.
Workspace contents are printed with:
\begin{verbatim}
./ROOTfit/fit.py --input ROOTFILENAME.root
\end{verbatim}
A fit of one or multiple parameters of interest is performed, e.g., with:
\begin{verbatim}
./ROOTfit/fit.py --input ROOTFILENAME.root --pois=cHu,cHd,cHj3,cHj1
\end{verbatim}
It is possible to specify \texttt{----poi=all} but this will usually not converge without extra constraints as the EWPD likelihood is degenerate.
One- and two-dimensional scans of the likelihood can be performed with the following commands, with scan points being stored in a textfile and optionally being plotted as \texttt{pdf} graphics.  Wilson coefficients other than the POIs can be required to ``float'' in the fit, in which case they are set to the value that maximized the likelihood at each scan point.
\begin{verbatim}
./ROOTfit/fit.py --input ROOTFILENAME.root --pois=cHu \
    --scan=-0.1:0.1 --outfolder=1Dscan --plot
./ROOTfit/fit.py --input ROOTFILENAME.root --pois=cHu \
    --scan=-0.5:0.5 --float=cHj3 --outfolder=1DscanProfiled --plot
./ROOTfit/fit.py --input ROOTFILENAME.root --pois=cHj3,cHu \
    --scan=-0.08:0.08,-0.3:0.3 --outfolder=2Dscan --plot --npoints=300
\end{verbatim}
\begin{figure}
	\includegraphics[width=0.5\textwidth]{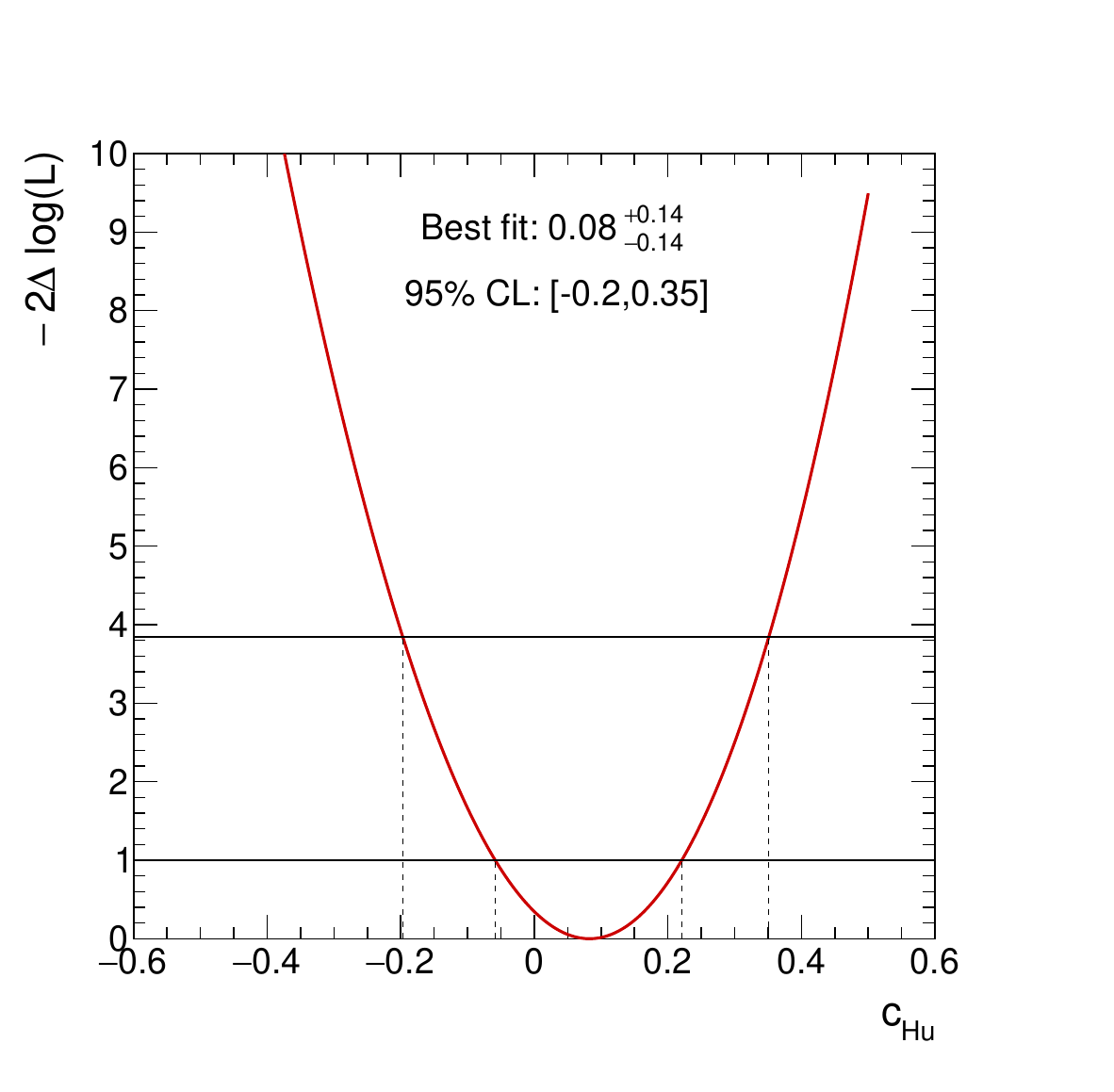}%
	\includegraphics[width=0.5\textwidth]{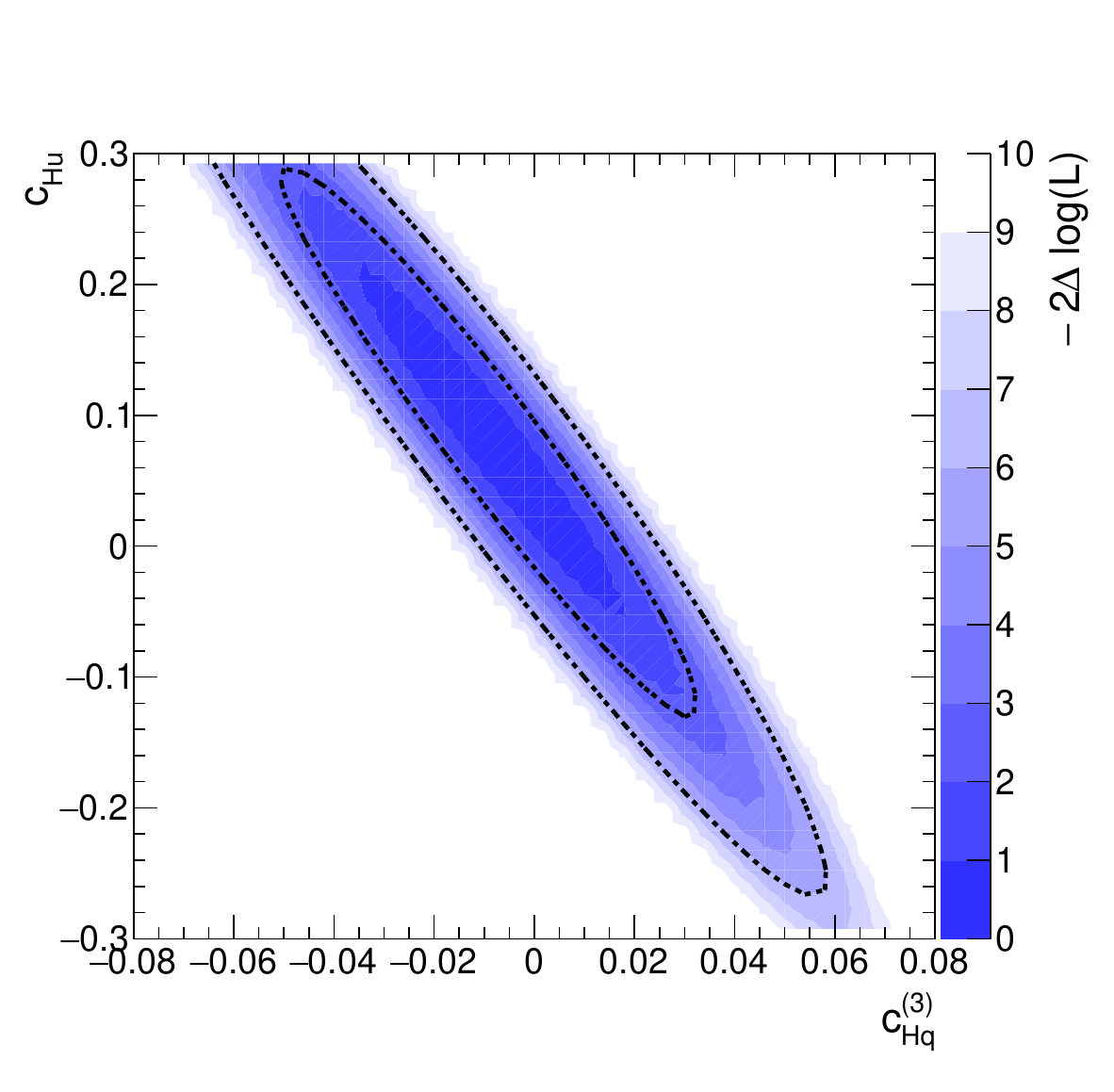}
	\caption{Examples of likelihood scans performed with \ewpdlhc. In the left plot, \cHu is varied and \cHjthree is left floating in the fit. In the right plot, both parameters are scanned. The corresponding commands can be found in the main text.}
	\label{fig:scans}
\end{figure}
The \texttt{pdf} output of the second and third command is shown in Figure~\ref{fig:scans}.
It can, for example, be compared to Figure~18 of the auxilary material of Ref.~\cite{ATLAS:2020jwz}, showing a complementarity between LHC Higgs measurements and EWPD -- that is however best explored by combining the \ewpdlhc workspace with the ATLAS workspace.

The low level classes like \texttt{SMcalculator}, \texttt{SMEFTlikelihood}, and  \texttt{LINAfit} can also be used directly in \texttt{python}. For example:
\begin{verbatim}import SMcalculator
sm=SMcalculator.EWPOcalculator(MH=125.25,
                               mt=172.69,
                               alphas=0.118,
                               MZ=91.1875,
                               MW=80.377)
print('AFBb:',sm.AFBb())
sm.update(MW=80.3)
print('AFBb(MW=80.3):',sm.AFBb())
sm.reset()
print('Also AFBb:',sm.get('AFBb'))
print('dAFBb/dMW:',sm.derivative('AFBb','MW'))
print('All observables:', sm.getall())
\end{verbatim}

Finally,  \texttt{MG5\_aMC}-based parametrizations can be generated using the code in the \texttt{ParaFactory} subfolder.

\end{document}